\documentclass[12pt,a4paper]{article}
\usepackage{jheppub}
\usepackage{bm}
\usepackage{ifsym}
\usepackage{amsthm}
\usepackage{amsfonts}
\usepackage{bbm}
\usepackage{ccaption}
\usepackage{mathrsfs}
\usepackage{booktabs}
\usepackage{color}
\usepackage{wrapfig}
\usepackage[labelfont={sf}]{caption}
\usepackage{multicol}
\usepackage{multirow}
\usepackage{graphics}
\usepackage{epstopdf}
\usepackage{rotating}

\newcommand{\hs}[1]{\hspace{#1 cm}}		\newcommand{\vs}[1]{\vspace{#1 cm}}
\newcommand{\beq}{\begin{equation}}		\newcommand{\enq}{\end{equation}}
\renewcommand{\(}{\left(}			\renewcommand{\)}{\right)}
\renewcommand{\[}{\left[}			\renewcommand{\]}{\right]}

\newcommand{\mc}[1]{\mathcal{#1}}

\newcommand{\parcial}[2][]{\frac{\partial #1}{\partial #2}}
\newcommand{\deriv}[2][]{\frac{\dd #1}{\dd #2}}

\newcommand{\riga}{\rightarrow}

\newcommand{\dd}{\mathrm{d}}		\newcommand{\nab}{\nabla}
\newcommand{\la}{\langle}		\newcommand{\ra}{\rangle}

\newcommand{\R}{\mathbb{R}}

\newcommand{\E}{\mathbb{E}}

  \captionnamefont{\bfseries}
  \captiontitlefont{\small\sffamily}
  \captiondelim{: }
  \hangcaption
  \renewcommand{\figurename}{Fig.}

\def\clock{{\count0=\time
           \divide\count0 60
           \ifnum\count0<10 0\fi\the\count0
           \multiply\count0 -60 \advance\count0 \time
           :\ifnum\count0<10 0\fi \the\count0
         }}
\newcommand{\timestamp}{{\small\vbox{\hbox{\tt\jobname.tex}
\hbox{\the\day/\the\month/\the\year, \clock}}}}

\usepackage{tocloft}

\definecolor{rust}{rgb}{0.8,0.2,0.2}
\definecolor{green}{rgb}{0.1,0.8,0.2}

\numberwithin{equation}{section}
\numberwithin{figure}{section}
\title{Critical Kaluza-Klein black holes and black strings in D = 10}

\author{Biel Cardona}
\author{and Pau Figueras}

\affiliation[]{School of Mathematical Sciences, \\
Queen Mary University of London, \\
Mile End Road, London E1 4NS, UK.}

\emailAdd{c.gabriel@qmul.ac.uk}
\emailAdd{p.figueras@qmul.ac.uk}

\abstract{We construct static vacuum localized black holes and non-uniform black strings in ten spacetime dimensions, where one of the dimension is compactified on a circle. We study the phase diagram of black objects with these boundary conditions, especially near the critical point where localized black holes and non-uniform black strings merge. Remarkably, we find that the merger happens at a cusp in the phase diagram. We verify that the critical geometry is controlled by a Ricci-flat double-cone as previously predicted. However, unlike the lower dimensional cases, we find that physical quantities approach to their critical values according to a power law plus a logarithmic correction. We extract the critical exponents and find very good agreement with the predictions from the double-cone geometry. According to holography, localized black holes and black strings are dual to thermal states of ($1+1$)-dimensional SU($N$) maximal Super-Yang Mills theory compactified on a circle; we recover and extend the details of the (recently found) 1st order phase transition in this system from the gravity side.}


\begin{document}

\maketitle

\flushbottom

\newpage

%~~~~~~~~~~~~~~~~~~~~~~~~~~~~~~~~~~~~~~~~~~~~~~~
\section{Introduction and results}
\label{intro}
%~~~~~~~~~~~~~~~~~~~~~~~~~~~~~~~~~~~~~~~~~~~~~~
In $D = 4$ spacetime dimensions, stationary asymptotically flat black hole solutions of the Einstein equation in vacuum of a given mass have spherical topology, are presumably unique and all the evidence suggests that they are dynamically stable. However, in $D>4$ these properties change radically. The physics of black objects turns out to be much richer, allowing for non-spherical topologies, instabilities and non-uniqueness (and thus phase transitions). Black holes are fundamental objects in general relativity. In recent years the study of such objects in non-astrophysical settings has received much attention due to the intrinsic interest in understanding fundamental aspects of gravity as described by general relativity (see \cite{Emparan:2008eg} for a review), and also because of the connections to string theory and the gauge/gravity duality \cite{Maldacena:1997re,Gubser:1998bc,Witten:1998qj,Aharony:1999ti,PhysRevD.58.046004}. In the latter context, it is natural to consider spacetimes that are asymptotic to $M_d\times N_n$, where $M_d$ is $d$-dimensional Minkwoski or Anti-de Sitter space and $N_n$ is an $n$-dimensional compact manifold so that the total number of spacetime dimensions is $D = d + n > 4$.

One of the most extensively studied models in this setting is that of $M_d = \text{Mink}_d$ and $N_1 = S^1$, a circle of length $L$. Since the compact dimension is flat, it is trivial to write down a black hole solution that is uniformly wrapped along the compact dimension: This is just given by the $(D-1)$-dimensional Schwarzschild solution times a (compact) flat direction, $\text{Schw}_{D-1}\times S^1$. Such a higher dimensional black hole is known as the uniform black string (UBS). In \cite{Gregory:1993vy}, Gregory and Laflamme (GL) famously showed that thin enough black strings are unstable under linear gravitational perturbations with a non-trivial dependence along the $S^1$-direction.\footnote{More precisely, the condition for the existence of a linear instability is that $r_0/L\lesssim \mc{O}(1)$, where $r_0$ is the mass parameter of the parent $\text{Schw}_{D-1}$ solution and $L$ is the asymptotic length of the compact circle.} Determining the endpoint of such an instability has been the subject of intense studies during the past few years.

At the onset of the instability, the linear GL mode is time-independent (i.e., a zero mode) and can be continued to the non-linear regime. This indicates that there exists a new branch of black strings which are non-uniform in the compact direction and are thus known as non-uniform black strings (NUBS). NUBS were first constructed perturbatively in $D=5$ by \cite{Gubser:2001ac}, subsequently constructed fully non-linearly in various spacetime dimensions using numerical methods \cite{Wiseman:2002zc,Sorkin:2006wp,Kleihaus:2006ee,Headrick:2009pv,Figueras:2012xj,Kalisch:2015via,Kalisch:2016fkm} and, more recently, using the large-$D$ expansion \cite{Emparan:2018bmi}. It turns out that in $D<D^* = 13$(.5), NUBS have less entropy than UBS with the same mass and hence they cannot be the endpoint of the GL instability \cite{Sorkin:2004qq}. In fact, based on entropic arguments, \cite{Gregory:1993vy} conjectured that unstable UBS would evolve into an array of localized black holes through a dynamical topology change transition; the latter can only happen through a singularity and hence the evolution of the GL instability of black strings could potentially constitute a counter-example of the weak cosmic censorship conjecture \cite{Penrose:1969pc,Christodoulou:1999ve} around such spacetimes. This scenario was recently confirmed by \cite{Lehner:2010pn,Lehner:2011wc}, using numerical relativity techniques.\footnote{Notice that this final fate is not exclusive of UBS and black holes with compact extra dimensions. Fully non-linear time evolutions of analogous instabilities in asymptotically flat black rings or ultra-spinning Myers-Perry black holes spacetimes,  similarly lead to violations of the weak cosmic censorship conjecture \cite{Figueras:2015hkb,Figueras:2017zwa}.}  On the other hand, for $D>D^*$, NUBS can be dynamically stable and hence be the endpoint of the GL instability, as \cite{Emparan:2015gva,Emparan:2018bmi} confirmed.

Apart from UBS and NUBS, spaces that are asymptotically Mink$_d\times S^1$ also admit static black hole solutions that are localized on the $S^1$. These localized black holes (LOC) have been constructed numerically \cite{Kudoh:2003ki,Sorkin:2003ka,Headrick:2009pv,Kalisch:2017bin} and perturbatively in the limit in which the black holes are small  compared to $L $ \cite{Harmark:2003yz,Gorbonos:2004uc,Karasik:2004ds,Gorbonos:2005px,Chu:2006ce,Kol:2007rx}. Motivated by geometrical considerations, \cite{Kol:2002xz} conjectured that the NUBS and LOC branches should merge at a topology changing critical solution governed by a Ricci-flat double-cone. This conjecture was tested from the black string side in \cite{Kol:2003ja}, and later in various dimensions in \cite{Figueras:2012xj}. However, the most non-uniform black strings in these early constructions were still too far from the critical regime to provide conclusive results (see however \cite{Kleihaus:2006ee}). Only recently, Kalisch et al.\ \cite{Kalisch:2016fkm,Kalisch:2017bin}, in an impressive numerical construction, have managed to obtain NUBS and LOC in $D = 5,6$ extremely close to the critical point, confirming the double-cone predictions to an unprecedented level of detail. 
 
The goal of the present work is to construct NUBS and LOC  in $D = 10$ very close to the critical point, where these branches of black holes merge. Critical solutions have only been previously constructed in $D = 5, 6$ \cite{Kalisch:2016fkm,Kalisch:2017bin}; for higher values of $D$, gravity becomes more localized near the horizon of the black object, which makes the numerical construction more challenging, especially very close to the critical point. Note that \cite{Dias:2017uyv} previously constructed both NUBS and LOC in $D = 10$, but their solutions were very far from the critical regime since the aim of that paper was different (see below).

\begin{figure}[h!]
\centering
\includegraphics[height = 5.4cm]{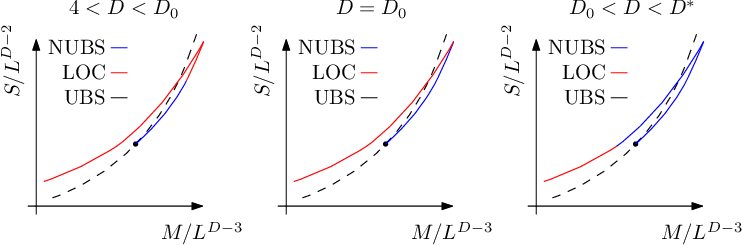}
\captionsetup{width=.9\linewidth}
\captionof{figure}{Schematic phase diagrams in the microcanonical ensemble for various $D$'s. Here $D_0 = 10$ is the critical dimension of the double-cone geometry \cite{Kol:2002xz} and $D^\ast = 13(.5)$ is the critical dimension in the microcanonical ensemble  \cite{Sorkin:2004qq}.}
\label{fig:conj}
\end{figure}

At the critical dimension $D^*$ the dynamical stability of weakly NUBS changes from being unstable for $D<D^*$ to being stable for $D>D^*$. However, for $D=12, 13$ \cite{Figueras:2012xj} found that NUBS with a sufficiently large non-uniformity can also be dynamically stable.\footnote{It is plausible that stable NUBS also exist in $D=11$ for larger values of the non-uniformity parameter than in \cite{Figueras:2012xj}.} This paper showed that there exists a turning point, i.e.\ a maximum of the mass/area, along the NUBS branch where the stability properties of the solutions change. On the other hand, in $D = 5,6$ such a turning point is present along the LOC branch. In $D = 10$, as we move along both the LOC and the NUBS branches and approach the critical solution from both sides, we do not find any turning points on either of the branches. Therefore, the simplest picture that emerges from our work is that, sufficiently far from the critical solution, in $D<10$ there should exist a turning point along the LOC branch, in $D>10$ the turning point occurs along the NUBS, and in $D=10$ there are no turning points at all. See Fig.\ \ref{fig:conj}. Recently, \cite{Emparan:2018bmi} confirmed the existence turning points in the phase diagram of NUBS in $D<14$ using the large-$D$ expansion, but the reliability of their approach breaks down at around $D\approx 9$. Notice, however, that the methods of \cite{Emparan:2018bmi} did not allow them to study critical solutions in detail and therefore our results complement theirs.

Our numerical data suggests that in $D=10$ the merger happens precisely at a cusp in the phase diagram. The study of the critical geometry in \cite{Kol:2002xz} showed that $D=10$ is the critical dimension of the cone geometry that governs the topology change. For $D<10$, the approach of physical quantities to their critical values is controlled by a (dimension-dependent) power law with infinitely many oscillations (i.e.\ turning points); this behavior has been beautifully confirmed in \cite{Kalisch:2017bin} for $D=5,6$. On the other hand, for $D>10$ the approach to the critical point should be given by two independent power laws, with no oscillations. $D=10$ is the marginal case and the approach to the critical point should be controlled by a power law with a logarithmic correction. In this paper we confirm this in $D=10$. 

In this paper we also compute the spectrum of negative modes of the Lichnerowicz operator, $\Delta_L$, around the LOC and NUBS branches, restricted to modes that preserve the isometries of the background. Just as in \cite{Headrick:2009pv,Figueras:2012xj}, we find that NUBS posses two negative modes: one is continuously connected (as the non-uniformity parameter goes to zero) to the negative mode of the parent Schwarzschild black hole \cite{Gross:1982cv}. This mode diverges as the NUBS approach the critical solution. The other negative mode is the continuation of the GL zero mode to non-zero values of the non-uniformity parameter and our data suggests that it tends to a finite value at the critical solution.  On the other hand, LOC have a single negative mode throughout the branch and it approaches the same finite value as the NUBS at the critical solution. See Fig.\ \ref{fig:negmodes}. Note that at an extremum of the temperature one can have zero modes corresponding to variations of the parameters of the solution that respect the boundary conditions (i.e.\ preserve the temperature). We do not find any evidence for new zero modes, which is consistent with the absence of extrema of the temperature along either branch of solutions.

Another motivation for the present work comes from the gauge/gravity duality \cite{Maldacena:1997re,Gubser:1998bc,Witten:1998qj,Aharony:1999ti,PhysRevD.58.046004}. The best well-understood example of this correspondence  is between maximally supersymmetric Yang-Mills (SYM) theory in $p+1$ dimensions and gauge group SU($N$), and Type IIA (even $p$) or Type IIB (odd $p$) superstring theory containing $N$ coincident D$p$-branes in the decoupling limit. For $p = 1$ the duality is between 2-dimensional SU($N$) SYM theory and type IIA or IIB string theory in the presence of D0- or D1-branes respectively \cite{PhysRevD.58.046004}. At large $N$, strong coupling and finite temperature, the gauge theory is described by black hole solutions with D0- or D1-charge in the supergravity approximation, depending on the temperature (type IIA at low temperatures and type IIB at high temperatures respectively). In this paper we are interested in vacuum black hole solutions of the Einstein equation in 10 spacetime dimensions, one of which is compactified on a circle of length $L$. After a series of standard U-duality transformations \cite{Dias:2017uyv,1126-6708-2002-05-032,Harmark:2004ws,Aharony:2004ig}, these vacuum black holes can be given D0- or D1-brane charges.

According to the AdS/CFT correspondence, the black hole phase structure should be reproduced by the thermal phases of SYM on a circle at strong coupling and large $N$. Lattice simulations of SYM on a torus, with one of the circles being the thermal circle and with periodic boundary conditions for the fermions on the other spatial circle, have been performed. Most of the previous works in the past have focused on the $p=0$ SYM quantum mechanics and agreement with the gravity predictions has been confirmed \cite{Hanada:2007ti,Catterall:2007fp,Anagnostopoulos:2007fw,Catterall:2008yz,Hanada:2008gy,Hanada:2008ez,Kadoh:2015mka,Filev:2015hia,Filev:2015cmz,Hanada:2016zxj,Berkowitz:2016tyy,Berkowitz:2016jlq,Asano:2016xsf,Rinaldi:2017mjl}. The case $p=1$ has received less attention in the past \cite{Catterall:2010fx}, until the recent of work of \cite{Catterall:2017lub}. This paper predicted the temperature at which a first order phase transition occurs from lattice simulations, in a regime where the latter should overlap with the supergravity calculations. The latter was only recently computed in \cite{Dias:2017uyv} and found very good agreement with the lattice result. In this paper, as a by-product of our calculations, we recompute the value of the phase transition temperature (or energy, in the canonical ensemble); the values that we obtain are $t_{\text{crit}} = 1.09257\hs{0.05}t_{\text{GL}}$ and $\varepsilon_{\text{crit}} = 1.24181\hs{0.05}\varepsilon_{\text{GL}}$ for the temperature and energy at the phase transition measured with respect to the GL point. Our values differ with those found in \cite{Dias:2017uyv} by less than a $0.25\%$. In addition, we have been able to locate the merger between the non-uniform and localized phases.

The rest of this paper is organized as follows. In \S\ref{ubs} we start by reviewing some general aspects of black holes in Kaluza-Klein spaces. In subsections \ref{nubs} and \ref{loc} we present our numerical construction of NUBS and LOC, respectively. In \S\ref{results} we present our results.  \S\ref{thermo0} contains the phase diagrams in the microcanonical and canonical ensembles, in \S\ref{geo} we consider the horizon geometry and in \S\ref{crit} we study in the detail the critical behavior of NUBS and LOC near the critical point and compare it with the predictions of the double-cone model. \S\ref{spec} is devoted to the computation of the spectrum of negative modes of the Lichnerowicz operator around the NUBS and LOC. \S\ref{sym} contains the results for the phase diagram of the supergravity solutions with D0-charge. We close the paper with a discussion in \S\ref{disc}. Some technical details are relegated to the Appendices. In appendix \ref{intdom} we give more details about the integration domain that we have used to construct the localized black holes and in appendix \ref{convtest} we present some convergence tests. The mapping from neutral solutions to charged ones is presented in detail in appendix \ref{d0charge}.

{\bf \textsf{Note added:}} while this paper was nearing completion, we became aware of \cite{Ammon:2018sin}, that has some overlap with ours and that has appeared on the arXiv on the same date.

%~~~~~~~~~~~~~~~~~~~~~~~~~~~~~~~~~~~~~~~~~~~~~~~
\section{Black objects in Kaluza-Klein theory}
\label{numconst}
%~~~~~~~~~~~~~~~~~~~~~~~~~~~~~~~~~~~~~~~~~~~~~~
Consider vacuum Einstein's gravity in $D=10$ spacetime dimensions with Kaluza-Klein (KK) asymptotic boundary conditions, i.e.\ $\text{Mink}_{9}\times S^1$. For (ultra)static spacetimes, this theory contains three different families of static black holes, namely, UBS, NUBS and LOC. After fixing the overall scale by fixing the length of the asymptotic $S^1$, these three different types of black holes can be parametrized by the temperature and one may distinguish them by the topology of the horizon and the isometries. Whilst UBS are translationally invariant along the $S^1$ and are known explicitly, for NUBS and LOC the translation invariance along the $S^1$ is broken and they have to be constructed numerically (or pertubatively). In this section we explain our numerical construction of such solutions. Since we are interested in studying the thermal phases, we will be working with the Euclidean form of the solutions where the Euclidean time $\tau$ is periodic, $\tau\sim \tau+\beta$, with $\beta$ being the inverse temperature.\footnote{Note that since we are considering static spacetimes, we can change to Lorentzian signature by a trivial change of coordinates $\tau \to \textrm{i}\,t$.}

%~~~~~~~~~~~~~~~~~~~~~~~~~~~~~~~~~~~~~~~~~~~~~~~
\subsection{Generic results and Uniform black strings}
\label{ubs}
%~~~~~~~~~~~~~~~~~~~~~~~~~~~~~~~~~~~~~~~~~~~~~~
In this paper we are interested in Einstein metrics that asymptote to the flat Euclidean metric, where one of the directions corresponds the Euclidean time $\tau$, times a KK circle of length $L$. As usual, the Euclidean time $\tau$ is compact and has period $\beta$. Ultimately we will consider ten dimensional spaces but for now we shall keep the total number of spacetime dimensions $D$ general.  Moreover, we will only consider spacetimes that preserve an SO$(D-2)$ subgroup of the full rotation group of the flat Euclidean metric in $D-1$ dimensions. Therefore the asymptotic isometry group of the spaces that we shall consider is U$(1)_\beta \times $SO$(D-2)\times$U$(1)_L$, which is made explicit in the asymptotic form of the flat metric on the product space $S^1_\beta \times \R^{D-2} \times S^1_L$,
\begin{equation}
\dd s^2 = \dd \tau^2 + \dd r^2 + r^2\,\dd\Omega_{D-3}^2+ \dd y^2\,,
\end{equation}
with $\tau\sim \tau + \beta$ and $y\sim y+L$. For more general spaces, from the asymptotic behavior of the metric components
\beq\label{gttgyy}
g_{\tau\tau} \simeq 1 - \frac{C_\tau}{r^{D-4}}, \hs{0.75}  g_{yy} \simeq 1+\frac{C_y}{r^{D-4}}\,,
\enq
one can extract two asymptotic charges, namely mass and tension, of the solution \cite{Harmark:2003dg}:\footnote{Throughout, we use units of $G_D = 1$, where $G_D$ is the Newton's constant in $D$ spacetime dimensions.} 
\beq\label{masstension}
M = \frac{\Omega_{D-3}L}{16\pi}\big((D-3)C_\tau - C_y\big), \hs{0.75} \mc{T} = \frac{\Omega_{D-3}}{16\pi}\big(C_\tau - (D-3)C_y\big).
\enq
From these quantities one can define the relative tension $n = \mc{T}L/M$, which is bounded: $n \leq 0 \leq D-3$. In addition to these charges, NUBS and LOC can be characterized using their own geometric quantities which are discussed in \S\ref{geo}. All neutral KK solutions with a single connected horizon have temperature $T = \kappa/(2\pi)$ and entropy $S = A_H/4$; they satisfy the 1st law of thermodynamics, $\dd M = T\dd S + \mc{T}\dd L$, and the Smarr's relation, $(D-3-n)M=(D-2)TS$. From the point of view of the numerics, the latter may be used as a consistency check, since the entropy and the mass are obtained from the metric. The free energy is given by $F = M-TS$.

UBS are known explicitly for all values of $D$: The metric is $\text{Schw}^E_{D-1}\times S^1$ ($E$ stands for Euclidean), 
\beq\label{unibs}
\dd s^2 = \(1-\frac{r_0^{D-4}}{r^{D-4}}\)\dd\tau^2 + \(1-\frac{r_0^{D-4}}{r^{D-4}}\)^{-1}\dd r^2 + r^2\dd\Omega_{D-3} + \dd y^2,
\enq
where $(\tau,r,\Omega_{D-3})$ are the usual $D-1$ Schwarzschild coordinates and $y$ is the $S^1$ coordinate.  The parameter $r_0$ labels each solution and it is directly related to the physical quantities: \beq\begin{split}
\kappa &= \frac{D-4}{2r_0}, \hs{1.57} M = \frac{\Omega_{D-3}L}{16\pi}(D-3)r_0^{D-4}, \\
 A_H &= Lr_0^{D-3}\Omega_{D-3}, \hs{0.75} \mc{T} = \frac{\Omega_{D-3}}{16\pi}r_0^{D-4}.
\end{split}\enq
(Notice that the uniform black string has $C_\tau = r_0^{D-4}$, $C_y = 0$ and constant relative tension $n = (D-3)^{-1}$.) Finally, recall that the topology of the horizon is $S^{D-3}\times S^1$.

Gregory and Laflamme \cite{Gregory:1993vy} famously discovered that thin enough UBS, i.e.\ $r_0/L\lesssim\mc{O}(1)$, are dynamically unstable to clumping along the compact direction. More precisely, for fixed $L$ there is a critical value $r_0^{\text{GL}}$ below which there exist regular (linear) perturbations that grow exponentially with time and that break the translational invariance along the $S^1$; at precisely this critical value, the perturbations are time-independent thus signaling the existence of a linear solution of the Einstein equation which is not uniform along the $S^1$. This linear solution can be continued into the fully non-linear regime, giving rise to the NUBS. For $D = 10$, the critical value of the horizon radius at the onset of the GL instability is: $r_0^{\text{GL}} = 0.36671(3)L$.

Our aim in this work is to numerically construct vacuum NUBS and LOC solutions in $D=10$. In practice, we will treat the different metrics as smooth Riemannian manifolds with a U$(1)_\beta$ Killing vector that vanishes at the horizon and solve the Einstein vacuum equation, $R_{ab} = 0$, subject to certain regularity and asymptotic boundary conditions. As is well-know, due to the underlying gauge invariance of the theory, this equation does not yield a well-posed boundary value problem. Instead, we solve the Einstein-DeTurck equation, $R_{ab}^H = 0$, which is manifestly elliptic \cite{Headrick:2009pv}, where \beq\label{EdT}
R_{ab}^H \equiv R_{ab} - \nab_{(a}\xi_{b)}, \hs{0.75} \xi^a = g^{bc}\(\Gamma^a_{bc} - \bar{\Gamma}^a_{bc}\).
\enq
$R_{ab}$ is the Ricci tensor and $\xi_a$ is the so-called DeTurck vector. The last is formed from the usual Levi-Civita connection $\Gamma$ compatible with the spacetime metric $g$, and a Levi-Civita connection $\bar{\Gamma}$ compatible with some reference metric $\bar{g}$ that we are free to prescribe. This has now become a standard approach in stationary numerical relativity and we refer the reader to the literature for more details \cite{Headrick:2009pv,Wiseman:2011by,Figueras:2011va,Figueras:2016nmo}.

The equations are always discretized using pseudo-spectral methods on a Chebyshev grid, and we solve them by an iterative Newton-Raphson method; at each step of the iterative process the linear system of equations is solved using LU decomposition implemented by subroutine \texttt{LinearSolve} in \texttt{Mathematica}.

%~~~~~~~~~~~~~~~~~~~~~~~~~~~~~~~~~~~~~~~~~~~~~~~
\subsection{Numerical construction of Non-uniform black strings}
\label{nubs}
%~~~~~~~~~~~~~~~~~~~~~~~~~~~~~~~~~~~~~~~~~~~~~~
NUBS wrap the KK circle, and, for regular solutions,  the horizon $S^{D-3}$ is finite everywhere. This implies that with our symmetry assumptions, the integration domain has the following effective boundaries: the horizon, asymptotic infinity and the periodic boundary. Due to the symmetry of first GL harmonic, NUBS have a $\mathbb Z_2$-symmetry and then one has an additional mirror boundary. Hence, a single coordinate patch is enough to cover the whole computational domain. In practice, to numerically construct highly non-uniform black strings near the critical point it is convenient to use more than one patch to get enough resolution in the regions of interest.

To find NUBS we consider the following ansatz for the metric:
\beq\label{ansatznubs}
\dd s^2 = 4\,r_0^2\,\Delta^2\(x^2e^{Q_1}\,\dd\tau^2 + \frac{e^{Q_2}}{f(x)^{2(D-3)}}\,\dd x^2\) + e^{Q_3}\,\dd y^2 + 2\,Q_4\,\dd x\,\dd y + \frac{r_0^2\,e^{Q_5}}{f(x)^2}\,\dd\Omega_{D-3}^2,
\enq
with $\Delta = (D-4)^{-1}$, $f(x) = (1-x^2)^\Delta$, and unknowns $Q\equiv \{Q_1$, $Q_2$, $Q_3$, $Q_4$, $Q_5\}(x,y)$. For $Q = 0$, this ansatz reduces to the UBS in $D$ dimensions written in terms of the compact radial coordinate $x$, $x(r) = 1- r_0^{D-4}/r^{D-4}$. The UBS satisfies all the relevant boundary conditions that we will impose on our solutions (see below) and  we shall use it as the reference metric in the Einstein-DeTurck equation. The compact radial coordinate $x\in[0,1)$ covers the region from the horizon ($x= 0$) to infinity ($x =1$).  Note that NUBS posses reflection symmetry along the $S^1$ direction. This allows us to consider only one half of the KK circle subject to mirror boundary conditions. Therefore, we take $y\in[0,1]$, where $y = 0$ corresponds to the reflection plane and $y = 1$ the periodic boundary. This implies that the asymptotic length of the KK circle is kept fixed to be $L = 2$.

The radius of the round $S^{D-3}$ at the horizon is a good geometric invariant that can be used to describe NUBS; with our ansatz \eqref{ansatznubs}, this is given by
 \beq
R(y) = r_0\sqrt{e^{Q_5}}\Big|_H.
\label{eq:horSrad}
\enq
Black string solutions can be characterized with the non-uniformity parameter introduced in \cite{Gubser:2001ac}, $\lambda = \(R_{\text{max}}/R_{\text{min}} - 1\)/2$, where $R_{\text{max}} = \text{max}[R(y)]$ and $R_{\text{min}} = \text{min}[R(y)]$. UBS have $\lambda = 0$, whereas NUBS have $\lambda > 0$; the limit $\lambda\riga\infty$ corresponds to the merger point with the LOC branch, where $R_{\text{min}}\riga0$ while $R_\textrm{max}$ remains finite.

To obtain a well-posed boundary value problem that can be solved with elliptic methods we need to supplement the equations of motion with appropriate boundary conditions. These require regularity at the horizon, reflection symmetry, periodicity and KK asymptotics:
\begin{itemize}
\item{{\bf\textsf{Horizon}} at $x = 0$: smoothness of the metric at the horizon implies that all $Q$'s must be even in $x$ and therefore we impose Neumann boundary conditions on all $Q$'s, except the crossed term which must be Dirichlet. The condition $Q_1(0,y) = Q_2(0,y)$ ensures that the geometry is free of conical singularities and fixes the surface gravity of the solution to be that of our reference metric.}
\item{{\bf\textsf{Asymptotic boundary}} at $x = 1$: the metric must approach the KK space. This implies the Dirichlet boundary conditions, $Q_i(1,y) = 1$, $\forall i \neq 4$, and $Q_4(1,y) = 0$.}
\item{{\bf\textsf{Reflection plane and periodic boundary}} at $y = 0$ and $y = 1$ respectively: all $Q$'s must be even in the compact $S^1$ coordinate and thus we impose Neumann boundary conditions for all $Q$'s, except for the crossed term which must be Dirichlet there.}
\end{itemize}

To find NUBS, we start with the UBS close to the GL point and add a bit of the GL zero mode. This gives a good initial guess that allows us to find weakly non-uniform black strings. Once we have found a NUBS, we can move along the family varying the temperature; with our boundary conditions, the inverse temperature is given by \beq\label{THnubs}
\beta = \frac{4\,\pi\,r_0}{D-4}\,.
\enq
We move along the branch of NUBS by using the previous solution as a seed and varying the value of the parameter $r_0$; we start at $r_0 = 0.73450$ which corresponds to $\lambda = 0.04$ (recall that $r_0^{\text{GL}} = 0.73342(6)$ for $L = 2$) and, given our modest resources, we move up to a value of $r_0 = 0.79184$, corresponding to $\lambda = 5.05$.

For $\lambda \lesssim 1$, the NUBS are relatively weakly non-uniform, not much resolution is required to construct the solutions accurately and one single patch is suffices. At this point, the solutions satisfy $\xi_a\xi^a \equiv \xi^2 < 10^{-10}$, with estimated numerical error to be less than $0.01\%$. The Smarr's relation is satisfied up to the order $10^{-7}$. As we move along  the branch of NUBS to greater values of $\lambda$,  the function $Q_4$ develops very pronounced peaks near the origin, corresponding to the waist of the non-uniform black string, and some form of mesh-refinement there is needed to construct accurate solutions. We found that two conforming patches were enough to obtain good results, though the bound on the DeTurck vector goes up to $\xi^2 < 10^{-7}$ and the Smarr's relation is satisfied up to $10^{-6}$. Notice that our mesh-refinement introduces a new parameter $x_0$, which is the coordinate location where the two patches meet. We also considered $\tilde{y} = \text{mesh}(y;0,1,\chi)$, with the mesh-refinement function $\text{mesh}(\dots)$ given by (\ref{meshh}); here $\chi$ is just a parameter that controls the density of the new grid points. Since the steep gradients move towards the origin as $\lambda$ increases, we used two different setups with appropriate grid sizes $x_0$ ($\sim 10^{-1}, 10^{-2}$) and values of $\chi$ ($\sim 1, 10$). It is possible that by choosing a different reference metric for highly NUBS one can achieve larger values of $\lambda$ without losing accuracy.

%~~~~~~~~~~~~~~~~~~~~~~~~~~~~~~~~~~~~~~~~~~~~~~~
\subsection{Numerical construction of Localized black holes}
\label{loc}
%~~~~~~~~~~~~~~~~~~~~~~~~~~~~~~~~~~~~~~~~~~~~~~
To numerically construct LOC we follow the approach of Kalisch et al.\ \cite{Kalisch:2017bin} with minor modifications. Essentially, we considered a different compacitifaction of the radial coordinate so that we could extract the constants $C_\tau$ and $C_y$ appearing in the conserved charges \eqref{masstension} by calculating 1st derivatives of our unknown functions. In this section we superficially discuss the actual numerical construction of LOC and refer the reader to \cite{Kalisch:2017bin} for further details.

\begin{figure}[h!]
\centering
\includegraphics[height = 4.2cm]{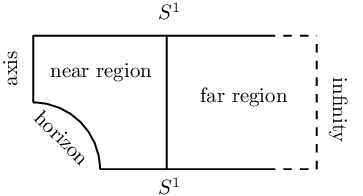}
\captionsetup{width=.9\linewidth}
\captionof{figure}{Sketch of the integration domain for localized black holes.}
\label{intdomain}
\end{figure}

We seek static axisymmetric black holes that are asymptotically KK and localized on a circle of (asymptotic) length $L$. We choose adapted coordinates so that symmetries of the spacetime become manifest. This implies that the actual boundaries of the computational domain are: the black hole horizon, the asymptotic infinity, the periodic boundary, the reflection plane and an axis of symmetry where the horizon $S^{D-2}$ smoothly shrinks to zero size, which is exposed because the localization on the $S^1$. From the point of view of finding these black holes numerically, since the integration domain has five boundaries, we naturally work with two coordinate patches: One patch adapted to a `near' region (containing the horizon), and another one adapted to a `far' region (containing the asymptotic infinity). The integration domain is schematically shown in Fig.\ \ref{intdomain}.

As in \cite{Headrick:2009pv}, one can work with cartesian coordinates ($x,y$) in the far patch and polar coordinates ($r,a$) in the near patch, and the relation between them is simply given by the polar map: $x = r\cos a$, $y = r\sin a$. To transfer information between the two coordinate patches, one can use two overlapping domains and impose uniqueness of the solution \cite{Headrick:2009pv}. This is simpler to implement if one uses finite differences. On the other hand, if one uses spectral methods, one can deform the two domains using some transfinite transformation and ensure that the two domains match along a curve; along this common boundary, one then imposes continuity of the functions and their normal derivatives. Alternatively, in the near region \cite{Kalisch:2017bin} introduce polar-like coordinates with a modified radial coordinate which naturally matches with the Cartesian coordinates sufficiently far from the black hole. This is the approach we follow in the near region. We recall the details of the integration domain and introduce the new compactification in appendix \ref{intdom}.

The ansatz for the metric in the far patch is: \beq
\dd s^2_{\text{Far}} = Q_1\,\dd\tau^2 + x^2\,Q_2\,\dd\Omega_{D-3}^2 + Q_3\,\dd x^2 + Q_4\,\dd y^2 + 2\,Q_5\,\dd x\dd y\,,
\enq
where the functions $Q\equiv\{Q_1,Q_2,Q_3,Q_4,Q_5\}(x,y)$ are our unknowns. As illustrated in the integration domain \ref{lattice}, the coordinate $x$ ranges from $L/2$, which is the boundary between the far and near regions, to infinity; on the other hand, $y\in[0,L/2]$, where $y=0$ is the reflection plane and $L/2$ is the periodic boundary. The boundary conditions we impose on the unknown functions $Q$ in this patch are: \begin{itemize}
\item{{\bf\textsf{Asymptotic boundary}} at $x = \infty$: the metric must approach the KK space. This implies the Dirichlet boundary conditions, $Q_i(\infty,y) = 1$, $\forall i \neq 5$, and $Q_5(\infty,y) = 0$.}
\item{{\bf\textsf{Matching boundary}}} at $x = L/2$: we impose continuity of the metric and its normal derivative.
\item{{\bf\textsf{Reflection plane and periodic boundary}} at $y = 0$ and $y = L/2$ respectively: all $Q$'s must be even in the compact coordinate $y$ and thus we impose Neumann boundary conditions on all them except the crossed term $Q_5$, which must be Dirichlet there.}
\end{itemize}

The near horizon region ansatz covers the horizon and the symmetry axis; at the horizon, the Killing $\partial_\tau$ becomes null and at the symmetry axis the round $S^{D-3}$ (and in fact the whole horizon $S^{D-2})$ smoothly shrinks to zero. The ansatz we consider is: \beq
\dd s^2_{\text{Near}} = \kappa^2\,(r-r_0)^2\,Q'_1\,\dd\tau^2 + r^2\,\cos^2a\ Q'_2\,\dd\Omega_{D-3}^2 + Q'_3\,\dd r^2 + r^2\,Q'_4\,\dd a^2 + 2\,r\,Q'_5\,\dd r\dd a,
\enq
where $Q'\equiv\{Q_1',Q_2',Q_3',Q_4',Q_5'\}(r,a)$ are the unknowns in this patch. This metric has a Killing horizon located at $r = r_0$ with surface gravity $\kappa$, and an axis at $a = \pi/2$; $a=0$ is the reflection plane.  Although the horizon is at $r = r_0$, $r_0$ is simply a parameter in our ansatz and we keep it fixed throughout the calculation (we choose $r_0=0.8$ for convenience); the physical parameter labelling each solution is the surface gravity $\kappa$, and this the parameter that we vary to move along the branch of LOC. With the definitions given in Appendix \ref{intdom}, the boundary conditions that we impose on the unknown functions $Q'$ in this region are: 
\begin{itemize}
\item{{\bf\textsf{Horizon}} at $r = r_0$: smoothness of the metric at the horizon implies that all $Q'$'s must be even in $r$ and therefore we impose Neumann boundary conditions for $Q_1'$, $r^2Q_2'$, $Q_3'$, $r^2Q_4'$ and Dirichlet for the crossed term $Q'_5$. The condition $Q'_1(r_0,a) = Q'_3(r_0,a)$ ensures that the geometry is free of conical singularities and fixes the surface gravity of the solution to be that of the reference metric (see below).}
\item{{\bf\textsf{Axis of symmetry}} at $a = \pi/2$: regularity requires that all functions ${Q'}$'s are Neumann, except the crossed term which is Dirichlet there. In addition we impose $Q'_2(r,\pi/2) = Q'_4(r,\pi/2)$ to avoid conical singularities.}
\item{{\bf\textsf{Reflection plane }} at $a = 0$: all functions are Neumann except for $Q_5'$, that vanishes there.}
\item{{\bf\textsf{Periodic boundary}}  at $r_3(L/2,a)\sin a = L/2$: using the relation between the far and near coordinates and the relation between the far and near unknown functions, one can find the boundary conditions for the near horizon functions from the boundary conditions that the far region functions satisfy there.}
\item{{\bf\textsf{Matching boundary}} at $r_2(L/2,a)\cos a = L/2$: we impose continuity of the metric and its normal derivative.}
\end{itemize}

In addition to the ansatz and the boundary conditions, the DeTurck scheme requires a global reference metric as part of the gauge fixing procedure. The reference metric must satisfy the same boundary conditions as the solution we seek. For the LOC, there is no known Einstein metric in closed analytic form that satisfies the required boundary conditions and hence one has to design it. In this paper we follow \cite{Headrick:2009pv,Kalisch:2017bin}, and smoothly glue together two metrics, each of which satisfy the desired boundary conditions in each region, i.e.\ asymptotic Kaluza-Klein space in the far region and, for instance, the asymptotically flat Schwarzschild black hole in $D$ dimensions:
\beq\label{refmetricloc}
\dd \bar{s}^2 = H(r)\dd\tau^2 + \dd r^2 + G(r)\(\dd a^2 + \cos^2a\  \dd\Omega_{D-3}^2\)
\enq
with \beq
H(r) = \left\{\begin{matrix}
&H_{\text{hor}}(r) \hs{0.25} \text{for} \hs{0.25} r < r_1 \\
&1 \hs{1.35} \text{for} \hs{0.25} r \geq r_1
\end{matrix}\right., \hs{0.75} G(r) = \left\{\begin{matrix}
&G_{\text{hor}}(r) \hs{0.25} \text{for} \hs{0.25} r < r_1 \\
&r^2 \hs{1.2} \text{for} \hs{0.25} r \geq r_1
\end{matrix}\right.,
\enq
and \beq\begin{split}
H_{\text{hor}}(r) &= 1 - E(r), \\
G_{\text{hor}}(r) &= r^2 - E(r)\(r^2 -\frac{(D-3)^2}{4\kappa^2}-(r-r_0)^2\[\frac{D^2}{4}-D+\frac{3}{4}-\kappa^2r_0^2\]\),
\end{split}\enq
where the function $E(r)$ is given by \beq
E(r) = \exp\(-\kappa^2\frac{(r-r_0)^2}{1-(r-r_0)^2/(r_1-r_0)^2}\).
\enq
The reference metric depends on $\kappa$, $r_0$, and $r_1$, which is an additional parameter that can be adjusted (see appendix \ref{intdom}). The function $E(r)$ vanishes exponentially fast for $r\riga r_1$, and the reference metric (\ref{refmetricloc}) tends to the KK space written in polar coordinates. For $r\riga r_0$,  $E(r) \simeq 1 - \kappa^2(r-r_0)^2$ and (\ref{refmetricloc}) takes the form of the near horizon metric of the Schwarzschild black hole in $D$ dimensions. 

To find localized black holes we start with the reference metric as a seed with $\kappa = 2.4$ and $L = 6$ (we keep this value of $L$ for all solutions). Recall that the convergence of Newton's method strongly depends on the choice of the initial seed and finding a first solution may be difficult. To stay within the basin of attraction at each iteration, we introduce a parameter $\alpha\in\R^+$ in the iteration loop so that the update is $Q^{(n+1)} \sim Q^{(n)} +\alpha\delta Q$, where $\alpha \sim \mc{O}(1/100)$ or $\mc{O}(1/10)$ during the first iterations and is $\mc{O}(1)$ towards the end. Once we have found the first solution, we use it as the initial guess to find the next solution with a slightly different $\kappa$ while keeping  $\alpha = 1$. We kept the parameters of the integration domain and the coordinates fixed throughout the calculation and they are specified in Fig.\ \ref{lattice}. 

The most critical solution we found corresponds to $\kappa = 1.262768$. In this critical regime, the functions $Q'_2$ and $Q'_4$ develop steep gradients near the axis and the horizon; to resolve them, we redefine these two functions in the pure polar patch (blue and green dots in Fig.\ \ref{lattice}): $Q_i^c(r,a) = 1/Q_i'(r,a)$, $i = 2,4$ \cite{Kalisch:2017bin}. The boundary conditions for these redefined functions can be easily found from the original ones for $Q_2'$ and $Q_4'$. All solutions we found satisfy $\xi^2 < 10^{-10}$ with numerical error less than $0.01\%$ and the Smarr's relation is satisfied up to the order $10^{-6}$.

%~~~~~~~~~~~~~~~~~~~~~~~~~~~~~~~~~~~~~~~~~~~~~~~
\section{Results}
\label{results}
%~~~~~~~~~~~~~~~~~~~~~~~~~~~~~~~~~~~~~~~~~~~~~~
In this section we present our results for both NUBS and LOC. We first consider the behavior of the various thermodynamic quantities along each branch of solutions and the phase diagrams, and then we focus on the horizon geometry. We then study the critical behavior near the merger point, and provide evidence that the double-cone geometry proposed by \cite{Kol:2002xz} does indeed control the merger. We finally compute their spectrum of negative modes of the Lichnerowicz operator.

%~~~~~~~~~~~~~~~~~~~~~~~~~~~~~~~~~~~~~~~~~~~~~~~
\subsection{Thermodynamics}
\label{thermo0}
%~~~~~~~~~~~~~~~~~~~~~~~~~~~~~~~~~~~~~~~~~~~~~~
The horizon temperature labels both NUBS and LOC and is given by (\ref{THnubs}) for non-uniform black strings and by $\kappa/(2\pi)$ for localized solutions. The mass and the tension follow from (\ref{masstension}). For the NUBS the asymptotic charges are computed by \beq C_\tau = \frac{r_0^{D-4}}{2}\(\frac{L}{2}\)^{-1}\int_0^{L/2}\dd y\(2 + \parcial[Q_1]{x}\bigg|_{x = 1}\), \hs{0.75} C_y = -\frac{r_0^{D-4}}{2}\(\frac{L}{2}\)^{-1}\int_0^{L/2}\dd y\parcial[Q_3]{x}\bigg|_{x =1},
\enq
where in these expressions we first interpolate the numerical data and then perform the integration. For the LOC, these quantities are given in (\ref{ctcyloc}). The horizon area is found to be: \beq\begin{split}
\text{NUBS: } \hs{0.5}& A_H = 2r_0^{D-3}\Omega_{D-3}\int_0^{L/2}\dd y\sqrt{e^{Q_3+(D-3)Q_5}}\Big|_H, \\
\text{LOC: } \hs{0.5}& A_H =2r_0^{D-2}\Omega_{D-3}\int_0^{\pi/2}\dd a(\cos a)^{D-3}\sqrt{{Q_2'}^{D-3}Q_4'}\Big|_H.
\end{split}\enq

In Fig.\ \ref{microcano} we display the phase diagram in the microcanonical (\textsl{top left}) and canonical  (\textsl{top right}) ensembles, and the behavior of the horizon area (\textsl{middle}) and tension (\textsl{bottom}) as a function of the inverse temperature (normalized by $L$). The behavior of the mass and the relative tension as a function of the inverse temperature is similar to that of the area and tension and we do not display the corresponding plots here. To make the microcanonical and canonical phase diagrams easier to visualize we plot the dimensionless differences $\Delta S/L^8 \equiv \(S(M) - S_{\text{UBS}}(M)\)/L^8$ and  $\Delta F/L^7 \equiv \(F(T)- F_{\text{UBS}}(T)\)/L^7$ respectively. 

NUBS, which exist beyond the GL point, never dominate any of these ensembles and they are presumably dynamically unstable. The localized black hole phase crosses the UBS branch at 
\beq
M_{\text{PT}} = 0.01375(4)L^7, \hs{0.25} \text{ or } \hs{0.25} T_{\text{PT}} = 1.26682(1)L^{-1}.
\enq
For lower masses, $M < M_{\text{PT}}$, or higher temperatures temperatures, $T < T_{\text{PT}}$, the LOC dominate the corresponding ensemble and the UBS are unstable, whilst for $M>M_{\text{PT}} $ or $T<T_{\text{PT}} $, UBS dominate; at $M = M_{\text{PT}}$ or $T=T_{\text{PT}}$, there is first order phase transition. The phase diagrams are consistent with a merger between NUBS and LOC at \beq
M_{\text{Merger}} = 0.020404(6)L^7, \hs{0.25} \text{ or } \hs{0.25} T_{\text{Merger}} = 1.20585(6)L^{-1}.
\enq

One of the remarkable features of the phase diagram in $D=10$ is the lack of turning points away from the merger along any of the branches, either LOC or NUBS. This should be contrasted with the phase diagram in $D=5,6$, which exhibits a turning point along the LOC branch at some maximum mass and then there is a minimum of the temperature \cite{Headrick:2009pv,Kalisch:2017bin}. It is reasonable to expect that such a turning point (away from the merger) exists on the LOC branch for any dimension $D<10$.  This turning point switches to the NUBS branch in $D=12$ (and presumably in $D=11$), as shown in \cite{Figueras:2012xj} and more recently in the large-$D$ expansion in \cite{Emparan:2018bmi}. As we will argue below, the lack of turning points away from the merger in the phase diagram in $D=10$ may be related to the nature of the merger in this specific number of spacetime dimensions.
 
\begin{center}
\begin{minipage}{\textwidth}
\begin{minipage}[h!]{0.5\textwidth}
% GNUPLOT: LaTeX picture with Postscript
\begingroup
  \makeatletter
  \providecommand\color[2][]{%
    \GenericError{(gnuplot) \space\space\space\@spaces}{%
      Package color not loaded in conjunction with
      terminal option `colourtext'%
    }{See the gnuplot documentation for explanation.%
    }{Either use 'blacktext' in gnuplot or load the package
      color.sty in LaTeX.}%
    \renewcommand\color[2][]{}%
  }%
  \providecommand\includegraphics[2][]{%
    \GenericError{(gnuplot) \space\space\space\@spaces}{%
      Package graphicx or graphics not loaded%
    }{See the gnuplot documentation for explanation.%
    }{The gnuplot epslatex terminal needs graphicx.sty or graphics.sty.}%
    \renewcommand\includegraphics[2][]{}%
  }%
  \providecommand\rotatebox[2]{#2}%
  \@ifundefined{ifGPcolor}{%
    \newif\ifGPcolor
    \GPcolortrue
  }{}%
  \@ifundefined{ifGPblacktext}{%
    \newif\ifGPblacktext
    \GPblacktexttrue
  }{}%
  % define a \g@addto@macro without @ in the name:
  \let\gplgaddtomacro\g@addto@macro
  % define empty templates for all commands taking text:
  \gdef\gplbacktext{}%
  \gdef\gplfronttext{}%
  \makeatother
  \ifGPblacktext
    % no textcolor at all
    \def\colorrgb#1{}%
    \def\colorgray#1{}%
  \else
    % gray or color?
    \ifGPcolor
      \def\colorrgb#1{\color[rgb]{#1}}%
      \def\colorgray#1{\color[gray]{#1}}%
      \expandafter\def\csname LTw\endcsname{\color{white}}%
      \expandafter\def\csname LTb\endcsname{\color{black}}%
      \expandafter\def\csname LTa\endcsname{\color{black}}%
      \expandafter\def\csname LT0\endcsname{\color[rgb]{1,0,0}}%
      \expandafter\def\csname LT1\endcsname{\color[rgb]{0,1,0}}%
      \expandafter\def\csname LT2\endcsname{\color[rgb]{0,0,1}}%
      \expandafter\def\csname LT3\endcsname{\color[rgb]{1,0,1}}%
      \expandafter\def\csname LT4\endcsname{\color[rgb]{0,1,1}}%
      \expandafter\def\csname LT5\endcsname{\color[rgb]{1,1,0}}%
      \expandafter\def\csname LT6\endcsname{\color[rgb]{0,0,0}}%
      \expandafter\def\csname LT7\endcsname{\color[rgb]{1,0.3,0}}%
      \expandafter\def\csname LT8\endcsname{\color[rgb]{0.5,0.5,0.5}}%
    \else
      % gray
      \def\colorrgb#1{\color{black}}%
      \def\colorgray#1{\color[gray]{#1}}%
      \expandafter\def\csname LTw\endcsname{\color{white}}%
      \expandafter\def\csname LTb\endcsname{\color{black}}%
      \expandafter\def\csname LTa\endcsname{\color{black}}%
      \expandafter\def\csname LT0\endcsname{\color{black}}%
      \expandafter\def\csname LT1\endcsname{\color{black}}%
      \expandafter\def\csname LT2\endcsname{\color{black}}%
      \expandafter\def\csname LT3\endcsname{\color{black}}%
      \expandafter\def\csname LT4\endcsname{\color{black}}%
      \expandafter\def\csname LT5\endcsname{\color{black}}%
      \expandafter\def\csname LT6\endcsname{\color{black}}%
      \expandafter\def\csname LT7\endcsname{\color{black}}%
      \expandafter\def\csname LT8\endcsname{\color{black}}%
    \fi
  \fi
    \setlength{\unitlength}{0.0500bp}%
    \ifx\gptboxheight\undefined%
      \newlength{\gptboxheight}%
      \newlength{\gptboxwidth}%
      \newsavebox{\gptboxtext}%
    \fi%
    \setlength{\fboxrule}{0.5pt}%
    \setlength{\fboxsep}{1pt}%
\begin{picture}(5102.00,3400.00)%
    \gplgaddtomacro\gplbacktext{%
      \csname LTb\endcsname%%
      \put(1474,704){\makebox(0,0)[r]{\strut{}$-0.00015$}}%
      \csname LTb\endcsname%%
      \put(1474,1242){\makebox(0,0)[r]{\strut{}$-0.0001$}}%
      \csname LTb\endcsname%%
      \put(1474,1780){\makebox(0,0)[r]{\strut{}$-5\cdot10^{-5}$}}%
      \csname LTb\endcsname%%
      \put(1474,2318){\makebox(0,0)[r]{\strut{}$0$}}%
      \csname LTb\endcsname%%
      \put(1474,2856){\makebox(0,0)[r]{\strut{}$5\cdot10^{-5}$}}%
      \csname LTb\endcsname%%
      \put(1741,484){\makebox(0,0){\strut{}$0$}}%
      \csname LTb\endcsname%%
      \put(2414,484){\makebox(0,0){\strut{}$0.005$}}%
      \csname LTb\endcsname%%
      \put(3088,484){\makebox(0,0){\strut{}$0.01$}}%
      \csname LTb\endcsname%%
      \put(3762,484){\makebox(0,0){\strut{}$0.015$}}%
      \csname LTb\endcsname%%
      \put(4436,484){\makebox(0,0){\strut{}$0.02$}}%
      \put(3253,2349){\makebox(0,0)[l]{\strut{}}}%
    }%
    \gplgaddtomacro\gplfronttext{%
      \csname LTb\endcsname%%
      \put(198,1941){\rotatebox{-270}{\makebox(0,0){\strut{}$\Delta S/L^8$}}}%
      \put(3155,154){\makebox(0,0){\strut{}$M/L^7 $}}%
      \csname LTb\endcsname%%
      \put(2398,1592){\makebox(0,0)[r]{\strut{}NUBS}}%
      \csname LTb\endcsname%%
      \put(2398,1262){\makebox(0,0)[r]{\strut{}LOC}}%
      \csname LTb\endcsname%%
      \put(2398,932){\makebox(0,0)[r]{\strut{} UBS}}%
    }%
    \gplbacktext
    \put(0,0){\includegraphics{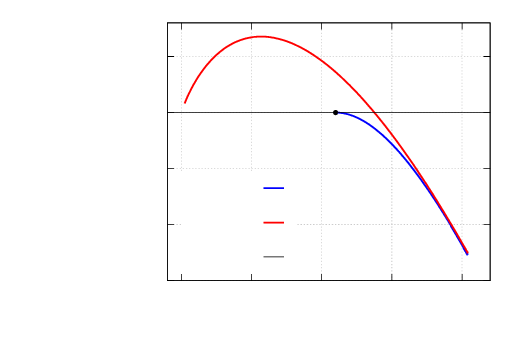}}%
    \gplfronttext
  \end{picture}%
\endgroup

\end{minipage}
\hfill
\begin{minipage}[h!]{0.5\textwidth}
% GNUPLOT: LaTeX picture with Postscript
\begingroup
  \makeatletter
  \providecommand\color[2][]{%
    \GenericError{(gnuplot) \space\space\space\@spaces}{%
      Package color not loaded in conjunction with
      terminal option `colourtext'%
    }{See the gnuplot documentation for explanation.%
    }{Either use 'blacktext' in gnuplot or load the package
      color.sty in LaTeX.}%
    \renewcommand\color[2][]{}%
  }%
  \providecommand\includegraphics[2][]{%
    \GenericError{(gnuplot) \space\space\space\@spaces}{%
      Package graphicx or graphics not loaded%
    }{See the gnuplot documentation for explanation.%
    }{The gnuplot epslatex terminal needs graphicx.sty or graphics.sty.}%
    \renewcommand\includegraphics[2][]{}%
  }%
  \providecommand\rotatebox[2]{#2}%
  \@ifundefined{ifGPcolor}{%
    \newif\ifGPcolor
    \GPcolortrue
  }{}%
  \@ifundefined{ifGPblacktext}{%
    \newif\ifGPblacktext
    \GPblacktexttrue
  }{}%
  % define a \g@addto@macro without @ in the name:
  \let\gplgaddtomacro\g@addto@macro
  % define empty templates for all commands taking text:
  \gdef\gplbacktext{}%
  \gdef\gplfronttext{}%
  \makeatother
  \ifGPblacktext
    % no textcolor at all
    \def\colorrgb#1{}%
    \def\colorgray#1{}%
  \else
    % gray or color?
    \ifGPcolor
      \def\colorrgb#1{\color[rgb]{#1}}%
      \def\colorgray#1{\color[gray]{#1}}%
      \expandafter\def\csname LTw\endcsname{\color{white}}%
      \expandafter\def\csname LTb\endcsname{\color{black}}%
      \expandafter\def\csname LTa\endcsname{\color{black}}%
      \expandafter\def\csname LT0\endcsname{\color[rgb]{1,0,0}}%
      \expandafter\def\csname LT1\endcsname{\color[rgb]{0,1,0}}%
      \expandafter\def\csname LT2\endcsname{\color[rgb]{0,0,1}}%
      \expandafter\def\csname LT3\endcsname{\color[rgb]{1,0,1}}%
      \expandafter\def\csname LT4\endcsname{\color[rgb]{0,1,1}}%
      \expandafter\def\csname LT5\endcsname{\color[rgb]{1,1,0}}%
      \expandafter\def\csname LT6\endcsname{\color[rgb]{0,0,0}}%
      \expandafter\def\csname LT7\endcsname{\color[rgb]{1,0.3,0}}%
      \expandafter\def\csname LT8\endcsname{\color[rgb]{0.5,0.5,0.5}}%
    \else
      % gray
      \def\colorrgb#1{\color{black}}%
      \def\colorgray#1{\color[gray]{#1}}%
      \expandafter\def\csname LTw\endcsname{\color{white}}%
      \expandafter\def\csname LTb\endcsname{\color{black}}%
      \expandafter\def\csname LTa\endcsname{\color{black}}%
      \expandafter\def\csname LT0\endcsname{\color{black}}%
      \expandafter\def\csname LT1\endcsname{\color{black}}%
      \expandafter\def\csname LT2\endcsname{\color{black}}%
      \expandafter\def\csname LT3\endcsname{\color{black}}%
      \expandafter\def\csname LT4\endcsname{\color{black}}%
      \expandafter\def\csname LT5\endcsname{\color{black}}%
      \expandafter\def\csname LT6\endcsname{\color{black}}%
      \expandafter\def\csname LT7\endcsname{\color{black}}%
      \expandafter\def\csname LT8\endcsname{\color{black}}%
    \fi
  \fi
    \setlength{\unitlength}{0.0500bp}%
    \ifx\gptboxheight\undefined%
      \newlength{\gptboxheight}%
      \newlength{\gptboxwidth}%
      \newsavebox{\gptboxtext}%
    \fi%
    \setlength{\fboxrule}{0.5pt}%
    \setlength{\fboxsep}{1pt}%
\begin{picture}(5102.00,3400.00)%
    \gplgaddtomacro\gplbacktext{%
      \csname LTb\endcsname%%
      \put(1342,910){\makebox(0,0)[r]{\strut{}$-0.0001$}}%
      \csname LTb\endcsname%%
      \put(1342,1426){\makebox(0,0)[r]{\strut{}$-5\cdot10^{-5}$}}%
      \csname LTb\endcsname%%
      \put(1342,1942){\makebox(0,0)[r]{\strut{}$0$}}%
      \csname LTb\endcsname%%
      \put(1342,2457){\makebox(0,0)[r]{\strut{}$5\cdot10^{-5}$}}%
      \csname LTb\endcsname%%
      \put(1342,2973){\makebox(0,0)[r]{\strut{}$0.0001$}}%
      \csname LTb\endcsname%%
      \put(1723,484){\makebox(0,0){\strut{}$1.2$}}%
      \csname LTb\endcsname%%
      \put(2220,484){\makebox(0,0){\strut{}$1.4$}}%
      \csname LTb\endcsname%%
      \put(2717,484){\makebox(0,0){\strut{}$1.6$}}%
      \csname LTb\endcsname%%
      \put(3214,484){\makebox(0,0){\strut{}$1.8$}}%
      \csname LTb\endcsname%%
      \put(3711,484){\makebox(0,0){\strut{}$2$}}%
      \csname LTb\endcsname%%
      \put(4208,484){\makebox(0,0){\strut{}$2.2$}}%
      \csname LTb\endcsname%%
      \put(4705,484){\makebox(0,0){\strut{}$2.4$}}%
      \put(2007,1973){\makebox(0,0)[l]{\strut{}}}%
    }%
    \gplgaddtomacro\gplfronttext{%
      \csname LTb\endcsname%%
      \put(198,1941){\rotatebox{-270}{\makebox(0,0){\strut{}$\Delta F/L^7$}}}%
      \put(3089,154){\makebox(0,0){\strut{}$TL$}}%
      \csname LTb\endcsname%%
      \put(4114,2951){\makebox(0,0)[r]{\strut{}NUBS}}%
      \csname LTb\endcsname%%
      \put(4114,2621){\makebox(0,0)[r]{\strut{}LOC}}%
      \csname LTb\endcsname%%
      \put(4114,2291){\makebox(0,0)[r]{\strut{}UBS}}%
    }%
    \gplbacktext
    \put(0,0){\includegraphics{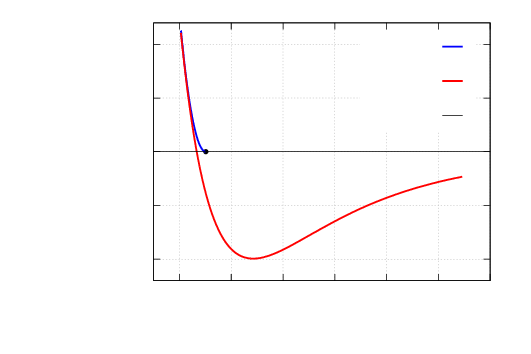}}%
    \gplfronttext
  \end{picture}%
\endgroup

\end{minipage}
\end{minipage}
\begin{minipage}{\textwidth}
\begin{minipage}[h!]{0.5\textwidth}
\hs{0.95}% GNUPLOT: LaTeX picture with Postscript
\begingroup
  \makeatletter
  \providecommand\color[2][]{%
    \GenericError{(gnuplot) \space\space\space\@spaces}{%
      Package color not loaded in conjunction with
      terminal option `colourtext'%
    }{See the gnuplot documentation for explanation.%
    }{Either use 'blacktext' in gnuplot or load the package
      color.sty in LaTeX.}%
    \renewcommand\color[2][]{}%
  }%
  \providecommand\includegraphics[2][]{%
    \GenericError{(gnuplot) \space\space\space\@spaces}{%
      Package graphicx or graphics not loaded%
    }{See the gnuplot documentation for explanation.%
    }{The gnuplot epslatex terminal needs graphicx.sty or graphics.sty.}%
    \renewcommand\includegraphics[2][]{}%
  }%
  \providecommand\rotatebox[2]{#2}%
  \@ifundefined{ifGPcolor}{%
    \newif\ifGPcolor
    \GPcolortrue
  }{}%
  \@ifundefined{ifGPblacktext}{%
    \newif\ifGPblacktext
    \GPblacktexttrue
  }{}%
  % define a \g@addto@macro without @ in the name:
  \let\gplgaddtomacro\g@addto@macro
  % define empty templates for all commands taking text:
  \gdef\gplbacktext{}%
  \gdef\gplfronttext{}%
  \makeatother
  \ifGPblacktext
    % no textcolor at all
    \def\colorrgb#1{}%
    \def\colorgray#1{}%
  \else
    % gray or color?
    \ifGPcolor
      \def\colorrgb#1{\color[rgb]{#1}}%
      \def\colorgray#1{\color[gray]{#1}}%
      \expandafter\def\csname LTw\endcsname{\color{white}}%
      \expandafter\def\csname LTb\endcsname{\color{black}}%
      \expandafter\def\csname LTa\endcsname{\color{black}}%
      \expandafter\def\csname LT0\endcsname{\color[rgb]{1,0,0}}%
      \expandafter\def\csname LT1\endcsname{\color[rgb]{0,1,0}}%
      \expandafter\def\csname LT2\endcsname{\color[rgb]{0,0,1}}%
      \expandafter\def\csname LT3\endcsname{\color[rgb]{1,0,1}}%
      \expandafter\def\csname LT4\endcsname{\color[rgb]{0,1,1}}%
      \expandafter\def\csname LT5\endcsname{\color[rgb]{1,1,0}}%
      \expandafter\def\csname LT6\endcsname{\color[rgb]{0,0,0}}%
      \expandafter\def\csname LT7\endcsname{\color[rgb]{1,0.3,0}}%
      \expandafter\def\csname LT8\endcsname{\color[rgb]{0.5,0.5,0.5}}%
    \else
      % gray
      \def\colorrgb#1{\color{black}}%
      \def\colorgray#1{\color[gray]{#1}}%
      \expandafter\def\csname LTw\endcsname{\color{white}}%
      \expandafter\def\csname LTb\endcsname{\color{black}}%
      \expandafter\def\csname LTa\endcsname{\color{black}}%
      \expandafter\def\csname LT0\endcsname{\color{black}}%
      \expandafter\def\csname LT1\endcsname{\color{black}}%
      \expandafter\def\csname LT2\endcsname{\color{black}}%
      \expandafter\def\csname LT3\endcsname{\color{black}}%
      \expandafter\def\csname LT4\endcsname{\color{black}}%
      \expandafter\def\csname LT5\endcsname{\color{black}}%
      \expandafter\def\csname LT6\endcsname{\color{black}}%
      \expandafter\def\csname LT7\endcsname{\color{black}}%
      \expandafter\def\csname LT8\endcsname{\color{black}}%
    \fi
  \fi
    \setlength{\unitlength}{0.0500bp}%
    \ifx\gptboxheight\undefined%
      \newlength{\gptboxheight}%
      \newlength{\gptboxwidth}%
      \newsavebox{\gptboxtext}%
    \fi%
    \setlength{\fboxrule}{0.5pt}%
    \setlength{\fboxsep}{1pt}%
\begin{picture}(8412.16,3400.00)%
    \gplgaddtomacro\gplbacktext{%
      \csname LTb\endcsname%%
      \put(946,881){\makebox(0,0)[r]{\strut{}$0$}}%
      \csname LTb\endcsname%%
      \put(946,1234){\makebox(0,0)[r]{\strut{}$0.01$}}%
      \csname LTb\endcsname%%
      \put(946,1588){\makebox(0,0)[r]{\strut{}$0.02$}}%
      \csname LTb\endcsname%%
      \put(946,1941){\makebox(0,0)[r]{\strut{}$0.03$}}%
      \csname LTb\endcsname%%
      \put(946,2295){\makebox(0,0)[r]{\strut{}$0.04$}}%
      \csname LTb\endcsname%%
      \put(946,2649){\makebox(0,0)[r]{\strut{}$0.05$}}%
      \csname LTb\endcsname%%
      \put(946,3002){\makebox(0,0)[r]{\strut{}$0.06$}}%
      \csname LTb\endcsname%%
      \put(1078,484){\makebox(0,0){\strut{}$0.42$}}%
      \csname LTb\endcsname%%
      \put(1592,484){\makebox(0,0){\strut{}$0.5$}}%
      \csname LTb\endcsname%%
      \put(2106,484){\makebox(0,0){\strut{}$0.58$}}%
      \csname LTb\endcsname%%
      \put(2619,484){\makebox(0,0){\strut{}$0.66$}}%
      \csname LTb\endcsname%%
      \put(3133,484){\makebox(0,0){\strut{}$0.74$}}%
      \csname LTb\endcsname%%
      \put(3647,484){\makebox(0,0){\strut{}$0.82$}}%
      \put(3344,1936){\makebox(0,0)[l]{\strut{}}}%
    }%
    \gplgaddtomacro\gplfronttext{%
      \csname LTb\endcsname%%
      \put(198,1941){\rotatebox{-270}{\makebox(0,0){\strut{}$A_H/L^8$}}}%
      \put(2523,154){\makebox(0,0){\strut{}$\beta/L$}}%
      \csname LTb\endcsname%%
      \put(1870,2951){\makebox(0,0)[r]{\strut{}NUBS}}%
      \csname LTb\endcsname%%
      \put(1870,2621){\makebox(0,0)[r]{\strut{}LOC}}%
      \csname LTb\endcsname%%
      \put(1870,2291){\makebox(0,0)[r]{\strut{}UBS}}%
    }%
    \gplgaddtomacro\gplbacktext{%
      \csname LTb\endcsname%%
      \put(5645,881){\makebox(0,0)[r]{\strut{}$0.05896$}}%
      \csname LTb\endcsname%%
      \put(5645,1617){\makebox(0,0)[r]{\strut{}$0.05901$}}%
      \csname LTb\endcsname%%
      \put(5645,2352){\makebox(0,0)[r]{\strut{}$0.05906$}}%
      \csname LTb\endcsname%%
      \put(5777,661){\makebox(0,0){\strut{}$0.829$}}%
      \csname LTb\endcsname%%
      \put(6856,661){\makebox(0,0){\strut{}$0.8292$}}%
      \csname LTb\endcsname%%
      \put(7935,661){\makebox(0,0){\strut{}$0.8294$}}%
    }%
    \gplgaddtomacro\gplfronttext{%
    }%
    \gplbacktext
    \put(0,0){\includegraphics{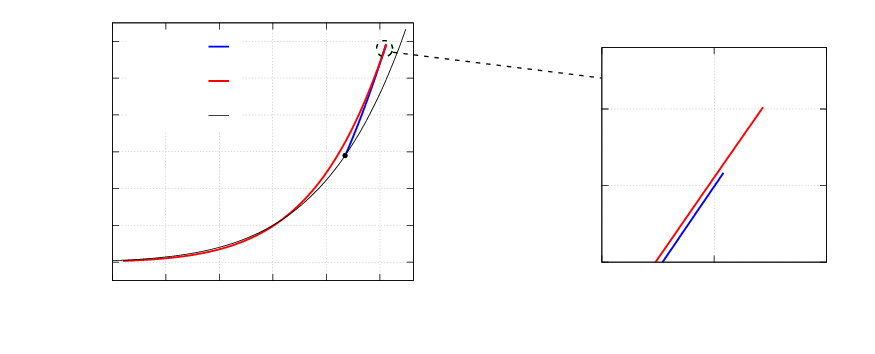}}%
    \gplfronttext
  \end{picture}%
\endgroup

\end{minipage}

\hfill

\begin{minipage}[h!]{0.5\textwidth}
\hs{0.}% GNUPLOT: LaTeX picture with Postscript
\begingroup
  \makeatletter
  \providecommand\color[2][]{%
    \GenericError{(gnuplot) \space\space\space\@spaces}{%
      Package color not loaded in conjunction with
      terminal option `colourtext'%
    }{See the gnuplot documentation for explanation.%
    }{Either use 'blacktext' in gnuplot or load the package
      color.sty in LaTeX.}%
    \renewcommand\color[2][]{}%
  }%
  \providecommand\includegraphics[2][]{%
    \GenericError{(gnuplot) \space\space\space\@spaces}{%
      Package graphicx or graphics not loaded%
    }{See the gnuplot documentation for explanation.%
    }{The gnuplot epslatex terminal needs graphicx.sty or graphics.sty.}%
    \renewcommand\includegraphics[2][]{}%
  }%
  \providecommand\rotatebox[2]{#2}%
  \@ifundefined{ifGPcolor}{%
    \newif\ifGPcolor
    \GPcolortrue
  }{}%
  \@ifundefined{ifGPblacktext}{%
    \newif\ifGPblacktext
    \GPblacktexttrue
  }{}%
  % define a \g@addto@macro without @ in the name:
  \let\gplgaddtomacro\g@addto@macro
  % define empty templates for all commands taking text:
  \gdef\gplbacktext{}%
  \gdef\gplfronttext{}%
  \makeatother
  \ifGPblacktext
    % no textcolor at all
    \def\colorrgb#1{}%
    \def\colorgray#1{}%
  \else
    % gray or color?
    \ifGPcolor
      \def\colorrgb#1{\color[rgb]{#1}}%
      \def\colorgray#1{\color[gray]{#1}}%
      \expandafter\def\csname LTw\endcsname{\color{white}}%
      \expandafter\def\csname LTb\endcsname{\color{black}}%
      \expandafter\def\csname LTa\endcsname{\color{black}}%
      \expandafter\def\csname LT0\endcsname{\color[rgb]{1,0,0}}%
      \expandafter\def\csname LT1\endcsname{\color[rgb]{0,1,0}}%
      \expandafter\def\csname LT2\endcsname{\color[rgb]{0,0,1}}%
      \expandafter\def\csname LT3\endcsname{\color[rgb]{1,0,1}}%
      \expandafter\def\csname LT4\endcsname{\color[rgb]{0,1,1}}%
      \expandafter\def\csname LT5\endcsname{\color[rgb]{1,1,0}}%
      \expandafter\def\csname LT6\endcsname{\color[rgb]{0,0,0}}%
      \expandafter\def\csname LT7\endcsname{\color[rgb]{1,0.3,0}}%
      \expandafter\def\csname LT8\endcsname{\color[rgb]{0.5,0.5,0.5}}%
    \else
      % gray
      \def\colorrgb#1{\color{black}}%
      \def\colorgray#1{\color[gray]{#1}}%
      \expandafter\def\csname LTw\endcsname{\color{white}}%
      \expandafter\def\csname LTb\endcsname{\color{black}}%
      \expandafter\def\csname LTa\endcsname{\color{black}}%
      \expandafter\def\csname LT0\endcsname{\color{black}}%
      \expandafter\def\csname LT1\endcsname{\color{black}}%
      \expandafter\def\csname LT2\endcsname{\color{black}}%
      \expandafter\def\csname LT3\endcsname{\color{black}}%
      \expandafter\def\csname LT4\endcsname{\color{black}}%
      \expandafter\def\csname LT5\endcsname{\color{black}}%
      \expandafter\def\csname LT6\endcsname{\color{black}}%
      \expandafter\def\csname LT7\endcsname{\color{black}}%
      \expandafter\def\csname LT8\endcsname{\color{black}}%
    \fi
  \fi
    \setlength{\unitlength}{0.0500bp}%
    \ifx\gptboxheight\undefined%
      \newlength{\gptboxheight}%
      \newlength{\gptboxwidth}%
      \newsavebox{\gptboxtext}%
    \fi%
    \setlength{\fboxrule}{0.5pt}%
    \setlength{\fboxsep}{1pt}%
\begin{picture}(8904.19,3400.00)%
    \gplgaddtomacro\gplbacktext{%
      \csname LTb\endcsname%%
      \put(1474,704){\makebox(0,0)[r]{\strut{}$-0.00015$}}%
      \csname LTb\endcsname%%
      \put(1474,1339){\makebox(0,0)[r]{\strut{}$0.00035$}}%
      \csname LTb\endcsname%%
      \put(1474,1973){\makebox(0,0)[r]{\strut{}$0.00085$}}%
      \csname LTb\endcsname%%
      \put(1474,2608){\makebox(0,0)[r]{\strut{}$0.00135$}}%
      \csname LTb\endcsname%%
      \put(1606,484){\makebox(0,0){\strut{}$0.42$}}%
      \csname LTb\endcsname%%
      \put(2273,484){\makebox(0,0){\strut{}$0.52$}}%
      \csname LTb\endcsname%%
      \put(2939,484){\makebox(0,0){\strut{}$0.62$}}%
      \csname LTb\endcsname%%
      \put(3606,484){\makebox(0,0){\strut{}$0.72$}}%
      \csname LTb\endcsname%%
      \put(4272,484){\makebox(0,0){\strut{}$0.82$}}%
      \put(3957,2919){\makebox(0,0)[l]{\strut{}}}%
    }%
    \gplgaddtomacro\gplfronttext{%
      \csname LTb\endcsname%%
      \put(198,1941){\rotatebox{-270}{\makebox(0,0){\strut{}$\mc{T}/L^6$}}}%
      \put(3072,154){\makebox(0,0){\strut{}$\beta/L$}}%
      \csname LTb\endcsname%%
      \put(2398,2951){\makebox(0,0)[r]{\strut{}NUBS}}%
      \csname LTb\endcsname%%
      \put(2398,2621){\makebox(0,0)[r]{\strut{}LOC}}%
      \csname LTb\endcsname%%
      \put(2398,2291){\makebox(0,0)[r]{\strut{}UBS}}%
    }%
    \gplgaddtomacro\gplbacktext{%
      \csname LTb\endcsname%%
      \put(6190,881){\makebox(0,0)[r]{\strut{}0.000400}}%
      \csname LTb\endcsname%%
      \put(6190,1525){\makebox(0,0)[r]{\strut{}0.000403}}%
      \csname LTb\endcsname%%
      \put(6190,2169){\makebox(0,0)[r]{\strut{}0.000405}}%
      \csname LTb\endcsname%%
      \put(6190,2812){\makebox(0,0)[r]{\strut{}0.000408}}%
      \csname LTb\endcsname%%
      \put(6322,661){\makebox(0,0){\strut{}$0.8289$}}%
      \csname LTb\endcsname%%
      \put(7414,661){\makebox(0,0){\strut{}$0.82915$}}%
      \csname LTb\endcsname%%
      \put(8507,661){\makebox(0,0){\strut{}$0.8294$}}%
    }%
    \gplgaddtomacro\gplfronttext{%
    }%
    \gplbacktext
    \put(0,0){\includegraphics{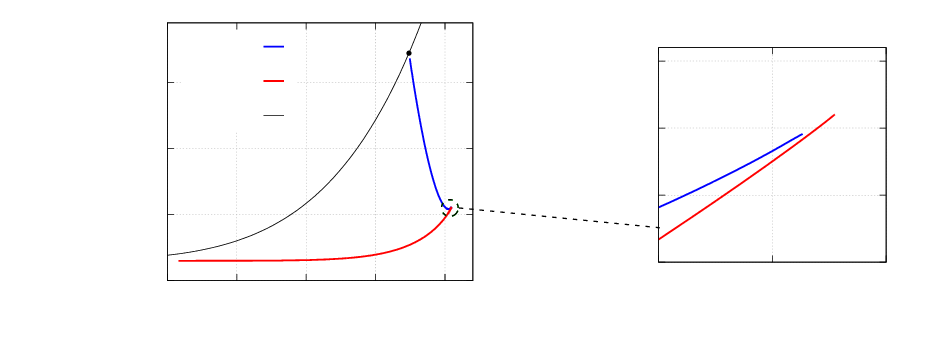}}%
    \gplfronttext
  \end{picture}%
\endgroup

\end{minipage}
\end{minipage}
\captionsetup{width=0.9\textwidth}
\captionof{figure}{\textsl{Phase diagram in the microcanonical (top left) and canonical (top right) ensembles respectively for the three different families of KK black objects in $D = 10$. These plots reproduce and complete those shown in the appendix of \cite{Dias:2017uyv}. Dimensionless horizon area $A_H/L^8$ (middle) and tension $\mc{T}/L^6$ (bottom) as a function of the dimensionless ratio $\beta/L$. The GL critical point is indicated with a solid black disc. The dimensionless mass and relative tension plots are very similar to the ones shown above.}}
\label{microcano}
\end{center}

In Fig.\ \ref{allthermo} we plot various physical quantities, normalized by their value at the GL point, against the normalized relative tension $n/n_\textrm{GL}$. Close to the merger point, our results in $D=10$ show that the physical quantities do not approach their critical values following a spiraling behavior, with presumably infinitely many turning points, as in $D=5,6$ \cite{Kleihaus:2006ee,Kalisch:2017bin}. Instead, the physical quantities of the NUBS and LOC branches merge at a cusp in the phase diagram, with no oscillations. As we discuss in \S\ref{crit}, this behavior is precisely what the double-cone model of \cite{Kol:2002xz} for the merger predicts in $D=10$. Notice that the physical quantities corresponding to both branches emerge from the cusp in the `same direction'.

\begin{center}
\begin{minipage}{\textwidth}
\begin{center}
% GNUPLOT: LaTeX picture with Postscript
\begingroup
  \makeatletter
  \providecommand\color[2][]{%
    \GenericError{(gnuplot) \space\space\space\@spaces}{%
      Package color not loaded in conjunction with
      terminal option `colourtext'%
    }{See the gnuplot documentation for explanation.%
    }{Either use 'blacktext' in gnuplot or load the package
      color.sty in LaTeX.}%
    \renewcommand\color[2][]{}%
  }%
  \providecommand\includegraphics[2][]{%
    \GenericError{(gnuplot) \space\space\space\@spaces}{%
      Package graphicx or graphics not loaded%
    }{See the gnuplot documentation for explanation.%
    }{The gnuplot epslatex terminal needs graphicx.sty or graphics.sty.}%
    \renewcommand\includegraphics[2][]{}%
  }%
  \providecommand\rotatebox[2]{#2}%
  \@ifundefined{ifGPcolor}{%
    \newif\ifGPcolor
    \GPcolortrue
  }{}%
  \@ifundefined{ifGPblacktext}{%
    \newif\ifGPblacktext
    \GPblacktexttrue
  }{}%
  % define a \g@addto@macro without @ in the name:
  \let\gplgaddtomacro\g@addto@macro
  % define empty templates for all commands taking text:
  \gdef\gplbacktext{}%
  \gdef\gplfronttext{}%
  \makeatother
  \ifGPblacktext
    % no textcolor at all
    \def\colorrgb#1{}%
    \def\colorgray#1{}%
  \else
    % gray or color?
    \ifGPcolor
      \def\colorrgb#1{\color[rgb]{#1}}%
      \def\colorgray#1{\color[gray]{#1}}%
      \expandafter\def\csname LTw\endcsname{\color{white}}%
      \expandafter\def\csname LTb\endcsname{\color{black}}%
      \expandafter\def\csname LTa\endcsname{\color{black}}%
      \expandafter\def\csname LT0\endcsname{\color[rgb]{1,0,0}}%
      \expandafter\def\csname LT1\endcsname{\color[rgb]{0,1,0}}%
      \expandafter\def\csname LT2\endcsname{\color[rgb]{0,0,1}}%
      \expandafter\def\csname LT3\endcsname{\color[rgb]{1,0,1}}%
      \expandafter\def\csname LT4\endcsname{\color[rgb]{0,1,1}}%
      \expandafter\def\csname LT5\endcsname{\color[rgb]{1,1,0}}%
      \expandafter\def\csname LT6\endcsname{\color[rgb]{0,0,0}}%
      \expandafter\def\csname LT7\endcsname{\color[rgb]{1,0.3,0}}%
      \expandafter\def\csname LT8\endcsname{\color[rgb]{0.5,0.5,0.5}}%
    \else
      % gray
      \def\colorrgb#1{\color{black}}%
      \def\colorgray#1{\color[gray]{#1}}%
      \expandafter\def\csname LTw\endcsname{\color{white}}%
      \expandafter\def\csname LTb\endcsname{\color{black}}%
      \expandafter\def\csname LTa\endcsname{\color{black}}%
      \expandafter\def\csname LT0\endcsname{\color{black}}%
      \expandafter\def\csname LT1\endcsname{\color{black}}%
      \expandafter\def\csname LT2\endcsname{\color{black}}%
      \expandafter\def\csname LT3\endcsname{\color{black}}%
      \expandafter\def\csname LT4\endcsname{\color{black}}%
      \expandafter\def\csname LT5\endcsname{\color{black}}%
      \expandafter\def\csname LT6\endcsname{\color{black}}%
      \expandafter\def\csname LT7\endcsname{\color{black}}%
      \expandafter\def\csname LT8\endcsname{\color{black}}%
    \fi
  \fi
    \setlength{\unitlength}{0.0500bp}%
    \ifx\gptboxheight\undefined%
      \newlength{\gptboxheight}%
      \newlength{\gptboxwidth}%
      \newsavebox{\gptboxtext}%
    \fi%
    \setlength{\fboxrule}{0.5pt}%
    \setlength{\fboxsep}{1pt}%
\begin{picture}(8640.00,6048.00)%
    \gplgaddtomacro\gplbacktext{%
      \csname LTb\endcsname%%
      \put(594,704){\makebox(0,0)[r]{\strut{}$0$}}%
      \csname LTb\endcsname%%
      \put(594,1868){\makebox(0,0)[r]{\strut{}$0.5$}}%
      \csname LTb\endcsname%%
      \put(594,3033){\makebox(0,0)[r]{\strut{}$1$}}%
      \csname LTb\endcsname%%
      \put(594,4197){\makebox(0,0)[r]{\strut{}$1.5$}}%
      \csname LTb\endcsname%%
      \put(594,5361){\makebox(0,0)[r]{\strut{}$2$}}%
      \csname LTb\endcsname%%
      \put(822,484){\makebox(0,0){\strut{}$0$}}%
      \csname LTb\endcsname%%
      \put(1586,484){\makebox(0,0){\strut{}$0.2$}}%
      \csname LTb\endcsname%%
      \put(2350,484){\makebox(0,0){\strut{}$0.4$}}%
      \csname LTb\endcsname%%
      \put(3114,484){\makebox(0,0){\strut{}$0.6$}}%
      \csname LTb\endcsname%%
      \put(3879,484){\makebox(0,0){\strut{}$0.8$}}%
      \csname LTb\endcsname%%
      \put(4643,484){\makebox(0,0){\strut{}$1$}}%
      \csname LTb\endcsname%%
      \put(1013,1286){\makebox(0,0)[l]{\strut{}LOC}}%
      \put(3191,3918){\makebox(0,0)[l]{\strut{}NUBS}}%
      \put(4674,3064){\makebox(0,0)[l]{\strut{}}}%
    }%
    \gplgaddtomacro\gplfronttext{%
      \csname LTb\endcsname%%
      \put(2684,154){\makebox(0,0){\strut{}$n/n_{\text{GL}}$}}%
      \csname LTb\endcsname%%
      \put(4052,1592){\makebox(0,0)[r]{\strut{}$S/S_{\text{GL}}$}}%
      \csname LTb\endcsname%%
      \put(4052,1262){\makebox(0,0)[r]{\strut{}$M/M_{\text{GL}}$}}%
      \csname LTb\endcsname%%
      \put(4052,932){\makebox(0,0)[r]{\strut{}$T/T_{\text{GL}}$}}%
    }%
    \gplgaddtomacro\gplbacktext{%
      \csname LTb\endcsname%%
      \put(594,704){\makebox(0,0)[r]{\strut{}$0$}}%
      \csname LTb\endcsname%%
      \put(594,1868){\makebox(0,0)[r]{\strut{}$0.5$}}%
      \csname LTb\endcsname%%
      \put(594,3033){\makebox(0,0)[r]{\strut{}$1$}}%
      \csname LTb\endcsname%%
      \put(594,4197){\makebox(0,0)[r]{\strut{}$1.5$}}%
      \csname LTb\endcsname%%
      \put(594,5361){\makebox(0,0)[r]{\strut{}$2$}}%
      \csname LTb\endcsname%%
      \put(822,484){\makebox(0,0){\strut{}$0$}}%
      \csname LTb\endcsname%%
      \put(1586,484){\makebox(0,0){\strut{}$0.2$}}%
      \csname LTb\endcsname%%
      \put(2350,484){\makebox(0,0){\strut{}$0.4$}}%
      \csname LTb\endcsname%%
      \put(3114,484){\makebox(0,0){\strut{}$0.6$}}%
      \csname LTb\endcsname%%
      \put(3879,484){\makebox(0,0){\strut{}$0.8$}}%
      \csname LTb\endcsname%%
      \put(4643,484){\makebox(0,0){\strut{}$1$}}%
      \csname LTb\endcsname%%
      \put(1013,1286){\makebox(0,0)[l]{\strut{}LOC}}%
      \put(3191,3918){\makebox(0,0)[l]{\strut{}NUBS}}%
      \put(4674,3064){\makebox(0,0)[l]{\strut{}}}%
    }%
    \gplgaddtomacro\gplfronttext{%
      \csname LTb\endcsname%%
      \put(2684,154){\makebox(0,0){\strut{}$n/n_{\text{GL}}$}}%
    }%
    \gplgaddtomacro\gplbacktext{%
      \csname LTb\endcsname%%
      \put(5754,440){\makebox(0,0)[r]{\strut{}$0.924$}}%
      \csname LTb\endcsname%%
      \put(5754,1344){\makebox(0,0)[r]{\strut{}$0.928$}}%
      \csname LTb\endcsname%%
      \put(5886,220){\makebox(0,0){\strut{}$0.132$}}%
      \csname LTb\endcsname%%
      \put(6828,220){\makebox(0,0){\strut{}$0.136$}}%
      \csname LTb\endcsname%%
      \put(7771,220){\makebox(0,0){\strut{}$0.14$}}%
      \csname LTb\endcsname%%
      \put(-13431,-151882){\makebox(0,0)[l]{\strut{}LOC}}%
      \put(120859,103496){\makebox(0,0)[l]{\strut{}NUBS}}%
      \put(210418,17647){\makebox(0,0)[l]{\strut{}}}%
    }%
    \gplgaddtomacro\gplfronttext{%
    }%
    \gplgaddtomacro\gplbacktext{%
      \csname LTb\endcsname%%
      \put(5766,2456){\makebox(0,0)[r]{\strut{}$1.82$}}%
      \csname LTb\endcsname%%
      \put(5766,3359){\makebox(0,0)[r]{\strut{}$1.86$}}%
      \csname LTb\endcsname%%
      \put(5898,2236){\makebox(0,0){\strut{}$0.132$}}%
      \csname LTb\endcsname%%
      \put(6836,2236){\makebox(0,0){\strut{}$0.136$}}%
      \csname LTb\endcsname%%
      \put(7773,2236){\makebox(0,0){\strut{}$0.14$}}%
      \csname LTb\endcsname%%
      \put(-13321,-32998){\makebox(0,0)[l]{\strut{}LOC}}%
      \put(120285,-7479){\makebox(0,0)[l]{\strut{}NUBS}}%
      \put(209388,-16029){\makebox(0,0)[l]{\strut{}}}%
    }%
    \gplgaddtomacro\gplfronttext{%
    }%
    \gplgaddtomacro\gplbacktext{%
      \csname LTb\endcsname%%
      \put(5766,4472){\makebox(0,0)[r]{\strut{}$2$}}%
      \csname LTb\endcsname%%
      \put(5766,5375){\makebox(0,0)[r]{\strut{}$2.04$}}%
      \csname LTb\endcsname%%
      \put(5898,4252){\makebox(0,0){\strut{}$0.132$}}%
      \csname LTb\endcsname%%
      \put(6836,4252){\makebox(0,0){\strut{}$0.136$}}%
      \csname LTb\endcsname%%
      \put(7773,4252){\makebox(0,0){\strut{}$0.14$}}%
      \csname LTb\endcsname%%
      \put(-13321,-35047){\makebox(0,0)[l]{\strut{}LOC}}%
      \put(120285,-9528){\makebox(0,0)[l]{\strut{}NUBS}}%
      \put(209388,-18078){\makebox(0,0)[l]{\strut{}}}%
    }%
    \gplgaddtomacro\gplfronttext{%
    }%
    \gplbacktext
    \put(0,0){\includegraphics{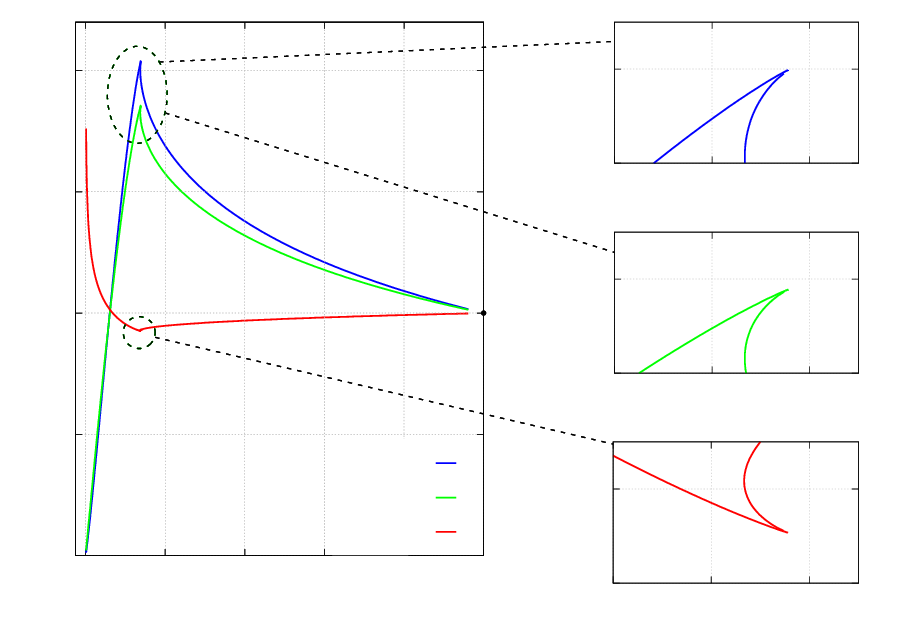}}%
    \gplfronttext
  \end{picture}%
\endgroup

\end{center}
\end{minipage}
\captionsetup{width=0.9\textwidth}
\captionof{figure}{\textsl{Entropy, mass and temperature normalized with respect to the values at the GL point, as a function of the relative tension (with the same normalization). The three mini-plots at the right hand side correspond to zooming at the merger point, as indicated by the dashed lines.}}
\label{allthermo}
\end{center}

%~~~~~~~~~~~~~~~~~~~~~~~~~~~~~~~~~~~~~~~~~~~~~~~
\subsection{Horizon geometry}
\label{geo}
%~~~~~~~~~~~~~~~~~~~~~~~~~~~~~~~~~~~~~~~~~~~~~~
In this subsection we display the behavior of various geometric quantities defined on the horizon along the branches of solutions. Then we present the embeddings of the horizon geometry into flat space to help to visualize the geometry of NUBS and LOC.

We characterize the NUBS using the non-uniformity parameter $\lambda$ defined in  \cite{Gubser:2001ac}, \eqref{eq:horSrad}. In addition, we consider the proper length of the horizon along the $S^1$: 
\beq
L_{\text{hor}} = 2\int_0^{L/2}\dd y\,\sqrt{e^{Q_3}}\Big|_{H}.
\enq

Following \cite{Headrick:2009pv}, for LOC, one can define $R_{\text{eq}}$ as the equatorial radius of the horizon round $S^{D-3}$, $R_{\text{eq}} = r_0\sqrt{Q'_2(r_0,0)}$. Similarly, one defines $L_{\textrm{polar}}$ to be the proper distance from the `south' pole to the `north' pole along the horizon $S^{D-2}$,
\begin{equation}
L_{\textrm{polar}} = 2r_0\int_0^{\pi/2}\dd a\sqrt{Q'_4}\Big|_H,
\end{equation}
and $L_{\textrm{axis}}$ to be the proper distance between the poles along the axis: \beq\label{Lss}
L_{\text{axis}} = 2\int_{r_0}^{L/2}\dd r\sqrt{Q_3'}\Big|_A\,.
\enq
Recall that the asymptotically flat Schwarzschild solution in $D$ dimensions is spherically symmetric and hence it enjoys the symmetry of the full rotation group SO$(D-1)$. On the other hand, LOC break the SO$(D-1)$ symmetry down to SO$(D-2)$, and  only for very small localized black holes, i.e.\ high temperatures, the full SO$(D-1)$ is approximately recovered. We can characterize the deformation of the horizon geometry by comparing the area of the round equatorial horizon $S^{D-3}$, $A_{\text{eq}} \propto R_{\text{eq}}^{D-3}$, and the area of the geodesic $S^{D-3}$ on the horizon that contains both poles, $A_{\text{pol}} \propto R_{\text{pol}}^{D-3}$ with $R_{\text{pol}} = r_0\sqrt{Q_2'(r_0,\pi/2)}$. We compare these two areas defining the eccentricity parameter, \begin{equation}
\epsilon =  (A_{\text{pol}}/A_{\text{eq}}) - 1\,. 
\end{equation}
A spherically symmetric black hole has zero eccentricity and $\epsilon$ diverges for the critical solution. 

The behavior of these geometric quantities along each branch of solutions is displayed in Fig.\ \ref{geometry}. In the top row we display $\epsilon$ and $\lambda$ as functions of $\beta/L$. At high temperatures, LOC are nearly spherically symmetric and the eccentricity is very small. In fact, $\epsilon$ remains quite small until pretty close to the merger with NUBS, where it diverges (notice that the vertical axis is in a log-scale). This explains why perturbation theory works so well for localized black holes in $D=10$ \cite{Dias:2017uyv}, and it is another manifestation of the fact that gravity becomes more localized near the horizon as $D$ grows. The behavior of the non-uniformity parameter $\lambda$ for the NUBS is qualitatively similar. From these two plots it is clear that we managed to get closer to the merger from the LOC branch. From the behavior of $\epsilon$ and $\lambda$ we can estimate that the merger occurs at $\beta_{\text{Merger}} \simeq 0.829L$.

At the bottom of Fig.\ \ref{geometry} we display the remaining geometric quantities as functions of the relative tension $n$ normalized by its value at the GL point. We have added zooms of these plots to better appreciate the region where the various curves merge. In $D=10$ the merger happens at a cusp, with the physical quantities of both the NUBS and the LOC coming out of the cusp in the same direction. This behavior should be contrasted with the $D=5,6$ case, in which a part from the shrinking spirals, the physical quantities for the NUBS and the LOC approach the merger from opposite sides. It would be nice to understand this behavior from the double-cone geometry. From the behavior of $L_{\text{axis}}/L$ and $R_{\text{min}}/L$ as they approach zero, we estimate the value of $n/n_{\text{GL}}$ at the merger to be $n_{\text{Merger}} \simeq 0.139n_{\text{GL}}$.

A useful way to visualize the geometry of $\tau = \textrm{const.}$ sections of the horizon is by embedding them into $(D-1)$-dimensional Euclidean space $\E^{D-1}$, with a flat metric
\begin{equation}
 \dd s^2_{\E^{D-1}} = \dd X^2 + \dd Y^2 + Y^2\dd\Omega_{D-3}^2\,.
 \end{equation}
For NUBS, the horizon geometry can be described as a surface $X = X(y)$, $Y(y) = R(y)$ in $\E^{D-1}$, whilst for LOC one has $X = X(a)$, $Y(a) = r_0\cos a\sqrt{Q_2'}\big|_H$. In each case, the embedding coordinate is given by \beq\begin{split}
\text{NUBS: } \hs{0.5}& X(y) = \int_0^y\dd y'\sqrt{e^{Q_3} - \frac{r_0^2}{4}e^{Q_5}\(\deriv[Q_5]{y'}\)^2}\bigg|_H, \\
\text{LOC: } \hs{0.5}& X(a) = r_0\int_0^a\dd a'\sqrt{Q'_4 - \bigg(\sin a\sqrt{Q'_2} - \frac{\cos a}{2\sqrt{Q_2'}}\deriv[Q_2']{a'}\bigg)^2}\bigg|_H.
\end{split}\enq
In Fig.\ \ref{emb} we plot $Y/L$ vs $X/L$ for some representative solutions, including the most critical ones. We postpone the detailed comparison with the double-cone metric to the next subsection.

\begin{center}
\begin{minipage}{\textwidth}
\begin{minipage}[h!]{0.5\textwidth}
\hs{0.2}% GNUPLOT: LaTeX picture with Postscript
\begingroup
  \makeatletter
  \providecommand\color[2][]{%
    \GenericError{(gnuplot) \space\space\space\@spaces}{%
      Package color not loaded in conjunction with
      terminal option `colourtext'%
    }{See the gnuplot documentation for explanation.%
    }{Either use 'blacktext' in gnuplot or load the package
      color.sty in LaTeX.}%
    \renewcommand\color[2][]{}%
  }%
  \providecommand\includegraphics[2][]{%
    \GenericError{(gnuplot) \space\space\space\@spaces}{%
      Package graphicx or graphics not loaded%
    }{See the gnuplot documentation for explanation.%
    }{The gnuplot epslatex terminal needs graphicx.sty or graphics.sty.}%
    \renewcommand\includegraphics[2][]{}%
  }%
  \providecommand\rotatebox[2]{#2}%
  \@ifundefined{ifGPcolor}{%
    \newif\ifGPcolor
    \GPcolortrue
  }{}%
  \@ifundefined{ifGPblacktext}{%
    \newif\ifGPblacktext
    \GPblacktexttrue
  }{}%
  % define a \g@addto@macro without @ in the name:
  \let\gplgaddtomacro\g@addto@macro
  % define empty templates for all commands taking text:
  \gdef\gplbacktext{}%
  \gdef\gplfronttext{}%
  \makeatother
  \ifGPblacktext
    % no textcolor at all
    \def\colorrgb#1{}%
    \def\colorgray#1{}%
  \else
    % gray or color?
    \ifGPcolor
      \def\colorrgb#1{\color[rgb]{#1}}%
      \def\colorgray#1{\color[gray]{#1}}%
      \expandafter\def\csname LTw\endcsname{\color{white}}%
      \expandafter\def\csname LTb\endcsname{\color{black}}%
      \expandafter\def\csname LTa\endcsname{\color{black}}%
      \expandafter\def\csname LT0\endcsname{\color[rgb]{1,0,0}}%
      \expandafter\def\csname LT1\endcsname{\color[rgb]{0,1,0}}%
      \expandafter\def\csname LT2\endcsname{\color[rgb]{0,0,1}}%
      \expandafter\def\csname LT3\endcsname{\color[rgb]{1,0,1}}%
      \expandafter\def\csname LT4\endcsname{\color[rgb]{0,1,1}}%
      \expandafter\def\csname LT5\endcsname{\color[rgb]{1,1,0}}%
      \expandafter\def\csname LT6\endcsname{\color[rgb]{0,0,0}}%
      \expandafter\def\csname LT7\endcsname{\color[rgb]{1,0.3,0}}%
      \expandafter\def\csname LT8\endcsname{\color[rgb]{0.5,0.5,0.5}}%
    \else
      % gray
      \def\colorrgb#1{\color{black}}%
      \def\colorgray#1{\color[gray]{#1}}%
      \expandafter\def\csname LTw\endcsname{\color{white}}%
      \expandafter\def\csname LTb\endcsname{\color{black}}%
      \expandafter\def\csname LTa\endcsname{\color{black}}%
      \expandafter\def\csname LT0\endcsname{\color{black}}%
      \expandafter\def\csname LT1\endcsname{\color{black}}%
      \expandafter\def\csname LT2\endcsname{\color{black}}%
      \expandafter\def\csname LT3\endcsname{\color{black}}%
      \expandafter\def\csname LT4\endcsname{\color{black}}%
      \expandafter\def\csname LT5\endcsname{\color{black}}%
      \expandafter\def\csname LT6\endcsname{\color{black}}%
      \expandafter\def\csname LT7\endcsname{\color{black}}%
      \expandafter\def\csname LT8\endcsname{\color{black}}%
    \fi
  \fi
    \setlength{\unitlength}{0.0500bp}%
    \ifx\gptboxheight\undefined%
      \newlength{\gptboxheight}%
      \newlength{\gptboxwidth}%
      \newsavebox{\gptboxtext}%
    \fi%
    \setlength{\fboxrule}{0.5pt}%
    \setlength{\fboxsep}{1pt}%
\begin{picture}(4818.00,3118.00)%
    \gplgaddtomacro\gplbacktext{%
      \csname LTb\endcsname%%
      \put(1078,816){\makebox(0,0)[r]{\strut{}$0.001$}}%
      \csname LTb\endcsname%%
      \put(1078,1187){\makebox(0,0)[r]{\strut{}$0.01$}}%
      \csname LTb\endcsname%%
      \put(1078,1559){\makebox(0,0)[r]{\strut{}$0.1$}}%
      \csname LTb\endcsname%%
      \put(1078,1930){\makebox(0,0)[r]{\strut{}$1$}}%
      \csname LTb\endcsname%%
      \put(1078,2302){\makebox(0,0)[r]{\strut{}$10$}}%
      \csname LTb\endcsname%%
      \put(1078,2673){\makebox(0,0)[r]{\strut{}$100$}}%
      \csname LTb\endcsname%%
      \put(1210,484){\makebox(0,0){\strut{}$0.42$}}%
      \csname LTb\endcsname%%
      \put(1807,484){\makebox(0,0){\strut{}$0.5$}}%
      \csname LTb\endcsname%%
      \put(2405,484){\makebox(0,0){\strut{}$0.58$}}%
      \csname LTb\endcsname%%
      \put(3002,484){\makebox(0,0){\strut{}$0.66$}}%
      \csname LTb\endcsname%%
      \put(3600,484){\makebox(0,0){\strut{}$0.74$}}%
      \csname LTb\endcsname%%
      \put(4197,484){\makebox(0,0){\strut{}$0.82$}}%
    }%
    \gplgaddtomacro\gplfronttext{%
      \csname LTb\endcsname%%
      \put(198,1800){\rotatebox{-270}{\makebox(0,0){\strut{}$\epsilon$}}}%
      \put(2815,154){\makebox(0,0){\strut{}$\beta/L$}}%
    }%
    \gplbacktext
    \put(0,0){\includegraphics{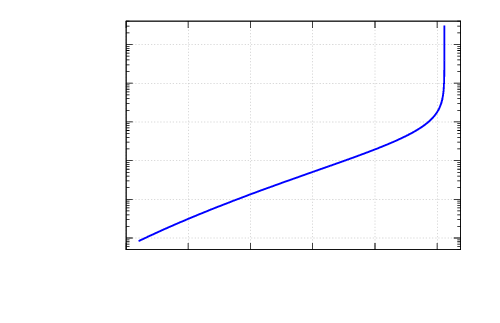}}%
    \gplfronttext
  \end{picture}%
\endgroup

\end{minipage}
\hfill
\begin{minipage}[h!]{0.5\textwidth}
\hs{0.05}% GNUPLOT: LaTeX picture with Postscript
\begingroup
  \makeatletter
  \providecommand\color[2][]{%
    \GenericError{(gnuplot) \space\space\space\@spaces}{%
      Package color not loaded in conjunction with
      terminal option `colourtext'%
    }{See the gnuplot documentation for explanation.%
    }{Either use 'blacktext' in gnuplot or load the package
      color.sty in LaTeX.}%
    \renewcommand\color[2][]{}%
  }%
  \providecommand\includegraphics[2][]{%
    \GenericError{(gnuplot) \space\space\space\@spaces}{%
      Package graphicx or graphics not loaded%
    }{See the gnuplot documentation for explanation.%
    }{The gnuplot epslatex terminal needs graphicx.sty or graphics.sty.}%
    \renewcommand\includegraphics[2][]{}%
  }%
  \providecommand\rotatebox[2]{#2}%
  \@ifundefined{ifGPcolor}{%
    \newif\ifGPcolor
    \GPcolortrue
  }{}%
  \@ifundefined{ifGPblacktext}{%
    \newif\ifGPblacktext
    \GPblacktexttrue
  }{}%
  % define a \g@addto@macro without @ in the name:
  \let\gplgaddtomacro\g@addto@macro
  % define empty templates for all commands taking text:
  \gdef\gplbacktext{}%
  \gdef\gplfronttext{}%
  \makeatother
  \ifGPblacktext
    % no textcolor at all
    \def\colorrgb#1{}%
    \def\colorgray#1{}%
  \else
    % gray or color?
    \ifGPcolor
      \def\colorrgb#1{\color[rgb]{#1}}%
      \def\colorgray#1{\color[gray]{#1}}%
      \expandafter\def\csname LTw\endcsname{\color{white}}%
      \expandafter\def\csname LTb\endcsname{\color{black}}%
      \expandafter\def\csname LTa\endcsname{\color{black}}%
      \expandafter\def\csname LT0\endcsname{\color[rgb]{1,0,0}}%
      \expandafter\def\csname LT1\endcsname{\color[rgb]{0,1,0}}%
      \expandafter\def\csname LT2\endcsname{\color[rgb]{0,0,1}}%
      \expandafter\def\csname LT3\endcsname{\color[rgb]{1,0,1}}%
      \expandafter\def\csname LT4\endcsname{\color[rgb]{0,1,1}}%
      \expandafter\def\csname LT5\endcsname{\color[rgb]{1,1,0}}%
      \expandafter\def\csname LT6\endcsname{\color[rgb]{0,0,0}}%
      \expandafter\def\csname LT7\endcsname{\color[rgb]{1,0.3,0}}%
      \expandafter\def\csname LT8\endcsname{\color[rgb]{0.5,0.5,0.5}}%
    \else
      % gray
      \def\colorrgb#1{\color{black}}%
      \def\colorgray#1{\color[gray]{#1}}%
      \expandafter\def\csname LTw\endcsname{\color{white}}%
      \expandafter\def\csname LTb\endcsname{\color{black}}%
      \expandafter\def\csname LTa\endcsname{\color{black}}%
      \expandafter\def\csname LT0\endcsname{\color{black}}%
      \expandafter\def\csname LT1\endcsname{\color{black}}%
      \expandafter\def\csname LT2\endcsname{\color{black}}%
      \expandafter\def\csname LT3\endcsname{\color{black}}%
      \expandafter\def\csname LT4\endcsname{\color{black}}%
      \expandafter\def\csname LT5\endcsname{\color{black}}%
      \expandafter\def\csname LT6\endcsname{\color{black}}%
      \expandafter\def\csname LT7\endcsname{\color{black}}%
      \expandafter\def\csname LT8\endcsname{\color{black}}%
    \fi
  \fi
    \setlength{\unitlength}{0.0500bp}%
    \ifx\gptboxheight\undefined%
      \newlength{\gptboxheight}%
      \newlength{\gptboxwidth}%
      \newsavebox{\gptboxtext}%
    \fi%
    \setlength{\fboxrule}{0.5pt}%
    \setlength{\fboxsep}{1pt}%
\begin{picture}(4250.00,3118.00)%
    \gplgaddtomacro\gplbacktext{%
      \csname LTb\endcsname%%
      \put(550,704){\makebox(0,0)[r]{\strut{}$0$}}%
      \csname LTb\endcsname%%
      \put(550,1122){\makebox(0,0)[r]{\strut{}$1$}}%
      \csname LTb\endcsname%%
      \put(550,1539){\makebox(0,0)[r]{\strut{}$2$}}%
      \csname LTb\endcsname%%
      \put(550,1957){\makebox(0,0)[r]{\strut{}$3$}}%
      \csname LTb\endcsname%%
      \put(550,2375){\makebox(0,0)[r]{\strut{}$4$}}%
      \csname LTb\endcsname%%
      \put(550,2793){\makebox(0,0)[r]{\strut{}$5$}}%
      \csname LTb\endcsname%%
      \put(682,484){\makebox(0,0){\strut{}$0.767$}}%
      \csname LTb\endcsname%%
      \put(1403,484){\makebox(0,0){\strut{}$0.782$}}%
      \csname LTb\endcsname%%
      \put(2123,484){\makebox(0,0){\strut{}$0.797$}}%
      \csname LTb\endcsname%%
      \put(2844,484){\makebox(0,0){\strut{}$0.812$}}%
      \csname LTb\endcsname%%
      \put(3565,484){\makebox(0,0){\strut{}$0.827$}}%
    }%
    \gplgaddtomacro\gplfronttext{%
      \csname LTb\endcsname%%
      \put(198,1800){\rotatebox{-270}{\makebox(0,0){\strut{}$\lambda$}}}%
      \put(2267,154){\makebox(0,0){\strut{}$\beta/L$}}%
    }%
    \gplbacktext
    \put(0,0){\includegraphics{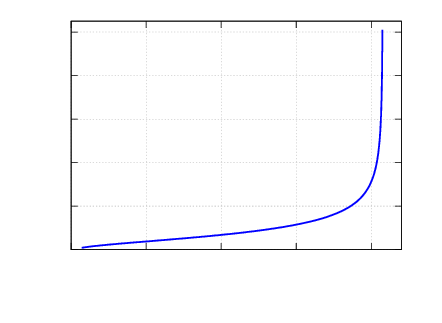}}%
    \gplfronttext
  \end{picture}%
\endgroup

\end{minipage}
\begin{center}
% GNUPLOT: LaTeX picture with Postscript
\begingroup
  \makeatletter
  \providecommand\color[2][]{%
    \GenericError{(gnuplot) \space\space\space\@spaces}{%
      Package color not loaded in conjunction with
      terminal option `colourtext'%
    }{See the gnuplot documentation for explanation.%
    }{Either use 'blacktext' in gnuplot or load the package
      color.sty in LaTeX.}%
    \renewcommand\color[2][]{}%
  }%
  \providecommand\includegraphics[2][]{%
    \GenericError{(gnuplot) \space\space\space\@spaces}{%
      Package graphicx or graphics not loaded%
    }{See the gnuplot documentation for explanation.%
    }{The gnuplot epslatex terminal needs graphicx.sty or graphics.sty.}%
    \renewcommand\includegraphics[2][]{}%
  }%
  \providecommand\rotatebox[2]{#2}%
  \@ifundefined{ifGPcolor}{%
    \newif\ifGPcolor
    \GPcolortrue
  }{}%
  \@ifundefined{ifGPblacktext}{%
    \newif\ifGPblacktext
    \GPblacktexttrue
  }{}%
  % define a \g@addto@macro without @ in the name:
  \let\gplgaddtomacro\g@addto@macro
  % define empty templates for all commands taking text:
  \gdef\gplbacktext{}%
  \gdef\gplfronttext{}%
  \makeatother
  \ifGPblacktext
    % no textcolor at all
    \def\colorrgb#1{}%
    \def\colorgray#1{}%
  \else
    % gray or color?
    \ifGPcolor
      \def\colorrgb#1{\color[rgb]{#1}}%
      \def\colorgray#1{\color[gray]{#1}}%
      \expandafter\def\csname LTw\endcsname{\color{white}}%
      \expandafter\def\csname LTb\endcsname{\color{black}}%
      \expandafter\def\csname LTa\endcsname{\color{black}}%
      \expandafter\def\csname LT0\endcsname{\color[rgb]{1,0,0}}%
      \expandafter\def\csname LT1\endcsname{\color[rgb]{0,1,0}}%
      \expandafter\def\csname LT2\endcsname{\color[rgb]{0,0,1}}%
      \expandafter\def\csname LT3\endcsname{\color[rgb]{1,0,1}}%
      \expandafter\def\csname LT4\endcsname{\color[rgb]{0,1,1}}%
      \expandafter\def\csname LT5\endcsname{\color[rgb]{1,1,0}}%
      \expandafter\def\csname LT6\endcsname{\color[rgb]{0,0,0}}%
      \expandafter\def\csname LT7\endcsname{\color[rgb]{1,0.3,0}}%
      \expandafter\def\csname LT8\endcsname{\color[rgb]{0.5,0.5,0.5}}%
    \else
      % gray
      \def\colorrgb#1{\color{black}}%
      \def\colorgray#1{\color[gray]{#1}}%
      \expandafter\def\csname LTw\endcsname{\color{white}}%
      \expandafter\def\csname LTb\endcsname{\color{black}}%
      \expandafter\def\csname LTa\endcsname{\color{black}}%
      \expandafter\def\csname LT0\endcsname{\color{black}}%
      \expandafter\def\csname LT1\endcsname{\color{black}}%
      \expandafter\def\csname LT2\endcsname{\color{black}}%
      \expandafter\def\csname LT3\endcsname{\color{black}}%
      \expandafter\def\csname LT4\endcsname{\color{black}}%
      \expandafter\def\csname LT5\endcsname{\color{black}}%
      \expandafter\def\csname LT6\endcsname{\color{black}}%
      \expandafter\def\csname LT7\endcsname{\color{black}}%
      \expandafter\def\csname LT8\endcsname{\color{black}}%
    \fi
  \fi
    \setlength{\unitlength}{0.0500bp}%
    \ifx\gptboxheight\undefined%
      \newlength{\gptboxheight}%
      \newlength{\gptboxwidth}%
      \newsavebox{\gptboxtext}%
    \fi%
    \setlength{\fboxrule}{0.5pt}%
    \setlength{\fboxsep}{1pt}%
\begin{picture}(8640.00,6048.00)%
    \gplgaddtomacro\gplbacktext{%
      \csname LTb\endcsname%%
      \put(594,1005){\makebox(0,0)[r]{\strut{}$0$}}%
      \csname LTb\endcsname%%
      \put(594,1608){\makebox(0,0)[r]{\strut{}$0.2$}}%
      \csname LTb\endcsname%%
      \put(594,2211){\makebox(0,0)[r]{\strut{}$0.4$}}%
      \csname LTb\endcsname%%
      \put(594,2813){\makebox(0,0)[r]{\strut{}$0.6$}}%
      \csname LTb\endcsname%%
      \put(594,3416){\makebox(0,0)[r]{\strut{}$0.8$}}%
      \csname LTb\endcsname%%
      \put(594,4019){\makebox(0,0)[r]{\strut{}$1$}}%
      \csname LTb\endcsname%%
      \put(594,4622){\makebox(0,0)[r]{\strut{}$1.2$}}%
      \csname LTb\endcsname%%
      \put(594,5224){\makebox(0,0)[r]{\strut{}$1.4$}}%
      \csname LTb\endcsname%%
      \put(594,5827){\makebox(0,0)[r]{\strut{}$1.6$}}%
      \csname LTb\endcsname%%
      \put(1082,484){\makebox(0,0){\strut{}$0$}}%
      \csname LTb\endcsname%%
      \put(1794,484){\makebox(0,0){\strut{}$0.2$}}%
      \csname LTb\endcsname%%
      \put(2506,484){\makebox(0,0){\strut{}$0.4$}}%
      \csname LTb\endcsname%%
      \put(3219,484){\makebox(0,0){\strut{}$0.6$}}%
      \csname LTb\endcsname%%
      \put(3931,484){\makebox(0,0){\strut{}$0.8$}}%
      \csname LTb\endcsname%%
      \put(4643,484){\makebox(0,0){\strut{}$1$}}%
      \put(797,1186){\makebox(0,0)[l]{\strut{}$\displaystyle\frac{L_{\text{axis}}}{L}$}}%
      \put(1260,4019){\makebox(0,0)[l]{\strut{}$L_{\text{polar}}/L$}}%
      \put(1367,2753){\makebox(0,0)[l]{\strut{}$R_{\text{eq}}/L$}}%
      \put(3397,1729){\makebox(0,0)[l]{\strut{}$R_{\text{min}}/L$}}%
      \put(3397,2452){\makebox(0,0)[l]{\strut{}$R_{\text{max}}/L$}}%
      \put(3397,4471){\makebox(0,0)[l]{\strut{}$L_{\text{hor}}/L$}}%
    }%
    \gplgaddtomacro\gplfronttext{%
      \csname LTb\endcsname%%
      \put(2684,154){\makebox(0,0){\strut{}$n/n_{\text{GL}}$}}%
    }%
    \gplgaddtomacro\gplbacktext{%
      \csname LTb\endcsname%%
      \put(5766,440){\makebox(0,0)[r]{\strut{}$0$}}%
      \csname LTb\endcsname%%
      \put(5766,827){\makebox(0,0)[r]{\strut{}$0.04$}}%
      \csname LTb\endcsname%%
      \put(5766,1215){\makebox(0,0)[r]{\strut{}$0.08$}}%
      \csname LTb\endcsname%%
      \put(5766,1602){\makebox(0,0)[r]{\strut{}$0.12$}}%
      \csname LTb\endcsname%%
      \put(5898,220){\makebox(0,0){\strut{}$0.132$}}%
      \csname LTb\endcsname%%
      \put(6836,220){\makebox(0,0){\strut{}$0.136$}}%
      \csname LTb\endcsname%%
      \put(7773,220){\makebox(0,0){\strut{}$0.14$}}%
      \put(-43793,1021){\makebox(0,0)[l]{\strut{}$\displaystyle\frac{L_{\text{axis}}}{L}$}}%
      \put(-13321,10126){\makebox(0,0)[l]{\strut{}$L_{\text{polar}}/L$}}%
      \put(-6289,6058){\makebox(0,0)[l]{\strut{}$R_{\text{eq}}/L$}}%
      \put(127317,2765){\makebox(0,0)[l]{\strut{}$R_{\text{min}}/L$}}%
      \put(127317,5089){\makebox(0,0)[l]{\strut{}$R_{\text{max}}/L$}}%
      \put(127317,11579){\makebox(0,0)[l]{\strut{}$L_{\text{hor}}/L$}}%
    }%
    \gplgaddtomacro\gplfronttext{%
    }%
    \gplgaddtomacro\gplbacktext{%
      \csname LTb\endcsname%%
      \put(5754,2456){\makebox(0,0)[r]{\strut{}$0.455$}}%
      \csname LTb\endcsname%%
      \put(5754,3359){\makebox(0,0)[r]{\strut{}$0.457$}}%
      \csname LTb\endcsname%%
      \put(5886,2236){\makebox(0,0){\strut{}$0.132$}}%
      \csname LTb\endcsname%%
      \put(6828,2236){\makebox(0,0){\strut{}$0.136$}}%
      \csname LTb\endcsname%%
      \put(7771,2236){\makebox(0,0){\strut{}$0.14$}}%
      \put(-44059,-175950){\makebox(0,0)[l]{\strut{}$\displaystyle\frac{L_{\text{axis}}}{L}$}}%
      \put(-13431,248614){\makebox(0,0)[l]{\strut{}$L_{\text{polar}}/L$}}%
      \put(-6363,58914){\makebox(0,0)[l]{\strut{}$R_{\text{eq}}/L$}}%
      \put(127927,-94650){\makebox(0,0)[l]{\strut{}$R_{\text{min}}/L$}}%
      \put(127927,13748){\makebox(0,0)[l]{\strut{}$R_{\text{max}}/L$}}%
      \put(127927,316364){\makebox(0,0)[l]{\strut{}$L_{\text{hor}}/L$}}%
    }%
    \gplgaddtomacro\gplfronttext{%
    }%
    \gplgaddtomacro\gplbacktext{%
      \csname LTb\endcsname%%
      \put(5766,4472){\makebox(0,0)[r]{\strut{}$1.45$}}%
      \csname LTb\endcsname%%
      \put(5766,5014){\makebox(0,0)[r]{\strut{}$1.49$}}%
      \csname LTb\endcsname%%
      \put(5766,5556){\makebox(0,0)[r]{\strut{}$1.53$}}%
      \csname LTb\endcsname%%
      \put(5898,4252){\makebox(0,0){\strut{}$0.132$}}%
      \csname LTb\endcsname%%
      \put(6836,4252){\makebox(0,0){\strut{}$0.136$}}%
      \csname LTb\endcsname%%
      \put(7773,4252){\makebox(0,0){\strut{}$0.14$}}%
      \put(-43793,-14360){\makebox(0,0)[l]{\strut{}$\displaystyle\frac{L_{\text{axis}}}{L}$}}%
      \put(-13321,-1623){\makebox(0,0)[l]{\strut{}$L_{\text{polar}}/L$}}%
      \put(-6289,-7314){\makebox(0,0)[l]{\strut{}$R_{\text{eq}}/L$}}%
      \put(127317,-11921){\makebox(0,0)[l]{\strut{}$R_{\text{min}}/L$}}%
      \put(127317,-8669){\makebox(0,0)[l]{\strut{}$R_{\text{max}}/L$}}%
      \put(127317,407){\makebox(0,0)[l]{\strut{}$L_{\text{hor}}/L$}}%
    }%
    \gplgaddtomacro\gplfronttext{%
    }%
    \gplbacktext
    \put(0,0){\includegraphics{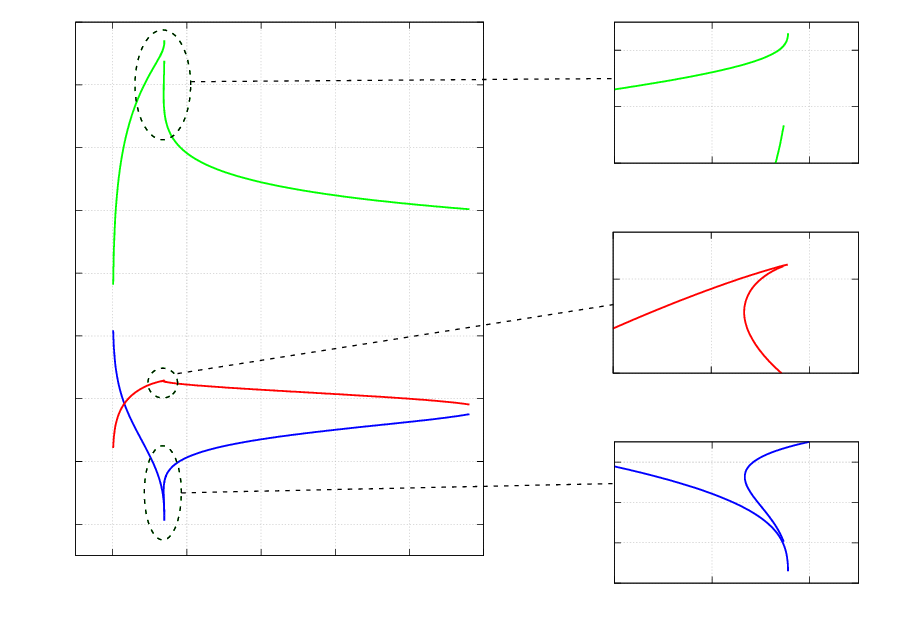}}%
    \gplfronttext
  \end{picture}%
\endgroup

\end{center}
\end{minipage}
\captionsetup{width=0.9\textwidth}
\captionof{figure}{\textsl{Eccentricity (top left) and non-uniformity parameter (top right) as a function of the dimensionless inverse temperature. These quantities give a direct measure of the deformation of LOC and NUBS respectively. Different geometrical lengths and radii for NUBS and LOC (bottom) as a function of the relative tension normalized at the GL threshold point.}}
\label{geometry}
\end{center}

{\begin{figure}[h!]
\centering
\input{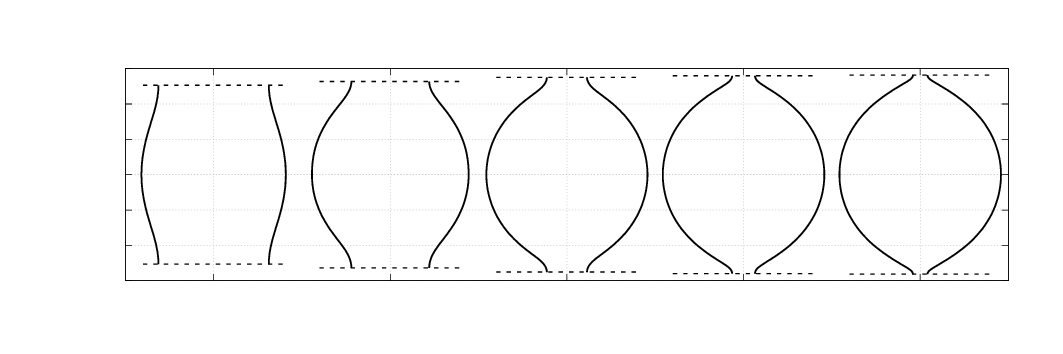}

\vs{0.25}

\input{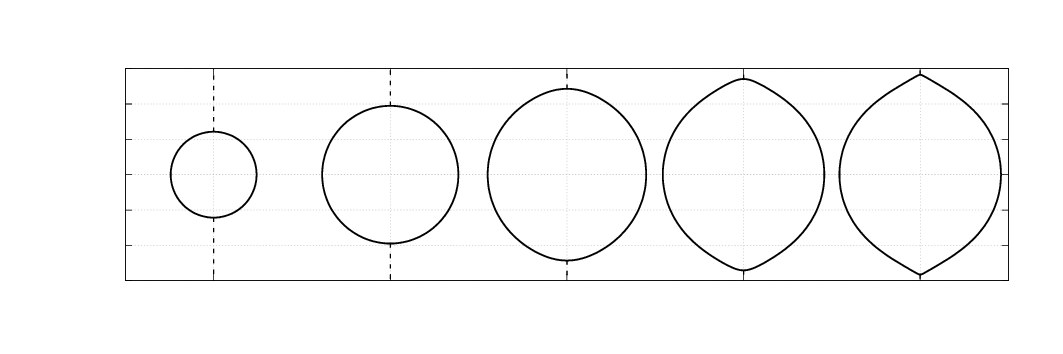}
\captionsetup{width=0.9\textwidth}
\captionof{figure}{\textsl{Embedding of the spatial cross-section of the horizon into Euclidean space for different NUBS (top) and LOC (bottom). For NUBS, from left to right, $r_0 = 0.74328$ ($\lambda\sim 0.1 $), $r_0 = 0.77216$ ($\lambda \sim 0.5$), $r_0 = 0.78950$ ($\lambda \sim 1.5$), $r_0 = 0.79156$ ($\lambda \sim 3$) and $r_0 = 0.79184$ ($\lambda \sim 5$). For LOC the axis is parallel to $X/L$ and represented by a dashed line starting at the poles. From left to right: $\kappa = 2.4$ ($\epsilon \sim 10^{-3}$),  $\kappa = 1.5$ ($\epsilon \sim 10^{-1}$),  $\kappa = 1.29$ ($\epsilon \sim 10^0$),  $\kappa = 1.26341$ ($\epsilon \sim 10$) and $\kappa = 1.26277$ ($\epsilon \sim 3\cdot10^2$). Note that the embeddings look `rounder' or `fatter' compared to the ones in lower dimensions; this is just a manifestation that gravity becomes more localized as $D$ increases.}}
\label{emb}
\end{figure}}

%~~~~~~~~~~~~~~~~~~~~~~~~~~~~~~~~~~~~~~~~~~~~~~~
\subsection{Critical behavior at the merger point}
\label{crit}
\begin{figure}[h!]
\centering
\input{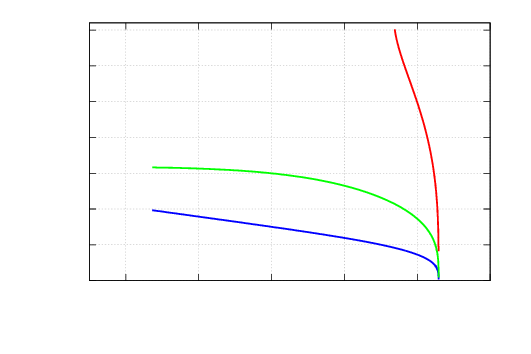}
\captionsetup{width=0.9\textwidth}
\captionof{figure}{\textsl{Euclidean time radius and axis radius as a function of the dimensionless inverse temperature. At the horizon points along the axis, i.e.\ at the extremes of the $S^1_{\text{axis}}$, the euclidean circle $S^1_{\beta}$ has zero size. Then the fibration of one circle on the other gives a topological $S^2$. According to this figure, as we approach the merger with the non-uniform branch, $R_\tau \sim R_{\text{axis}}/L$: the 2-sphere is round.}}
\label{RtauRaxis}
\end{figure}

%~~~~~~~~~~~~~~~~~~~~~~~~~~~~~~~~~~~~~~~~~~~~~~
Kol argued that the merger between the NUBS and the LOC implies a topology change not only of the horizon geometry but in fact of the whole Euclidean manifold \cite{Kol:2002xz}. This is a much stronger statement than simply considering the change of the topology of the horizon. Moreover, \cite{Kol:2002xz} conjectured that this topology change of the Euclidean manifold should locally be controlled by a Ricci-flat double-cone over $S^2\times S^{D-3}$:
\begin{equation}
\dd s^2 = \dd\rho^2 + \frac{\rho^2}{D-2}\(\dd\Omega_{(2)}^2 + (D-4)\dd\Omega_{(D-3)}^2\)\,.
\label{eqn:2cone}
\end{equation}
This double-cone arises as follows. Both the NUBS and the LOC possess an explicit SO$(D-2)$ spherical symmetry which must be inherited by the critical metric, i.e.\ it must contain a round $S^{D-3}$. The $S^2$ is less obvious and its origin is the following \cite{Kol:2002xz}: away from the waist, the Euclidean time, which is periodic to avoid a conical singularity at the horizon, is fibered over an interval whose endpoints are on the horizon, thus giving rise to a two-sphere. Such an $S^2$ is finite everywhere on the localized phase whilst it is contractible to zero size in the black string phase (see \cite{Kol:2004ww} for a nice depiction). On the localized phase, one can compute the radius of this sphere on the symmetry axis at the equidistant points from the poles of the horizon $S^{D-2}$. By symmetry, this corresponds to the equatorial radius of the $S^2$ and is given by
\begin{equation}
R_\tau = \frac{\kappa}{2\pi}\(\frac{L}{2}-r_0\)\sqrt{Q_1'(L/2,\pi/2)}\,.
\end{equation}  
One can compare it to the radius of this $S^2$ along the symmetry axis, $R_\textrm{axis} = L_\textrm{axis}/(2\pi)$, along the branch of LOC. See Fig.\ \ref{RtauRaxis}. From this plot we see that $R_\tau \sim R_\textrm{axis}$ as the solutions approach the merger and both radii tend to zero. This shows that the $S^2$ becomes round as it shrinks, just as the double-cone model of \cite{Kol:2002xz} predicts. Also shown in this plot is the minimum radius of the horizon $S^{D-3}$, $R_\textrm{min}$, on the NUBS. This quantity also shrinks to zero at the merger. 

\begin{figure}[h!]
\centering
\input{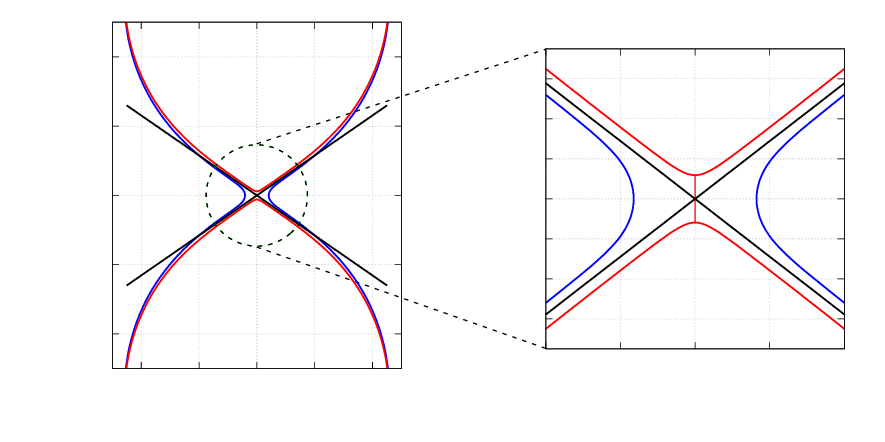}
\captionsetup{width=0.9\textwidth}
\captionof{figure}{\textsl{Comparison between the embeddings into $\mathbb{E}^9$ space of the most critical NUBS (blue line) and LOC (red line) that we have found and the Ricci-flat cone (black line). Clearly, both geometries approximate quite well the double-cone metric.}}
\label{fig:embcone}
\end{figure}

One can further test the double-cone model of the merger by considering the embedding of the $\tau = \textrm{const.}$ section of \eqref{eqn:2cone} into Euclidean $\mathbb E^{D-1}$ space. The embedding coordinates of the double-cone metric \eqref{eqn:2cone} are simply given by
\begin{equation}
X(\rho) = \rho\sqrt{\frac{2}{D-2}}, \hs{0.75} Y(\rho) = \rho\sqrt{\frac{D-4}{D-2}}\,.
\end{equation}
In Fig.\ \ref{fig:embcone} we compare the embedding of the double-cone in $D=10$ dimensions with the embeddings corresponding to the most critical LOC (red) and NUBS (blue) solutions that we have found. As this plot shows, the double-cone can be smoothed in two different ways, each one leading to one of the phases at each side of the transition.

One can consider deformations of the double-cone metric of the form \cite{Kol:2002xz}: 
\beq
\dd s^2 = \dd\rho^2 + \frac{\rho^2}{D-2}\(e^{\epsilon(\rho)}\dd\Omega_{(2)}^2 + (D-4)e^{-\frac{2}{D-3}\epsilon(\rho)}\dd\Omega_{(D-3)}^2\).
\enq
The linearized perturbations satisfy the following equation of motion: \beq
\epsilon''(\rho) + \frac{D-1}{\rho}\epsilon'(\rho) + \frac{2(D-2)}{\rho^2}\epsilon(\rho) = 0\,,
\enq
and, in for any $D\neq 10$, solutions of this equation are given by
\beq
\epsilon(\rho) =  c_+\rho^{s_+} + c_-\rho^{s_-},
\label{eqn:conelin1}
\enq
with \beq
s_\pm = \frac{D-2}{2}\bigg(-1\pm i\sqrt{\frac{8}{D-2}-1}\bigg).
\enq
For $D < 10$, the imaginary part of $s_\pm$ causes oscillations in $\epsilon(\rho)$, while for $D>10$ there are two independent (real) powers. Furthermore, \cite{Kol:2005vy} argued that the behavior of the deformations of the double-cone metric \eqref{eqn:conelin1} should be reflected in the behavior of the physical quantities of NUBS and LOC sufficiently close to criticality. The argument goes as follows: if the zero mode $\epsilon(\rho)$ measures the deviation from the double-cone, then any physical quantity $Q$ near the critical solution should behave as \begin{equation}
\delta Q = C_+ \left(\frac{\rho}{\rho_0}\right)^{s_+} + C_- \left(\frac{\rho}{\rho_0}\right)^{s_-} = \tilde C_+ \rho_0^{-s_+} + \tilde C_- \rho_0^{-s_-}\,,
\end{equation}
where $\delta Q \equiv Q-Q_c$ and $\rho_0$ is the typical length scale associated to the smooth cone. Recently, \cite{Kalisch:2017bin} has beautifully confirmed this prediction in $D=5,6$.

The linearized solutions \eqref{eqn:conelin1} degenerate in $D=10$. Hence this is the critical dimension of the double-cone metric \cite{Kol:2002xz}. In this degenerate case, Frobenius' method gives two independent solutions of the form: \beq\label{epsilon10d}
\epsilon^{D = 10}(\rho) \sim c_1 \,\rho^4 + c_2\,\rho^4\ln\rho\,.
\enq
In the remaining of this subsection, we fit the different physical quantities of the near critical solutions that we have constructed according to the double-cone's prediction \eqref{epsilon10d}. Without loss of generality, for any physical quantity near the merger we have
\begin{equation}
Q(x) = Q_c + a\,x^{b}\,(c+d\,\ln x)\,,
\label{eq:fits}
\end{equation}
where $\{a,b,c,d\}$ are the fitting parameters and $x$ measures the distance to the critical solution. We consider the following dimensionless quantities that tend to zero at the merger:
\beq
x_{\text{NUBS}} = \frac{R_{\text{min}}}{r_0^{\text{GL}}}, \hs{0.75} x_{\text{LOC}} = \frac{L_{\text{axis}}}{L},
\enq
where $L$ is the length of the KK circle and $r_0^{\text{GL}}$ is the horizon radius of the black string at the GL instability point given in \S\ref{ubs}. Any other definition of $x$ should give equivalent results up to a rescaling. We use \texttt{Mathematica}'s \texttt{FindFit} routine to carry out the fits.    

{In Fig.\ \ref{critdata} we present the fits for the mass (normalized with respect to the values of a UBS at the marginal GL point) for the NUBS and LOC branches. The other physical quantities behave in a qualitatively similar way and we do not present the fits here. Note that in contrast to the $D=5,6$ cases, in $D=10$ the physical quantities do not present any oscillations as they approach their critical values. In fact, the fits clearly show that the approach to the critical value is governed by a power law with a logarithmic correction, in very good agreement with the double-cone prediction \eqref{epsilon10d}. 

\vs{0.2}
\begin{center}
\begin{minipage}{\textwidth}
\begin{minipage}[h!]{0.5\textwidth}
\begin{center}
\hs{1.25}{\bf \textsf{Non-uniform black strings}}
\end{center}
\end{minipage}
\begin{minipage}[h!]{0.5\textwidth}
\begin{center}
\hs{1.25}{\bf \textsf{Localized black holes}}
\end{center}
\end{minipage}
\begin{minipage}[h!]{0.5\textwidth}
% GNUPLOT: LaTeX picture with Postscript
\begingroup
  \makeatletter
  \providecommand\color[2][]{%
    \GenericError{(gnuplot) \space\space\space\@spaces}{%
      Package color not loaded in conjunction with
      terminal option `colourtext'%
    }{See the gnuplot documentation for explanation.%
    }{Either use 'blacktext' in gnuplot or load the package
      color.sty in LaTeX.}%
    \renewcommand\color[2][]{}%
  }%
  \providecommand\includegraphics[2][]{%
    \GenericError{(gnuplot) \space\space\space\@spaces}{%
      Package graphicx or graphics not loaded%
    }{See the gnuplot documentation for explanation.%
    }{The gnuplot epslatex terminal needs graphicx.sty or graphics.sty.}%
    \renewcommand\includegraphics[2][]{}%
  }%
  \providecommand\rotatebox[2]{#2}%
  \@ifundefined{ifGPcolor}{%
    \newif\ifGPcolor
    \GPcolortrue
  }{}%
  \@ifundefined{ifGPblacktext}{%
    \newif\ifGPblacktext
    \GPblacktexttrue
  }{}%
  % define a \g@addto@macro without @ in the name:
  \let\gplgaddtomacro\g@addto@macro
  % define empty templates for all commands taking text:
  \gdef\gplbacktext{}%
  \gdef\gplfronttext{}%
  \makeatother
  \ifGPblacktext
    % no textcolor at all
    \def\colorrgb#1{}%
    \def\colorgray#1{}%
  \else
    % gray or color?
    \ifGPcolor
      \def\colorrgb#1{\color[rgb]{#1}}%
      \def\colorgray#1{\color[gray]{#1}}%
      \expandafter\def\csname LTw\endcsname{\color{white}}%
      \expandafter\def\csname LTb\endcsname{\color{black}}%
      \expandafter\def\csname LTa\endcsname{\color{black}}%
      \expandafter\def\csname LT0\endcsname{\color[rgb]{1,0,0}}%
      \expandafter\def\csname LT1\endcsname{\color[rgb]{0,1,0}}%
      \expandafter\def\csname LT2\endcsname{\color[rgb]{0,0,1}}%
      \expandafter\def\csname LT3\endcsname{\color[rgb]{1,0,1}}%
      \expandafter\def\csname LT4\endcsname{\color[rgb]{0,1,1}}%
      \expandafter\def\csname LT5\endcsname{\color[rgb]{1,1,0}}%
      \expandafter\def\csname LT6\endcsname{\color[rgb]{0,0,0}}%
      \expandafter\def\csname LT7\endcsname{\color[rgb]{1,0.3,0}}%
      \expandafter\def\csname LT8\endcsname{\color[rgb]{0.5,0.5,0.5}}%
    \else
      % gray
      \def\colorrgb#1{\color{black}}%
      \def\colorgray#1{\color[gray]{#1}}%
      \expandafter\def\csname LTw\endcsname{\color{white}}%
      \expandafter\def\csname LTb\endcsname{\color{black}}%
      \expandafter\def\csname LTa\endcsname{\color{black}}%
      \expandafter\def\csname LT0\endcsname{\color{black}}%
      \expandafter\def\csname LT1\endcsname{\color{black}}%
      \expandafter\def\csname LT2\endcsname{\color{black}}%
      \expandafter\def\csname LT3\endcsname{\color{black}}%
      \expandafter\def\csname LT4\endcsname{\color{black}}%
      \expandafter\def\csname LT5\endcsname{\color{black}}%
      \expandafter\def\csname LT6\endcsname{\color{black}}%
      \expandafter\def\csname LT7\endcsname{\color{black}}%
      \expandafter\def\csname LT8\endcsname{\color{black}}%
    \fi
  \fi
    \setlength{\unitlength}{0.0500bp}%
    \ifx\gptboxheight\undefined%
      \newlength{\gptboxheight}%
      \newlength{\gptboxwidth}%
      \newsavebox{\gptboxtext}%
    \fi%
    \setlength{\fboxrule}{0.5pt}%
    \setlength{\fboxsep}{1pt}%
\begin{picture}(5102.00,3400.00)%
    \gplgaddtomacro\gplbacktext{%
      \csname LTb\endcsname%%
      \put(814,704){\makebox(0,0)[r]{\strut{}$1$}}%
      \csname LTb\endcsname%%
      \put(814,1236){\makebox(0,0)[r]{\strut{}$1.2$}}%
      \csname LTb\endcsname%%
      \put(814,1769){\makebox(0,0)[r]{\strut{}$1.4$}}%
      \csname LTb\endcsname%%
      \put(814,2301){\makebox(0,0)[r]{\strut{}$1.6$}}%
      \csname LTb\endcsname%%
      \put(814,2833){\makebox(0,0)[r]{\strut{}$1.8$}}%
      \csname LTb\endcsname%%
      \put(946,484){\makebox(0,0){\strut{}$0$}}%
      \csname LTb\endcsname%%
      \put(1698,484){\makebox(0,0){\strut{}$0.2$}}%
      \csname LTb\endcsname%%
      \put(2450,484){\makebox(0,0){\strut{}$0.4$}}%
      \csname LTb\endcsname%%
      \put(3201,484){\makebox(0,0){\strut{}$0.6$}}%
      \csname LTb\endcsname%%
      \put(3953,484){\makebox(0,0){\strut{}$0.8$}}%
      \csname LTb\endcsname%%
      \put(4705,484){\makebox(0,0){\strut{}$1$}}%
    }%
    \gplgaddtomacro\gplfronttext{%
      \csname LTb\endcsname%%
      \put(198,1941){\rotatebox{-270}{\makebox(0,0){\strut{}$M/M_{\text{GL}}$}}}%
      \put(2825,154){\makebox(0,0){\strut{}$x_{\text{NUBS}}$}}%
      \csname LTb\endcsname%%
      \put(4114,2951){\makebox(0,0)[r]{\strut{}data}}%
      \csname LTb\endcsname%%
      \put(4114,2621){\makebox(0,0)[r]{\strut{}fit}}%
    }%
    \gplbacktext
    \put(0,0){\includegraphics{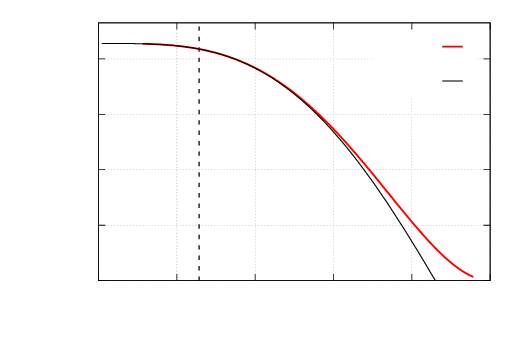}}%
    \gplfronttext
  \end{picture}%
\endgroup

\end{minipage}
\hfill
\begin{minipage}[h!]{0.5\textwidth}
% GNUPLOT: LaTeX picture with Postscript
\begingroup
  \makeatletter
  \providecommand\color[2][]{%
    \GenericError{(gnuplot) \space\space\space\@spaces}{%
      Package color not loaded in conjunction with
      terminal option `colourtext'%
    }{See the gnuplot documentation for explanation.%
    }{Either use 'blacktext' in gnuplot or load the package
      color.sty in LaTeX.}%
    \renewcommand\color[2][]{}%
  }%
  \providecommand\includegraphics[2][]{%
    \GenericError{(gnuplot) \space\space\space\@spaces}{%
      Package graphicx or graphics not loaded%
    }{See the gnuplot documentation for explanation.%
    }{The gnuplot epslatex terminal needs graphicx.sty or graphics.sty.}%
    \renewcommand\includegraphics[2][]{}%
  }%
  \providecommand\rotatebox[2]{#2}%
  \@ifundefined{ifGPcolor}{%
    \newif\ifGPcolor
    \GPcolortrue
  }{}%
  \@ifundefined{ifGPblacktext}{%
    \newif\ifGPblacktext
    \GPblacktexttrue
  }{}%
  % define a \g@addto@macro without @ in the name:
  \let\gplgaddtomacro\g@addto@macro
  % define empty templates for all commands taking text:
  \gdef\gplbacktext{}%
  \gdef\gplfronttext{}%
  \makeatother
  \ifGPblacktext
    % no textcolor at all
    \def\colorrgb#1{}%
    \def\colorgray#1{}%
  \else
    % gray or color?
    \ifGPcolor
      \def\colorrgb#1{\color[rgb]{#1}}%
      \def\colorgray#1{\color[gray]{#1}}%
      \expandafter\def\csname LTw\endcsname{\color{white}}%
      \expandafter\def\csname LTb\endcsname{\color{black}}%
      \expandafter\def\csname LTa\endcsname{\color{black}}%
      \expandafter\def\csname LT0\endcsname{\color[rgb]{1,0,0}}%
      \expandafter\def\csname LT1\endcsname{\color[rgb]{0,1,0}}%
      \expandafter\def\csname LT2\endcsname{\color[rgb]{0,0,1}}%
      \expandafter\def\csname LT3\endcsname{\color[rgb]{1,0,1}}%
      \expandafter\def\csname LT4\endcsname{\color[rgb]{0,1,1}}%
      \expandafter\def\csname LT5\endcsname{\color[rgb]{1,1,0}}%
      \expandafter\def\csname LT6\endcsname{\color[rgb]{0,0,0}}%
      \expandafter\def\csname LT7\endcsname{\color[rgb]{1,0.3,0}}%
      \expandafter\def\csname LT8\endcsname{\color[rgb]{0.5,0.5,0.5}}%
    \else
      % gray
      \def\colorrgb#1{\color{black}}%
      \def\colorgray#1{\color[gray]{#1}}%
      \expandafter\def\csname LTw\endcsname{\color{white}}%
      \expandafter\def\csname LTb\endcsname{\color{black}}%
      \expandafter\def\csname LTa\endcsname{\color{black}}%
      \expandafter\def\csname LT0\endcsname{\color{black}}%
      \expandafter\def\csname LT1\endcsname{\color{black}}%
      \expandafter\def\csname LT2\endcsname{\color{black}}%
      \expandafter\def\csname LT3\endcsname{\color{black}}%
      \expandafter\def\csname LT4\endcsname{\color{black}}%
      \expandafter\def\csname LT5\endcsname{\color{black}}%
      \expandafter\def\csname LT6\endcsname{\color{black}}%
      \expandafter\def\csname LT7\endcsname{\color{black}}%
      \expandafter\def\csname LT8\endcsname{\color{black}}%
    \fi
  \fi
    \setlength{\unitlength}{0.0500bp}%
    \ifx\gptboxheight\undefined%
      \newlength{\gptboxheight}%
      \newlength{\gptboxwidth}%
      \newsavebox{\gptboxtext}%
    \fi%
    \setlength{\fboxrule}{0.5pt}%
    \setlength{\fboxsep}{1pt}%
\begin{picture}(5102.00,3400.00)%
    \gplgaddtomacro\gplbacktext{%
      \csname LTb\endcsname%%
      \put(814,729){\makebox(0,0)[r]{\strut{}$0$}}%
      \csname LTb\endcsname%%
      \put(814,1341){\makebox(0,0)[r]{\strut{}$0.5$}}%
      \csname LTb\endcsname%%
      \put(814,1954){\makebox(0,0)[r]{\strut{}$1$}}%
      \csname LTb\endcsname%%
      \put(814,2566){\makebox(0,0)[r]{\strut{}$1.5$}}%
      \csname LTb\endcsname%%
      \put(814,3179){\makebox(0,0)[r]{\strut{}$2$}}%
      \csname LTb\endcsname%%
      \put(946,484){\makebox(0,0){\strut{}$0$}}%
      \csname LTb\endcsname%%
      \put(1533,484){\makebox(0,0){\strut{}$0.1$}}%
      \csname LTb\endcsname%%
      \put(2121,484){\makebox(0,0){\strut{}$0.2$}}%
      \csname LTb\endcsname%%
      \put(2708,484){\makebox(0,0){\strut{}$0.3$}}%
      \csname LTb\endcsname%%
      \put(3295,484){\makebox(0,0){\strut{}$0.4$}}%
      \csname LTb\endcsname%%
      \put(3883,484){\makebox(0,0){\strut{}$0.5$}}%
      \csname LTb\endcsname%%
      \put(4470,484){\makebox(0,0){\strut{}$0.6$}}%
    }%
    \gplgaddtomacro\gplfronttext{%
      \csname LTb\endcsname%%
      \put(198,1941){\rotatebox{-270}{\makebox(0,0){\strut{}$M/M_{\text{GL}}$}}}%
      \put(2825,154){\makebox(0,0){\strut{}$x_{\text{LOC}}$}}%
      \csname LTb\endcsname%%
      \put(4114,2951){\makebox(0,0)[r]{\strut{}data}}%
      \csname LTb\endcsname%%
      \put(4114,2621){\makebox(0,0)[r]{\strut{}fit}}%
    }%
    \gplbacktext
    \put(0,0){\includegraphics{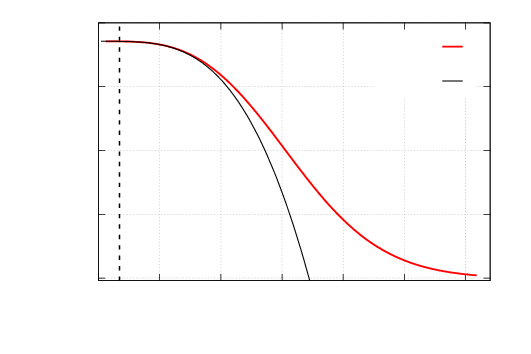}}%
    \gplfronttext
  \end{picture}%
\endgroup

\end{minipage}

\hfill

\begin{minipage}[h!]{0.5\textwidth}
% GNUPLOT: LaTeX picture with Postscript
\begingroup
  \makeatletter
  \providecommand\color[2][]{%
    \GenericError{(gnuplot) \space\space\space\@spaces}{%
      Package color not loaded in conjunction with
      terminal option `colourtext'%
    }{See the gnuplot documentation for explanation.%
    }{Either use 'blacktext' in gnuplot or load the package
      color.sty in LaTeX.}%
    \renewcommand\color[2][]{}%
  }%
  \providecommand\includegraphics[2][]{%
    \GenericError{(gnuplot) \space\space\space\@spaces}{%
      Package graphicx or graphics not loaded%
    }{See the gnuplot documentation for explanation.%
    }{The gnuplot epslatex terminal needs graphicx.sty or graphics.sty.}%
    \renewcommand\includegraphics[2][]{}%
  }%
  \providecommand\rotatebox[2]{#2}%
  \@ifundefined{ifGPcolor}{%
    \newif\ifGPcolor
    \GPcolortrue
  }{}%
  \@ifundefined{ifGPblacktext}{%
    \newif\ifGPblacktext
    \GPblacktexttrue
  }{}%
  % define a \g@addto@macro without @ in the name:
  \let\gplgaddtomacro\g@addto@macro
  % define empty templates for all commands taking text:
  \gdef\gplbacktext{}%
  \gdef\gplfronttext{}%
  \makeatother
  \ifGPblacktext
    % no textcolor at all
    \def\colorrgb#1{}%
    \def\colorgray#1{}%
  \else
    % gray or color?
    \ifGPcolor
      \def\colorrgb#1{\color[rgb]{#1}}%
      \def\colorgray#1{\color[gray]{#1}}%
      \expandafter\def\csname LTw\endcsname{\color{white}}%
      \expandafter\def\csname LTb\endcsname{\color{black}}%
      \expandafter\def\csname LTa\endcsname{\color{black}}%
      \expandafter\def\csname LT0\endcsname{\color[rgb]{1,0,0}}%
      \expandafter\def\csname LT1\endcsname{\color[rgb]{0,1,0}}%
      \expandafter\def\csname LT2\endcsname{\color[rgb]{0,0,1}}%
      \expandafter\def\csname LT3\endcsname{\color[rgb]{1,0,1}}%
      \expandafter\def\csname LT4\endcsname{\color[rgb]{0,1,1}}%
      \expandafter\def\csname LT5\endcsname{\color[rgb]{1,1,0}}%
      \expandafter\def\csname LT6\endcsname{\color[rgb]{0,0,0}}%
      \expandafter\def\csname LT7\endcsname{\color[rgb]{1,0.3,0}}%
      \expandafter\def\csname LT8\endcsname{\color[rgb]{0.5,0.5,0.5}}%
    \else
      % gray
      \def\colorrgb#1{\color{black}}%
      \def\colorgray#1{\color[gray]{#1}}%
      \expandafter\def\csname LTw\endcsname{\color{white}}%
      \expandafter\def\csname LTb\endcsname{\color{black}}%
      \expandafter\def\csname LTa\endcsname{\color{black}}%
      \expandafter\def\csname LT0\endcsname{\color{black}}%
      \expandafter\def\csname LT1\endcsname{\color{black}}%
      \expandafter\def\csname LT2\endcsname{\color{black}}%
      \expandafter\def\csname LT3\endcsname{\color{black}}%
      \expandafter\def\csname LT4\endcsname{\color{black}}%
      \expandafter\def\csname LT5\endcsname{\color{black}}%
      \expandafter\def\csname LT6\endcsname{\color{black}}%
      \expandafter\def\csname LT7\endcsname{\color{black}}%
      \expandafter\def\csname LT8\endcsname{\color{black}}%
    \fi
  \fi
    \setlength{\unitlength}{0.0500bp}%
    \ifx\gptboxheight\undefined%
      \newlength{\gptboxheight}%
      \newlength{\gptboxwidth}%
      \newsavebox{\gptboxtext}%
    \fi%
    \setlength{\fboxrule}{0.5pt}%
    \setlength{\fboxsep}{1pt}%
\begin{picture}(5102.00,3400.00)%
    \gplgaddtomacro\gplbacktext{%
      \csname LTb\endcsname%%
      \put(946,704){\makebox(0,0)[r]{\strut{}$0.2$}}%
      \csname LTb\endcsname%%
      \put(946,1136){\makebox(0,0)[r]{\strut{}$0.95$}}%
      \csname LTb\endcsname%%
      \put(946,1567){\makebox(0,0)[r]{\strut{}$1.7$}}%
      \csname LTb\endcsname%%
      \put(946,1999){\makebox(0,0)[r]{\strut{}$2.45$}}%
      \csname LTb\endcsname%%
      \put(946,2431){\makebox(0,0)[r]{\strut{}$3.2$}}%
      \csname LTb\endcsname%%
      \put(946,2862){\makebox(0,0)[r]{\strut{}$3.95$}}%
      \csname LTb\endcsname%%
      \put(1078,484){\makebox(0,0){\strut{}$-2.25$}}%
      \csname LTb\endcsname%%
      \put(1850,484){\makebox(0,0){\strut{}$-1.75$}}%
      \csname LTb\endcsname%%
      \put(2621,484){\makebox(0,0){\strut{}$-1.25$}}%
      \csname LTb\endcsname%%
      \put(3393,484){\makebox(0,0){\strut{}$-0.75$}}%
      \csname LTb\endcsname%%
      \put(4165,484){\makebox(0,0){\strut{}$-0.25$}}%
    }%
    \gplgaddtomacro\gplfronttext{%
      \csname LTb\endcsname%%
      \put(198,1941){\rotatebox{-270}{\makebox(0,0){\strut{}$(\delta M/M_{\text{GL}})/(ax^b_{\text{NUBS}})$}}}%
      \put(2891,154){\makebox(0,0){\strut{}$\log(x_{\text{NUBS}})$}}%
      \csname LTb\endcsname%%
      \put(4114,2951){\makebox(0,0)[r]{\strut{}data}}%
      \csname LTb\endcsname%%
      \put(4114,2621){\makebox(0,0)[r]{\strut{}fit}}%
    }%
    \gplbacktext
    \put(0,0){\includegraphics{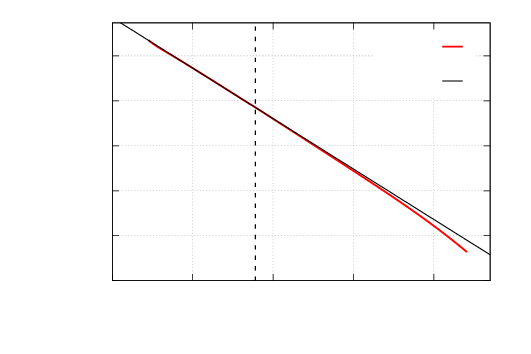}}%
    \gplfronttext
  \end{picture}%
\endgroup

\end{minipage}
\hfill
\begin{minipage}[h!]{0.5\textwidth}
\hs{0.15}% GNUPLOT: LaTeX picture with Postscript
\begingroup
  \makeatletter
  \providecommand\color[2][]{%
    \GenericError{(gnuplot) \space\space\space\@spaces}{%
      Package color not loaded in conjunction with
      terminal option `colourtext'%
    }{See the gnuplot documentation for explanation.%
    }{Either use 'blacktext' in gnuplot or load the package
      color.sty in LaTeX.}%
    \renewcommand\color[2][]{}%
  }%
  \providecommand\includegraphics[2][]{%
    \GenericError{(gnuplot) \space\space\space\@spaces}{%
      Package graphicx or graphics not loaded%
    }{See the gnuplot documentation for explanation.%
    }{The gnuplot epslatex terminal needs graphicx.sty or graphics.sty.}%
    \renewcommand\includegraphics[2][]{}%
  }%
  \providecommand\rotatebox[2]{#2}%
  \@ifundefined{ifGPcolor}{%
    \newif\ifGPcolor
    \GPcolortrue
  }{}%
  \@ifundefined{ifGPblacktext}{%
    \newif\ifGPblacktext
    \GPblacktexttrue
  }{}%
  % define a \g@addto@macro without @ in the name:
  \let\gplgaddtomacro\g@addto@macro
  % define empty templates for all commands taking text:
  \gdef\gplbacktext{}%
  \gdef\gplfronttext{}%
  \makeatother
  \ifGPblacktext
    % no textcolor at all
    \def\colorrgb#1{}%
    \def\colorgray#1{}%
  \else
    % gray or color?
    \ifGPcolor
      \def\colorrgb#1{\color[rgb]{#1}}%
      \def\colorgray#1{\color[gray]{#1}}%
      \expandafter\def\csname LTw\endcsname{\color{white}}%
      \expandafter\def\csname LTb\endcsname{\color{black}}%
      \expandafter\def\csname LTa\endcsname{\color{black}}%
      \expandafter\def\csname LT0\endcsname{\color[rgb]{1,0,0}}%
      \expandafter\def\csname LT1\endcsname{\color[rgb]{0,1,0}}%
      \expandafter\def\csname LT2\endcsname{\color[rgb]{0,0,1}}%
      \expandafter\def\csname LT3\endcsname{\color[rgb]{1,0,1}}%
      \expandafter\def\csname LT4\endcsname{\color[rgb]{0,1,1}}%
      \expandafter\def\csname LT5\endcsname{\color[rgb]{1,1,0}}%
      \expandafter\def\csname LT6\endcsname{\color[rgb]{0,0,0}}%
      \expandafter\def\csname LT7\endcsname{\color[rgb]{1,0.3,0}}%
      \expandafter\def\csname LT8\endcsname{\color[rgb]{0.5,0.5,0.5}}%
    \else
      % gray
      \def\colorrgb#1{\color{black}}%
      \def\colorgray#1{\color[gray]{#1}}%
      \expandafter\def\csname LTw\endcsname{\color{white}}%
      \expandafter\def\csname LTb\endcsname{\color{black}}%
      \expandafter\def\csname LTa\endcsname{\color{black}}%
      \expandafter\def\csname LT0\endcsname{\color{black}}%
      \expandafter\def\csname LT1\endcsname{\color{black}}%
      \expandafter\def\csname LT2\endcsname{\color{black}}%
      \expandafter\def\csname LT3\endcsname{\color{black}}%
      \expandafter\def\csname LT4\endcsname{\color{black}}%
      \expandafter\def\csname LT5\endcsname{\color{black}}%
      \expandafter\def\csname LT6\endcsname{\color{black}}%
      \expandafter\def\csname LT7\endcsname{\color{black}}%
      \expandafter\def\csname LT8\endcsname{\color{black}}%
    \fi
  \fi
    \setlength{\unitlength}{0.0500bp}%
    \ifx\gptboxheight\undefined%
      \newlength{\gptboxheight}%
      \newlength{\gptboxwidth}%
      \newsavebox{\gptboxtext}%
    \fi%
    \setlength{\fboxrule}{0.5pt}%
    \setlength{\fboxsep}{1pt}%
\begin{picture}(5102.00,3400.00)%
    \gplgaddtomacro\gplbacktext{%
      \csname LTb\endcsname%%
      \put(682,791){\makebox(0,0)[r]{\strut{}$0$}}%
      \csname LTb\endcsname%%
      \put(682,1225){\makebox(0,0)[r]{\strut{}$10$}}%
      \csname LTb\endcsname%%
      \put(682,1659){\makebox(0,0)[r]{\strut{}$20$}}%
      \csname LTb\endcsname%%
      \put(682,2093){\makebox(0,0)[r]{\strut{}$30$}}%
      \csname LTb\endcsname%%
      \put(682,2528){\makebox(0,0)[r]{\strut{}$40$}}%
      \csname LTb\endcsname%%
      \put(682,2962){\makebox(0,0)[r]{\strut{}$50$}}%
      \csname LTb\endcsname%%
      \put(814,484){\makebox(0,0){\strut{}$-4.65$}}%
      \csname LTb\endcsname%%
      \put(1651,484){\makebox(0,0){\strut{}$-3.65$}}%
      \csname LTb\endcsname%%
      \put(2488,484){\makebox(0,0){\strut{}$-2.65$}}%
      \csname LTb\endcsname%%
      \put(3324,484){\makebox(0,0){\strut{}$-1.65$}}%
      \csname LTb\endcsname%%
      \put(4161,484){\makebox(0,0){\strut{}$-0.65$}}%
    }%
    \gplgaddtomacro\gplfronttext{%
      \csname LTb\endcsname%%
      \put(198,1941){\rotatebox{-270}{\makebox(0,0){\strut{}$(\delta M/M_{\text{GL}})/(ax^b_{\text{LOC}})$}}}%
      \put(2759,154){\makebox(0,0){\strut{}$\log(x_{\text{LOC}})$}}%
      \csname LTb\endcsname%%
      \put(4114,2951){\makebox(0,0)[r]{\strut{}data}}%
      \csname LTb\endcsname%%
      \put(4114,2621){\makebox(0,0)[r]{\strut{}fit}}%
    }%
    \gplbacktext
    \put(0,0){\includegraphics{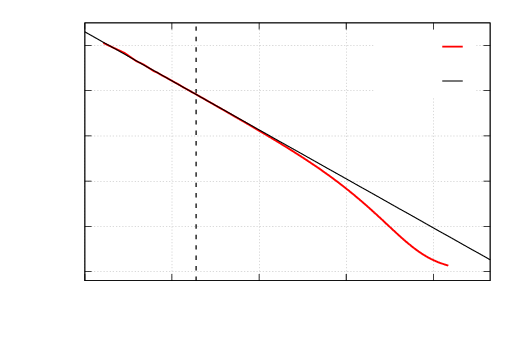}}%
    \gplfronttext
  \end{picture}%
\endgroup

\end{minipage}

\captionsetup{width=0.9\textwidth}
\captionof{figure}{\textsl{Normalized mass as a function of $x$ for NUBS and LOC (top row). Data points left to the dashed vertical line are the ones used for the fit. In contrast to $D = 5,6$ cases, in $D = 10$ our plots do not present any oscillations near the critical point, which agrees with the double-cone prediction of a real critical exponent. At the bottom row we represent $\delta M \equiv M - M_c$ normalized with respect $M_{\text{GL}} a x^b$, as a function of $\log x$. The relation is clearly lineal, in agreement with \eqref{epsilon10d}.}}
\label{critdata}
\end{minipage}
\end{center}\vs{0.2}

In Table \ref{tablecrit} we present the values of fitting parameters for the various physical quantities. To do the fits, we only have considered the solutions close enough to the merger, i.e.\ with small enough $x$; including more data points to perform the fit gives less accurate values of the critical thermodynamical values and exponent. For different physical quantities, the critical exponent coincides with the theoretical prediction of $4$ with deviations of less than $0.05\%$ in the worst case and the critical value of a given quantity coincide up to the $4^\textrm{th}$ or $5^\textrm{th}$ decimal number for both branches. We note that the critical values satisfy the Smarr's relation to the order $10^{-6}$ and $10^{-5}$ for NUBS and LOC respectively, which is consistent with the numerical error according to the values of $\xi^2$ we reached.}
 
{\begin{center}
\begin{tabular}{|r|r|r|r|r|r|r|}
\cline{3-7}
\multicolumn{1}{r}{} & \multicolumn{1}{r|}{} & $Q_c$ & $a$ & $b$ & $c$ & $d$ \\ \hline\hline
\multirow{2}{*}{$T/T_{\text{GL}}$} & NUBS & {\bf 0.92615} & $0.09539$ & {\bf 4.00001} & 0.76900 & $-2.08080$ \\
						   & LOC & {\bf 0.92615} & 2.41070 & {\bf 3.99967} & 0.21280 & $-3.43802$ \\ \hline\hline
\multirow{2}{*}{$M/M_{\text{GL}}$} & NUBS & {\bf 1.85551} & $-1.48224$ & {\bf 3.99975} & 0.80018 & $-1.68271$ \\
                                                      & LOC & {\bf 1.85551} & $-9.31138$ & {\bf 3.99814} & $2.62629$ & $-10.83657$ \\ \hline\hline
\multirow{2}{*}{$S/S_{\text{GL}}$} & NUBS & {\bf 2.03933} & $-0.76910$ & {\bf 4.00070} & 1.88490 & $-4.13116$ \\
                                                      & LOC & {\bf 2.03958} & $-15.75813$ & {\bf 4.00047} & $1.27332$ & $-8.34065$ \\ \hline\hline
\multirow{2}{*}{$\mc{T}/\mc{T}_{\text{GL}}$} & NUBS & {\bf 0.25816} & $-1.49842$ & {\bf 3.99996} & $-1.19917$ & $-1.53764$ \\
                                                     & LOC & {\bf 0.25813} & $-8.01729$ & {\bf 4.00188} & $-12.66599$ & $-11.94347$ \\ \hline\hline
\multirow{2}{*}{$n/n_{\text{GL}}$} & NUBS & {\bf 0.13913} & $-1.50564$ & {\bf 4.00000} & $-0.69536$ & $-0.69743$ \\
                                                     & LOC & {\bf 0.13912} & 6.72458 & {\bf 4.00172} & 8.18577 & 6.46396 \\ \hline
\end{tabular}
\vs{0.25}
\captionsetup{width=0.9\textwidth}
\captionof{table}{\textsl{Critical exponent and other parameters obtained from the fit of the non-uniform black strings (1st rows) and localized black holes (2nd rows) data points.}}
\label{tablecrit}
\end{center}}

{Only a couple of geometrical lengths do not follow the behavior \eqref{eq:fits}, as it may be seen from Fig.\ \ref{critdatalengths}. These are the horizon length $L_{\text{hor}}$ of the black string and the polar length $L_{\text{pol}}$ of the localized black holes. In lower dimensions this was also the case, and a linear term was introduced to get a proper fit \cite{Kalisch:2017bin}. In $D = 10$ the linear term appears naturally and the real critical exponent agrees to be one from both sides of the merger, just as in $D = 5,6$. The equivalent plots to Fig.\ \ref{critdata} for these lengths are shown in Fig.\ \ref{critdatalengths} and the extracted critical values and exponents are in Table \ref{tablecritlengths}. It would be interesting to better understand why these quantities do not follow the same critical behavior as the other physical quantities.

\vs{0.2}\begin{center}
\begin{tabular}{|r|r|r|r|r|r|r|}
\cline{3-7}
\multicolumn{1}{r}{} & \multicolumn{1}{r|}{} & $Q_c$ & $a$ & $b$ & $c$ & $d$ \\ \hline\hline
\multirow{2}{*}{$L_{\text{hor/polar}}/L$} & NUBS & {\bf 1.54505} & $-0.40768$ & {\bf 0.99955} & 1.54049 & 0.02272 \\
		                  & LOC & {\bf 1.54589} & 0.41840 & {\bf 1.00021} & $-0.84676$ & $-0.01620$ \\ \hline
\end{tabular}
\vs{0.25}
\captionsetup{width=0.9\textwidth}
\captionof{table}{\textsl{Critical exponent and other parameters obtained from the fit of the NUBS's horizon length (1st row) and LOC's polar length (2nd row).}}
\label{tablecritlengths}
\end{center}}

%~~~~~~~~~~~~~~~~~~~~~~~~~~~~~~~~~~~~~~~~~~~~~~~
\subsection{Spectrum of negative modes}
\label{spec}
%~~~~~~~~~~~~~~~~~~~~~~~~~~~~~~~~~~~~~~~~~~~~~~
In this subsection we present the spectrum of negative modes of the Lichnerowicz operator, $\Delta_L$, around the NUBS and LOC solutions that we have constructed. The negative eigenvalues of $\Delta_L$

\begin{center}
\begin{minipage}{\textwidth}
\begin{minipage}[h!]{0.5\textwidth}
\begin{center}
\hs{1.25}{\bf \textsf{Non-uniform black strings}}
\end{center}
\end{minipage}
\begin{minipage}[h!]{0.5\textwidth}
\begin{center}
\hs{1.25}{\bf \textsf{Localized black holes}}
\end{center}
\end{minipage}
\begin{minipage}[h!]{0.5\textwidth}
% GNUPLOT: LaTeX picture with Postscript
\begingroup
  \makeatletter
  \providecommand\color[2][]{%
    \GenericError{(gnuplot) \space\space\space\@spaces}{%
      Package color not loaded in conjunction with
      terminal option `colourtext'%
    }{See the gnuplot documentation for explanation.%
    }{Either use 'blacktext' in gnuplot or load the package
      color.sty in LaTeX.}%
    \renewcommand\color[2][]{}%
  }%
  \providecommand\includegraphics[2][]{%
    \GenericError{(gnuplot) \space\space\space\@spaces}{%
      Package graphicx or graphics not loaded%
    }{See the gnuplot documentation for explanation.%
    }{The gnuplot epslatex terminal needs graphicx.sty or graphics.sty.}%
    \renewcommand\includegraphics[2][]{}%
  }%
  \providecommand\rotatebox[2]{#2}%
  \@ifundefined{ifGPcolor}{%
    \newif\ifGPcolor
    \GPcolortrue
  }{}%
  \@ifundefined{ifGPblacktext}{%
    \newif\ifGPblacktext
    \GPblacktexttrue
  }{}%
  % define a \g@addto@macro without @ in the name:
  \let\gplgaddtomacro\g@addto@macro
  % define empty templates for all commands taking text:
  \gdef\gplbacktext{}%
  \gdef\gplfronttext{}%
  \makeatother
  \ifGPblacktext
    % no textcolor at all
    \def\colorrgb#1{}%
    \def\colorgray#1{}%
  \else
    % gray or color?
    \ifGPcolor
      \def\colorrgb#1{\color[rgb]{#1}}%
      \def\colorgray#1{\color[gray]{#1}}%
      \expandafter\def\csname LTw\endcsname{\color{white}}%
      \expandafter\def\csname LTb\endcsname{\color{black}}%
      \expandafter\def\csname LTa\endcsname{\color{black}}%
      \expandafter\def\csname LT0\endcsname{\color[rgb]{1,0,0}}%
      \expandafter\def\csname LT1\endcsname{\color[rgb]{0,1,0}}%
      \expandafter\def\csname LT2\endcsname{\color[rgb]{0,0,1}}%
      \expandafter\def\csname LT3\endcsname{\color[rgb]{1,0,1}}%
      \expandafter\def\csname LT4\endcsname{\color[rgb]{0,1,1}}%
      \expandafter\def\csname LT5\endcsname{\color[rgb]{1,1,0}}%
      \expandafter\def\csname LT6\endcsname{\color[rgb]{0,0,0}}%
      \expandafter\def\csname LT7\endcsname{\color[rgb]{1,0.3,0}}%
      \expandafter\def\csname LT8\endcsname{\color[rgb]{0.5,0.5,0.5}}%
    \else
      % gray
      \def\colorrgb#1{\color{black}}%
      \def\colorgray#1{\color[gray]{#1}}%
      \expandafter\def\csname LTw\endcsname{\color{white}}%
      \expandafter\def\csname LTb\endcsname{\color{black}}%
      \expandafter\def\csname LTa\endcsname{\color{black}}%
      \expandafter\def\csname LT0\endcsname{\color{black}}%
      \expandafter\def\csname LT1\endcsname{\color{black}}%
      \expandafter\def\csname LT2\endcsname{\color{black}}%
      \expandafter\def\csname LT3\endcsname{\color{black}}%
      \expandafter\def\csname LT4\endcsname{\color{black}}%
      \expandafter\def\csname LT5\endcsname{\color{black}}%
      \expandafter\def\csname LT6\endcsname{\color{black}}%
      \expandafter\def\csname LT7\endcsname{\color{black}}%
      \expandafter\def\csname LT8\endcsname{\color{black}}%
    \fi
  \fi
    \setlength{\unitlength}{0.0500bp}%
    \ifx\gptboxheight\undefined%
      \newlength{\gptboxheight}%
      \newlength{\gptboxwidth}%
      \newsavebox{\gptboxtext}%
    \fi%
    \setlength{\fboxrule}{0.5pt}%
    \setlength{\fboxsep}{1pt}%
\begin{picture}(5102.00,3400.00)%
    \gplgaddtomacro\gplbacktext{%
      \csname LTb\endcsname%%
      \put(946,704){\makebox(0,0)[r]{\strut{}$0.9$}}%
      \csname LTb\endcsname%%
      \put(946,1275){\makebox(0,0)[r]{\strut{}$1.05$}}%
      \csname LTb\endcsname%%
      \put(946,1846){\makebox(0,0)[r]{\strut{}$1.2$}}%
      \csname LTb\endcsname%%
      \put(946,2417){\makebox(0,0)[r]{\strut{}$1.35$}}%
      \csname LTb\endcsname%%
      \put(946,2989){\makebox(0,0)[r]{\strut{}$1.5$}}%
      \csname LTb\endcsname%%
      \put(1078,484){\makebox(0,0){\strut{}$0$}}%
      \csname LTb\endcsname%%
      \put(1803,484){\makebox(0,0){\strut{}$0.2$}}%
      \csname LTb\endcsname%%
      \put(2529,484){\makebox(0,0){\strut{}$0.4$}}%
      \csname LTb\endcsname%%
      \put(3254,484){\makebox(0,0){\strut{}$0.6$}}%
      \csname LTb\endcsname%%
      \put(3980,484){\makebox(0,0){\strut{}$0.8$}}%
      \csname LTb\endcsname%%
      \put(4705,484){\makebox(0,0){\strut{}$1$}}%
    }%
    \gplgaddtomacro\gplfronttext{%
      \csname LTb\endcsname%%
      \put(198,1941){\rotatebox{-270}{\makebox(0,0){\strut{}$L_{\text{hor}}/L$}}}%
      \put(2891,154){\makebox(0,0){\strut{}$x_{\text{NUBS}}$}}%
      \csname LTb\endcsname%%
      \put(4114,2951){\makebox(0,0)[r]{\strut{}data}}%
      \csname LTb\endcsname%%
      \put(4114,2621){\makebox(0,0)[r]{\strut{}fit}}%
    }%
    \gplbacktext
    \put(0,0){\includegraphics{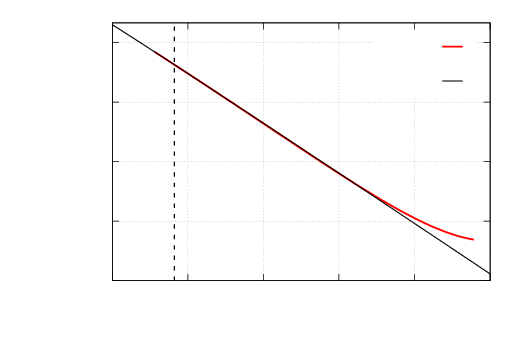}}%
    \gplfronttext
  \end{picture}%
\endgroup

\end{minipage}
\hfill
\begin{minipage}[h!]{0.5\textwidth}
% GNUPLOT: LaTeX picture with Postscript
\begingroup
  \makeatletter
  \providecommand\color[2][]{%
    \GenericError{(gnuplot) \space\space\space\@spaces}{%
      Package color not loaded in conjunction with
      terminal option `colourtext'%
    }{See the gnuplot documentation for explanation.%
    }{Either use 'blacktext' in gnuplot or load the package
      color.sty in LaTeX.}%
    \renewcommand\color[2][]{}%
  }%
  \providecommand\includegraphics[2][]{%
    \GenericError{(gnuplot) \space\space\space\@spaces}{%
      Package graphicx or graphics not loaded%
    }{See the gnuplot documentation for explanation.%
    }{The gnuplot epslatex terminal needs graphicx.sty or graphics.sty.}%
    \renewcommand\includegraphics[2][]{}%
  }%
  \providecommand\rotatebox[2]{#2}%
  \@ifundefined{ifGPcolor}{%
    \newif\ifGPcolor
    \GPcolortrue
  }{}%
  \@ifundefined{ifGPblacktext}{%
    \newif\ifGPblacktext
    \GPblacktexttrue
  }{}%
  % define a \g@addto@macro without @ in the name:
  \let\gplgaddtomacro\g@addto@macro
  % define empty templates for all commands taking text:
  \gdef\gplbacktext{}%
  \gdef\gplfronttext{}%
  \makeatother
  \ifGPblacktext
    % no textcolor at all
    \def\colorrgb#1{}%
    \def\colorgray#1{}%
  \else
    % gray or color?
    \ifGPcolor
      \def\colorrgb#1{\color[rgb]{#1}}%
      \def\colorgray#1{\color[gray]{#1}}%
      \expandafter\def\csname LTw\endcsname{\color{white}}%
      \expandafter\def\csname LTb\endcsname{\color{black}}%
      \expandafter\def\csname LTa\endcsname{\color{black}}%
      \expandafter\def\csname LT0\endcsname{\color[rgb]{1,0,0}}%
      \expandafter\def\csname LT1\endcsname{\color[rgb]{0,1,0}}%
      \expandafter\def\csname LT2\endcsname{\color[rgb]{0,0,1}}%
      \expandafter\def\csname LT3\endcsname{\color[rgb]{1,0,1}}%
      \expandafter\def\csname LT4\endcsname{\color[rgb]{0,1,1}}%
      \expandafter\def\csname LT5\endcsname{\color[rgb]{1,1,0}}%
      \expandafter\def\csname LT6\endcsname{\color[rgb]{0,0,0}}%
      \expandafter\def\csname LT7\endcsname{\color[rgb]{1,0.3,0}}%
      \expandafter\def\csname LT8\endcsname{\color[rgb]{0.5,0.5,0.5}}%
    \else
      % gray
      \def\colorrgb#1{\color{black}}%
      \def\colorgray#1{\color[gray]{#1}}%
      \expandafter\def\csname LTw\endcsname{\color{white}}%
      \expandafter\def\csname LTb\endcsname{\color{black}}%
      \expandafter\def\csname LTa\endcsname{\color{black}}%
      \expandafter\def\csname LT0\endcsname{\color{black}}%
      \expandafter\def\csname LT1\endcsname{\color{black}}%
      \expandafter\def\csname LT2\endcsname{\color{black}}%
      \expandafter\def\csname LT3\endcsname{\color{black}}%
      \expandafter\def\csname LT4\endcsname{\color{black}}%
      \expandafter\def\csname LT5\endcsname{\color{black}}%
      \expandafter\def\csname LT6\endcsname{\color{black}}%
      \expandafter\def\csname LT7\endcsname{\color{black}}%
      \expandafter\def\csname LT8\endcsname{\color{black}}%
    \fi
  \fi
    \setlength{\unitlength}{0.0500bp}%
    \ifx\gptboxheight\undefined%
      \newlength{\gptboxheight}%
      \newlength{\gptboxwidth}%
      \newsavebox{\gptboxtext}%
    \fi%
    \setlength{\fboxrule}{0.5pt}%
    \setlength{\fboxsep}{1pt}%
\begin{picture}(5102.00,3400.00)%
    \gplgaddtomacro\gplbacktext{%
      \csname LTb\endcsname%%
      \put(946,704){\makebox(0,0)[r]{\strut{}$0.65$}}%
      \csname LTb\endcsname%%
      \put(946,1199){\makebox(0,0)[r]{\strut{}$0.85$}}%
      \csname LTb\endcsname%%
      \put(946,1694){\makebox(0,0)[r]{\strut{}$1.05$}}%
      \csname LTb\endcsname%%
      \put(946,2189){\makebox(0,0)[r]{\strut{}$1.25$}}%
      \csname LTb\endcsname%%
      \put(946,2684){\makebox(0,0)[r]{\strut{}$1.45$}}%
      \csname LTb\endcsname%%
      \put(946,3179){\makebox(0,0)[r]{\strut{}$1.65$}}%
      \csname LTb\endcsname%%
      \put(1078,484){\makebox(0,0){\strut{}$0$}}%
      \csname LTb\endcsname%%
      \put(1645,484){\makebox(0,0){\strut{}$0.1$}}%
      \csname LTb\endcsname%%
      \put(2211,484){\makebox(0,0){\strut{}$0.2$}}%
      \csname LTb\endcsname%%
      \put(2778,484){\makebox(0,0){\strut{}$0.3$}}%
      \csname LTb\endcsname%%
      \put(3345,484){\makebox(0,0){\strut{}$0.4$}}%
      \csname LTb\endcsname%%
      \put(3912,484){\makebox(0,0){\strut{}$0.5$}}%
      \csname LTb\endcsname%%
      \put(4478,484){\makebox(0,0){\strut{}$0.6$}}%
    }%
    \gplgaddtomacro\gplfronttext{%
      \csname LTb\endcsname%%
      \put(198,1941){\rotatebox{-270}{\makebox(0,0){\strut{}$L_{\text{polar}}/L$}}}%
      \put(2891,154){\makebox(0,0){\strut{}$x_{\text{LOC}}$}}%
      \csname LTb\endcsname%%
      \put(4114,2951){\makebox(0,0)[r]{\strut{}data}}%
      \csname LTb\endcsname%%
      \put(4114,2621){\makebox(0,0)[r]{\strut{}fit}}%
    }%
    \gplbacktext
    \put(0,0){\includegraphics{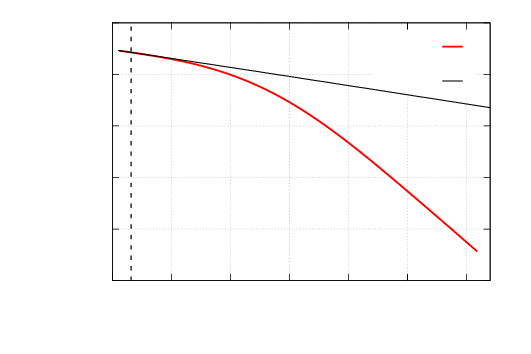}}%
    \gplfronttext
  \end{picture}%
\endgroup

\end{minipage}

\hfill

\begin{minipage}[h!]{0.5\textwidth}
% GNUPLOT: LaTeX picture with Postscript
\begingroup
  \makeatletter
  \providecommand\color[2][]{%
    \GenericError{(gnuplot) \space\space\space\@spaces}{%
      Package color not loaded in conjunction with
      terminal option `colourtext'%
    }{See the gnuplot documentation for explanation.%
    }{Either use 'blacktext' in gnuplot or load the package
      color.sty in LaTeX.}%
    \renewcommand\color[2][]{}%
  }%
  \providecommand\includegraphics[2][]{%
    \GenericError{(gnuplot) \space\space\space\@spaces}{%
      Package graphicx or graphics not loaded%
    }{See the gnuplot documentation for explanation.%
    }{The gnuplot epslatex terminal needs graphicx.sty or graphics.sty.}%
    \renewcommand\includegraphics[2][]{}%
  }%
  \providecommand\rotatebox[2]{#2}%
  \@ifundefined{ifGPcolor}{%
    \newif\ifGPcolor
    \GPcolortrue
  }{}%
  \@ifundefined{ifGPblacktext}{%
    \newif\ifGPblacktext
    \GPblacktexttrue
  }{}%
  % define a \g@addto@macro without @ in the name:
  \let\gplgaddtomacro\g@addto@macro
  % define empty templates for all commands taking text:
  \gdef\gplbacktext{}%
  \gdef\gplfronttext{}%
  \makeatother
  \ifGPblacktext
    % no textcolor at all
    \def\colorrgb#1{}%
    \def\colorgray#1{}%
  \else
    % gray or color?
    \ifGPcolor
      \def\colorrgb#1{\color[rgb]{#1}}%
      \def\colorgray#1{\color[gray]{#1}}%
      \expandafter\def\csname LTw\endcsname{\color{white}}%
      \expandafter\def\csname LTb\endcsname{\color{black}}%
      \expandafter\def\csname LTa\endcsname{\color{black}}%
      \expandafter\def\csname LT0\endcsname{\color[rgb]{1,0,0}}%
      \expandafter\def\csname LT1\endcsname{\color[rgb]{0,1,0}}%
      \expandafter\def\csname LT2\endcsname{\color[rgb]{0,0,1}}%
      \expandafter\def\csname LT3\endcsname{\color[rgb]{1,0,1}}%
      \expandafter\def\csname LT4\endcsname{\color[rgb]{0,1,1}}%
      \expandafter\def\csname LT5\endcsname{\color[rgb]{1,1,0}}%
      \expandafter\def\csname LT6\endcsname{\color[rgb]{0,0,0}}%
      \expandafter\def\csname LT7\endcsname{\color[rgb]{1,0.3,0}}%
      \expandafter\def\csname LT8\endcsname{\color[rgb]{0.5,0.5,0.5}}%
    \else
      % gray
      \def\colorrgb#1{\color{black}}%
      \def\colorgray#1{\color[gray]{#1}}%
      \expandafter\def\csname LTw\endcsname{\color{white}}%
      \expandafter\def\csname LTb\endcsname{\color{black}}%
      \expandafter\def\csname LTa\endcsname{\color{black}}%
      \expandafter\def\csname LT0\endcsname{\color{black}}%
      \expandafter\def\csname LT1\endcsname{\color{black}}%
      \expandafter\def\csname LT2\endcsname{\color{black}}%
      \expandafter\def\csname LT3\endcsname{\color{black}}%
      \expandafter\def\csname LT4\endcsname{\color{black}}%
      \expandafter\def\csname LT5\endcsname{\color{black}}%
      \expandafter\def\csname LT6\endcsname{\color{black}}%
      \expandafter\def\csname LT7\endcsname{\color{black}}%
      \expandafter\def\csname LT8\endcsname{\color{black}}%
    \fi
  \fi
    \setlength{\unitlength}{0.0500bp}%
    \ifx\gptboxheight\undefined%
      \newlength{\gptboxheight}%
      \newlength{\gptboxwidth}%
      \newsavebox{\gptboxtext}%
    \fi%
    \setlength{\fboxrule}{0.5pt}%
    \setlength{\fboxsep}{1pt}%
\begin{picture}(5102.00,3400.00)%
    \gplgaddtomacro\gplbacktext{%
      \csname LTb\endcsname%%
      \put(946,704){\makebox(0,0)[r]{\strut{}$1.39$}}%
      \csname LTb\endcsname%%
      \put(946,1168){\makebox(0,0)[r]{\strut{}$1.42$}}%
      \csname LTb\endcsname%%
      \put(946,1632){\makebox(0,0)[r]{\strut{}$1.45$}}%
      \csname LTb\endcsname%%
      \put(946,2096){\makebox(0,0)[r]{\strut{}$1.48$}}%
      \csname LTb\endcsname%%
      \put(946,2560){\makebox(0,0)[r]{\strut{}$1.51$}}%
      \csname LTb\endcsname%%
      \put(946,3024){\makebox(0,0)[r]{\strut{}$1.54$}}%
      \csname LTb\endcsname%%
      \put(1541,484){\makebox(0,0){\strut{}$-2$}}%
      \csname LTb\endcsname%%
      \put(2313,484){\makebox(0,0){\strut{}$-1.5$}}%
      \csname LTb\endcsname%%
      \put(3084,484){\makebox(0,0){\strut{}$-1$}}%
      \csname LTb\endcsname%%
      \put(3856,484){\makebox(0,0){\strut{}$-0.5$}}%
      \csname LTb\endcsname%%
      \put(4628,484){\makebox(0,0){\strut{}$0$}}%
    }%
    \gplgaddtomacro\gplfronttext{%
      \csname LTb\endcsname%%
      \put(198,1941){\rotatebox{-270}{\makebox(0,0){\strut{}$(\delta L_{\text{hor}}/L)/(ax^b_{\text{NUBS}})$}}}%
      \put(2891,154){\makebox(0,0){\strut{}$\log(x_{\text{NUBS}})$}}%
      \csname LTb\endcsname%%
      \put(2926,1262){\makebox(0,0)[r]{\strut{}data}}%
      \csname LTb\endcsname%%
      \put(2926,932){\makebox(0,0)[r]{\strut{}fit}}%
    }%
    \gplbacktext
    \put(0,0){\includegraphics{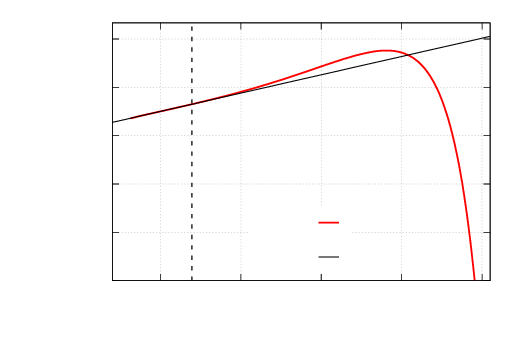}}%
    \gplfronttext
  \end{picture}%
\endgroup

\end{minipage}
\hfill
\begin{minipage}[h!]{0.5\textwidth}
% GNUPLOT: LaTeX picture with Postscript
\begingroup
  \makeatletter
  \providecommand\color[2][]{%
    \GenericError{(gnuplot) \space\space\space\@spaces}{%
      Package color not loaded in conjunction with
      terminal option `colourtext'%
    }{See the gnuplot documentation for explanation.%
    }{Either use 'blacktext' in gnuplot or load the package
      color.sty in LaTeX.}%
    \renewcommand\color[2][]{}%
  }%
  \providecommand\includegraphics[2][]{%
    \GenericError{(gnuplot) \space\space\space\@spaces}{%
      Package graphicx or graphics not loaded%
    }{See the gnuplot documentation for explanation.%
    }{The gnuplot epslatex terminal needs graphicx.sty or graphics.sty.}%
    \renewcommand\includegraphics[2][]{}%
  }%
  \providecommand\rotatebox[2]{#2}%
  \@ifundefined{ifGPcolor}{%
    \newif\ifGPcolor
    \GPcolortrue
  }{}%
  \@ifundefined{ifGPblacktext}{%
    \newif\ifGPblacktext
    \GPblacktexttrue
  }{}%
  % define a \g@addto@macro without @ in the name:
  \let\gplgaddtomacro\g@addto@macro
  % define empty templates for all commands taking text:
  \gdef\gplbacktext{}%
  \gdef\gplfronttext{}%
  \makeatother
  \ifGPblacktext
    % no textcolor at all
    \def\colorrgb#1{}%
    \def\colorgray#1{}%
  \else
    % gray or color?
    \ifGPcolor
      \def\colorrgb#1{\color[rgb]{#1}}%
      \def\colorgray#1{\color[gray]{#1}}%
      \expandafter\def\csname LTw\endcsname{\color{white}}%
      \expandafter\def\csname LTb\endcsname{\color{black}}%
      \expandafter\def\csname LTa\endcsname{\color{black}}%
      \expandafter\def\csname LT0\endcsname{\color[rgb]{1,0,0}}%
      \expandafter\def\csname LT1\endcsname{\color[rgb]{0,1,0}}%
      \expandafter\def\csname LT2\endcsname{\color[rgb]{0,0,1}}%
      \expandafter\def\csname LT3\endcsname{\color[rgb]{1,0,1}}%
      \expandafter\def\csname LT4\endcsname{\color[rgb]{0,1,1}}%
      \expandafter\def\csname LT5\endcsname{\color[rgb]{1,1,0}}%
      \expandafter\def\csname LT6\endcsname{\color[rgb]{0,0,0}}%
      \expandafter\def\csname LT7\endcsname{\color[rgb]{1,0.3,0}}%
      \expandafter\def\csname LT8\endcsname{\color[rgb]{0.5,0.5,0.5}}%
    \else
      % gray
      \def\colorrgb#1{\color{black}}%
      \def\colorgray#1{\color[gray]{#1}}%
      \expandafter\def\csname LTw\endcsname{\color{white}}%
      \expandafter\def\csname LTb\endcsname{\color{black}}%
      \expandafter\def\csname LTa\endcsname{\color{black}}%
      \expandafter\def\csname LT0\endcsname{\color{black}}%
      \expandafter\def\csname LT1\endcsname{\color{black}}%
      \expandafter\def\csname LT2\endcsname{\color{black}}%
      \expandafter\def\csname LT3\endcsname{\color{black}}%
      \expandafter\def\csname LT4\endcsname{\color{black}}%
      \expandafter\def\csname LT5\endcsname{\color{black}}%
      \expandafter\def\csname LT6\endcsname{\color{black}}%
      \expandafter\def\csname LT7\endcsname{\color{black}}%
      \expandafter\def\csname LT8\endcsname{\color{black}}%
    \fi
  \fi
    \setlength{\unitlength}{0.0500bp}%
    \ifx\gptboxheight\undefined%
      \newlength{\gptboxheight}%
      \newlength{\gptboxwidth}%
      \newsavebox{\gptboxtext}%
    \fi%
    \setlength{\fboxrule}{0.5pt}%
    \setlength{\fboxsep}{1pt}%
\begin{picture}(5102.00,3400.00)%
    \gplgaddtomacro\gplbacktext{%
      \csname LTb\endcsname%%
      \put(946,929){\makebox(0,0)[r]{\strut{}$-3$}}%
      \csname LTb\endcsname%%
      \put(946,1379){\makebox(0,0)[r]{\strut{}$-2.5$}}%
      \csname LTb\endcsname%%
      \put(946,1829){\makebox(0,0)[r]{\strut{}$-2$}}%
      \csname LTb\endcsname%%
      \put(946,2279){\makebox(0,0)[r]{\strut{}$-1.5$}}%
      \csname LTb\endcsname%%
      \put(946,2729){\makebox(0,0)[r]{\strut{}$-1$}}%
      \csname LTb\endcsname%%
      \put(946,3179){\makebox(0,0)[r]{\strut{}$-0.5$}}%
      \csname LTb\endcsname%%
      \put(1078,484){\makebox(0,0){\strut{}$-4.65$}}%
      \csname LTb\endcsname%%
      \put(1730,484){\makebox(0,0){\strut{}$-3.85$}}%
      \csname LTb\endcsname%%
      \put(2382,484){\makebox(0,0){\strut{}$-3.05$}}%
      \csname LTb\endcsname%%
      \put(3034,484){\makebox(0,0){\strut{}$-2.25$}}%
      \csname LTb\endcsname%%
      \put(3686,484){\makebox(0,0){\strut{}$-1.45$}}%
      \csname LTb\endcsname%%
      \put(4338,484){\makebox(0,0){\strut{}$-0.65$}}%
    }%
    \gplgaddtomacro\gplfronttext{%
      \csname LTb\endcsname%%
      \put(198,1941){\rotatebox{-270}{\makebox(0,0){\strut{}$(\delta L_{\text{polar}}/L)/(ax^b_{\text{LOC}})$}}}%
      \put(2891,154){\makebox(0,0){\strut{}$\log(x_{\text{LOC}})$}}%
      \csname LTb\endcsname%%
      \put(2926,1262){\makebox(0,0)[r]{\strut{}data}}%
      \csname LTb\endcsname%%
      \put(2926,932){\makebox(0,0)[r]{\strut{}fit}}%
    }%
    \gplbacktext
    \put(0,0){\includegraphics{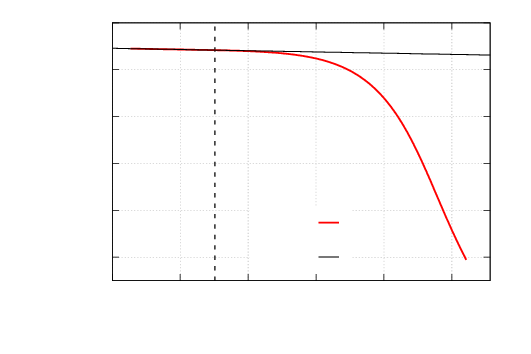}}%
    \gplfronttext
  \end{picture}%
\endgroup

\end{minipage}

\captionsetup{width=0.9\textwidth}
\captionof{figure}{\textsl{Normalized horizon length and polar length as a function of $x$ for NUBS and LOC respectively (top row). Data points left to the dashed vertical line are the ones used for the fit. At the bottom row we represent $\delta L \equiv L_{\text{hor/polar}} - L_c$ normalized with respect to $L a x^b$, as a function of $\log x$. In both cases the relation is lineal.}}
\label{critdatalengths}
\end{minipage}
\end{center}

\noindent are an invariant feature of the geometry and hence they provide another way to characterize the merger between NUBS and LOC. To compute the negative modes of $\Delta_L$, we take advantage of the fact that, when using Newton's method to construct the solutions numerically, we have to linearize the Einstein-DeTurck operator as part of the iterative process. Around an Einstein metric, the linearized Einstein-DeTurck operator coincides with the Lichnerowicz operator \cite{Headrick:2009pv}. It is then easy to readapt the code to find the low lying eigenvalues and eigenmodes of $\Delta_L$, associated to (physical) metric fluctuations. Notice that with this approach we only find perturbations that are singlets under the action of U$(1)_\beta\times\,$SO$(D-3)$.

We display the results in Fig.\ \ref{fig:negmodes}. We found that NUBS have two negative modes, as in lower $D$ \cite{Headrick:2009pv}.\footnote{In $D\geq 13$ NUBS have only one negative mode, and in $D=12$ a mode disappears at a minimum of the temperature along the NUBS branch \cite{Figueras:2012xj}. This is related to the fact that $D=12$ is the critical dimension for the canonical ensemble for this system.} The first one (green line in Fig.\ \ref{fig:negmodes}) corresponds to the continuation of the GL zero mode to a negative mode as one moves along the branch to larger non-uniformities. The other one (blue line in Fig.\ \ref{fig:negmodes}) is continuously connected to the negative mode of the UBS, which arises from the negative mode of Schwarzschild \cite{Gross:1982cv}. For the explored range of solutions in this work, no further negative modes appear on this branch. On the other hand, LOC have only one negative mode (red line in Fig.\ \ref{fig:negmodes}). For small black holes, this coincides with the negative mode of the asymptotically flat Schwarzschild solution in $D=10$, as expected. No further negative modes appear or disappear along this branch. As we approach the critical solution from both sides, one of the modes of the NUBS diverges while the other appears to tend to a finite value; the latter seems to match the limiting value of the negative mode on the LOC branch. Notice that in $D=10$, NUBS and LOC seem to have a different number of negative modes near the critical region. The reason is that there are no turning points along either branches, so modes cannot appear or disappear at a minimum of the temperature. However, modes can diverge as they approach the critical solution since it is singular. In $D = 5,6$, there is a minimum of the temperature on the LOC branch, at which point $\Delta_L$ has a zero mode that continues to a (second) negative mode near the merger. At the merger, these two negative modes approach those of the NUBS and, in particular, a pair of them diverge at the critical solution \cite{Headrick:2009pv}.

\begin{center}
\begin{minipage}{\textwidth}
\begin{center}
% GNUPLOT: LaTeX picture with Postscript
\begingroup
  \makeatletter
  \providecommand\color[2][]{%
    \GenericError{(gnuplot) \space\space\space\@spaces}{%
      Package color not loaded in conjunction with
      terminal option `colourtext'%
    }{See the gnuplot documentation for explanation.%
    }{Either use 'blacktext' in gnuplot or load the package
      color.sty in LaTeX.}%
    \renewcommand\color[2][]{}%
  }%
  \providecommand\includegraphics[2][]{%
    \GenericError{(gnuplot) \space\space\space\@spaces}{%
      Package graphicx or graphics not loaded%
    }{See the gnuplot documentation for explanation.%
    }{The gnuplot epslatex terminal needs graphicx.sty or graphics.sty.}%
    \renewcommand\includegraphics[2][]{}%
  }%
  \providecommand\rotatebox[2]{#2}%
  \@ifundefined{ifGPcolor}{%
    \newif\ifGPcolor
    \GPcolortrue
  }{}%
  \@ifundefined{ifGPblacktext}{%
    \newif\ifGPblacktext
    \GPblacktexttrue
  }{}%
  % define a \g@addto@macro without @ in the name:
  \let\gplgaddtomacro\g@addto@macro
  % define empty templates for all commands taking text:
  \gdef\gplbacktext{}%
  \gdef\gplfronttext{}%
  \makeatother
  \ifGPblacktext
    % no textcolor at all
    \def\colorrgb#1{}%
    \def\colorgray#1{}%
  \else
    % gray or color?
    \ifGPcolor
      \def\colorrgb#1{\color[rgb]{#1}}%
      \def\colorgray#1{\color[gray]{#1}}%
      \expandafter\def\csname LTw\endcsname{\color{white}}%
      \expandafter\def\csname LTb\endcsname{\color{black}}%
      \expandafter\def\csname LTa\endcsname{\color{black}}%
      \expandafter\def\csname LT0\endcsname{\color[rgb]{1,0,0}}%
      \expandafter\def\csname LT1\endcsname{\color[rgb]{0,1,0}}%
      \expandafter\def\csname LT2\endcsname{\color[rgb]{0,0,1}}%
      \expandafter\def\csname LT3\endcsname{\color[rgb]{1,0,1}}%
      \expandafter\def\csname LT4\endcsname{\color[rgb]{0,1,1}}%
      \expandafter\def\csname LT5\endcsname{\color[rgb]{1,1,0}}%
      \expandafter\def\csname LT6\endcsname{\color[rgb]{0,0,0}}%
      \expandafter\def\csname LT7\endcsname{\color[rgb]{1,0.3,0}}%
      \expandafter\def\csname LT8\endcsname{\color[rgb]{0.5,0.5,0.5}}%
    \else
      % gray
      \def\colorrgb#1{\color{black}}%
      \def\colorgray#1{\color[gray]{#1}}%
      \expandafter\def\csname LTw\endcsname{\color{white}}%
      \expandafter\def\csname LTb\endcsname{\color{black}}%
      \expandafter\def\csname LTa\endcsname{\color{black}}%
      \expandafter\def\csname LT0\endcsname{\color{black}}%
      \expandafter\def\csname LT1\endcsname{\color{black}}%
      \expandafter\def\csname LT2\endcsname{\color{black}}%
      \expandafter\def\csname LT3\endcsname{\color{black}}%
      \expandafter\def\csname LT4\endcsname{\color{black}}%
      \expandafter\def\csname LT5\endcsname{\color{black}}%
      \expandafter\def\csname LT6\endcsname{\color{black}}%
      \expandafter\def\csname LT7\endcsname{\color{black}}%
      \expandafter\def\csname LT8\endcsname{\color{black}}%
    \fi
  \fi
    \setlength{\unitlength}{0.0500bp}%
    \ifx\gptboxheight\undefined%
      \newlength{\gptboxheight}%
      \newlength{\gptboxwidth}%
      \newsavebox{\gptboxtext}%
    \fi%
    \setlength{\fboxrule}{0.5pt}%
    \setlength{\fboxsep}{1pt}%
\begin{picture}(7200.00,5040.00)%
    \gplgaddtomacro\gplbacktext{%
      \csname LTb\endcsname%%
      \put(814,704){\makebox(0,0)[r]{\strut{}$0.1$}}%
      \csname LTb\endcsname%%
      \put(814,1951){\makebox(0,0)[r]{\strut{}$1$}}%
      \csname LTb\endcsname%%
      \put(814,3197){\makebox(0,0)[r]{\strut{}$10$}}%
      \csname LTb\endcsname%%
      \put(814,4444){\makebox(0,0)[r]{\strut{}$100$}}%
      \csname LTb\endcsname%%
      \put(946,484){\makebox(0,0){\strut{}$0.42$}}%
      \csname LTb\endcsname%%
      \put(1627,484){\makebox(0,0){\strut{}$0.47$}}%
      \csname LTb\endcsname%%
      \put(2308,484){\makebox(0,0){\strut{}$0.52$}}%
      \csname LTb\endcsname%%
      \put(2989,484){\makebox(0,0){\strut{}$0.57$}}%
      \csname LTb\endcsname%%
      \put(3670,484){\makebox(0,0){\strut{}$0.62$}}%
      \csname LTb\endcsname%%
      \put(4351,484){\makebox(0,0){\strut{}$0.67$}}%
      \csname LTb\endcsname%%
      \put(5032,484){\makebox(0,0){\strut{}$0.72$}}%
      \csname LTb\endcsname%%
      \put(5713,484){\makebox(0,0){\strut{}$0.77$}}%
      \csname LTb\endcsname%%
      \put(6394,484){\makebox(0,0){\strut{}$0.82$}}%
      \csname LTb\endcsname%%
      \put(5718,3686){\makebox(0,0)[l]{\strut{}}}%
    }%
    \gplgaddtomacro\gplfronttext{%
      \csname LTb\endcsname%%
      \put(198,2761){\rotatebox{-270}{\makebox(0,0){\strut{}$\beta^2\lambda^2$}}}%
      \put(3874,154){\makebox(0,0){\strut{}$\beta/L$}}%
      \csname LTb\endcsname%%
      \put(2398,1592){\makebox(0,0)[r]{\strut{}$\text{NUBS }i=1$}}%
      \csname LTb\endcsname%%
      \put(2398,1262){\makebox(0,0)[r]{\strut{}$\text{NUBS }i=2$}}%
      \csname LTb\endcsname%%
      \put(2398,932){\makebox(0,0)[r]{\strut{}$\text{LOC}$}}%
    }%
    \gplgaddtomacro\gplbacktext{%
      \csname LTb\endcsname%%
      \put(3174,1548){\makebox(0,0)[r]{\strut{}$20$}}%
      \csname LTb\endcsname%%
      \put(3174,2097){\makebox(0,0)[r]{\strut{}$20.5$}}%
      \csname LTb\endcsname%%
      \put(3174,2633){\makebox(0,0)[r]{\strut{}$21$}}%
      \csname LTb\endcsname%%
      \put(3174,3156){\makebox(0,0)[r]{\strut{}$21.5$}}%
      \csname LTb\endcsname%%
      \put(3306,1328){\makebox(0,0){\strut{}$0.8$}}%
      \csname LTb\endcsname%%
      \put(4299,1328){\makebox(0,0){\strut{}$0.82$}}%
      \csname LTb\endcsname%%
      \put(5291,1328){\makebox(0,0){\strut{}$0.84$}}%
    }%
    \gplgaddtomacro\gplfronttext{%
      \csname LTb\endcsname%%
      \put(3174,1548){\makebox(0,0)[r]{\strut{}$20$}}%
      \put(3174,2097){\makebox(0,0)[r]{\strut{}$20.5$}}%
      \put(3174,2633){\makebox(0,0)[r]{\strut{}$21$}}%
      \put(3174,3156){\makebox(0,0)[r]{\strut{}$21.5$}}%
      \put(3306,1328){\makebox(0,0){\strut{}$0.8$}}%
      \put(4299,1328){\makebox(0,0){\strut{}$0.82$}}%
      \put(5291,1328){\makebox(0,0){\strut{}$0.84$}}%
    }%
    \gplbacktext
    \put(0,0){\includegraphics{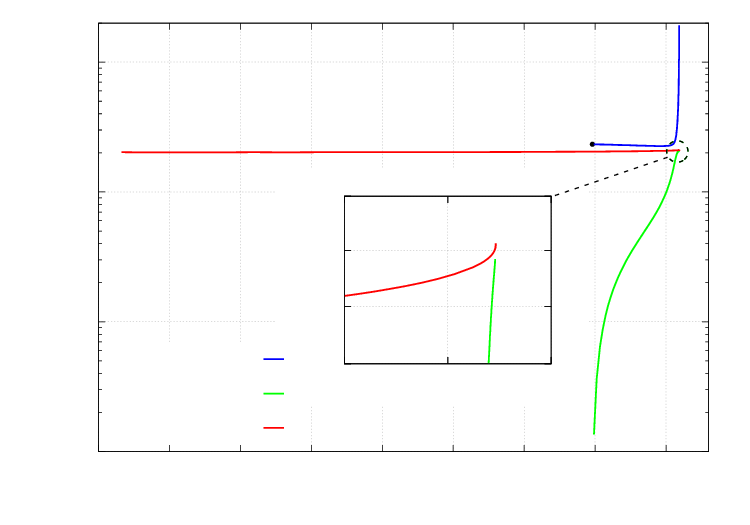}}%
    \gplfronttext
  \end{picture}%
\endgroup

\end{center}
\end{minipage}
\captionsetup{width=0.9\textwidth}
\captionof{figure}{\textsl{Absolute value of the negative modes of NUBS (green and blue lines) and LOC (red line) normalized with respect to the inverse temperature of the NUBS and LOC as a function of the dimensionless ratio $\beta/L$. The solid black discs show the zero mode and the negative mode of the UBS at the marginal GL point.}}
\label{fig:negmodes}
\end{center}

%~~~~~~~~~~~~~~~~~~~~~~~~~~~~~~~~~~~~~~~~~~~~~~~
\section{Implications for Super-Yang Mills on $\mathbb{T}^{\text{2}}$}
\label{sym}
%~~~~~~~~~~~~~~~~~~~~~~~~~~~~~~~~~~~~~~~~~~~~~~ 
In this section we (re)derive the thermodynamics of SYM on $\mathbb{T}^2$, see also \cite{Aharony:2004ig,Harmark:2004ws,Catterall:2010fx,Catterall:2017lub,Dias:2017uyv}, using the (neutral) KK black holes. We start with a lightening review of the different limits under which string theory can be described by its supergravity sector and then we use the solutions previously found to obtain the thermodynamics of those carrying D0-brane charge. Our results extend those in \cite{Dias:2017uyv} and allow us to find the merger point.

%~~~~~~~~~~~~~~~~~~~~~~~~~~~~~~~~~~~~~~~~~~~~~~~
\subsection{Toroidal limits and type IIB/IIA supergravity duals}
\label{sugralimits}
%~~~~~~~~~~~~~~~~~~~~~~~~~~~~~~~~~~~~~~~~~~~~~~
Consider (1+1)-dimensional SU$(N)$ SYM at large $N$ with 't Hooft coupling $\lambda = Ng^2_{\text{\tiny{YM}}}$. If the theory is at finite temperature $T$ so that $\beta = 1/T$ is the period of the thermal circle, and the spatial direction is also compactified on a circle of length $L$, then we can think of the theory as being defined on a 2-torus, $\mathbb{T}^2 = S^1_\beta\times S^1_L$. In these circumstances,  we can define dimensionless quantities, $t = TL$, $\lambda' = \lambda L^2$ to study different regimes of the theory. From the gauge theory perspective, phase transitions can be inferred by studying the expectation values of Wilson loops, $P_\beta$ and $ P_L$, wrapping the temporal and spatial circle respectively, which serve as order parameters. For SYM on $\mathbb{T}^2$ the expectation value $\la P_L\ra$ changes from zero (confined phase) to non-zero values (deconfined phase) upon heating the system, whereas $\la P_\beta\ra$ is always non-zero at all temperatures (see \cite{Aharony:2004ig,Catterall:2010fx,Catterall:2017lub} and references therein). 
 
Now we consider the dual gravity description. This is given by the near-horizon geometry of the spacetime sourced by a stack of $N$ coincident D1-branes of type IIB string theory \cite{PhysRevD.58.046004}, with a periodic identification on one spatial coordinate. In particular, one is interested in near-extremal black D1-brane configurations of the gravitational theory in the decoupling limit, which were studied in \cite{Aharony:2004ig}. The classical type IIB supergravity description is valid provided that $N$ is large, to suppress string quantum corrections; $\alpha'$-corrections are negligible when $t\ll \sqrt{\lambda'}$, while winding modes around the circle can be ignored when $t\gg 1/\sqrt{\lambda'}$. The two conditions imply the window of validity of IIB supergravity description \beq
\frac{1}{\sqrt{\lambda'}} \ll t \ll \sqrt{\lambda'}.
\enq
In this range one may use the type IIB supergravity solution to derive the thermodynamics of 2-dimensional SYM. In this regime, the thermal vacuum of type IIB supergravity is a black hole carrying D1-brane charge that uniformly wraps the compact circle. This solution is thought to be stable and corresponds to the uniform phase.

At temperatures $t\sim 1/\sqrt{\lambda'}$, stringy winding modes become unstable and the type IIB supergravity description is no longer valid. However, one can perform a T-duality transformation acting on the spatial circle, exchanging the theories IIB $\leftrightarrow$ IIA and hence the charges $D1\leftrightarrow D0$, so we can use type IIA black brane solutions to describe thermal states of SYM on $\mathbb{T}^2$ on that range of temperatures. In this case, the requirement that the supergravity solution is valid gives the conditions 
\beq\label{IIAok}
t \ll\sqrt{\lambda'}, \hs{0.75} t \ll \lambda'^{-1/6}.
\enq
Since D0-branes are point-like (instead of string-like, as the previous D1-branes), they can distribute the charge over the circle in various ways, either being uniformly or non-uniformly distributed, or localized on the compact circle. These three possibilities give rise to a non-trivial phase diagram, and it is then a dynamical question which case is preferred.

Ref.\ \cite{Aharony:2004ig} showed that the IIA supergravity solution with uniformly distributed D0-brane charge along the compact circle suffers a GL instability at the threshold temperature
\beq\label{gld0}
t_{\text{GL}} = \frac{3}{4\sqrt{\pi}}\frac{(2\pi a)^2}{\sqrt{\lambda'}},
\enq
where $a \equiv r_0^{\text{GL}}/\bar{L} = 0.36671(3)$, i.e.\ $t_{\text{GL}} = 2.24646(1/\sqrt{\lambda'})$. In the microcanonical ensemble the instability occurs at $\varepsilon_{\text{GL}} = 78.34939(N/\lambda')^2$.

For higher temperatures (or lower energies) the charged UBS is thought to be dynamically stable. However, there exists a range of temperatures $t_{\text{GL}}<t<t_{\text{PT}}$ for some $t_{\text{PT}}$, where it is thought (and now known) that the uniform solution becomes globally thermodynamically less favored than the localized black hole solution. Then the temperature $t_{\text{PT}}$ represents a first order thermal phase transition between the uniform and localized phase. In the literature this has been termed the Gregory-Laflamme phase transition \cite{Catterall:2010fx}. The natural interpretation of this picture on the dual gauge theory is a confinement/deconfinement phase transition.

In this section we find the temperature $t_{\text{PT}}$ and also $t_{\text{Merger}}$, at which the non-uniform and the localized phase merge. Note, however, that the non-uniform phase never dominates any ensemble. So far, lattice simulations on the gauge side estimated the ratio $t_{\text{PT}}/t_{\text{GL}} \sim 1.5$ \cite{Catterall:2010fx}. To determine the precise ratio from the gravity dual theory one would need to construct the near-extremal charged solutions, take the near-horizon limit and extract their thermodynamic quantities. Clearly, solving the full system of supergravity equations is a formidable numerical task. However, it is possible to generate charged solutions from uncharged ones via a process of uplifting + boosting + KK reduction \cite{Dias:2017uyv,1126-6708-2002-05-032,Harmark:2004ws,Aharony:2004ig}. Therefore, we can consider the neutral (vacuum) solutions we have previously found and from these obtain the thermodynamics that determine the phase structure of SYM theory under consideration.

Recent construction of KK black holes in $D = 10$ determined the energy or temperature to be $\varepsilon_{\text{PT}} = 97.067(N^2/{\lambda'}^2) = 1.245\varepsilon_{\text{GL}}$ or $t_{\text{PT}} = 2.451/\sqrt{\lambda'} = 1.093 t_{\text{GL}}$ \cite{Dias:2017uyv}. Since the derivation of the mapping \{Black hole thermodynamics on $\R^{1,8}\times S^1$\} $\riga$ \{SYM thermodynamics on $\mathbbm{T}^2$\} was derived in there we do not include it here; for completeness, it is rederived in detail in the appendix \ref{d0charge}. Using (\ref{SYMmap}), we have:
\beq\label{mapT2}
\varepsilon = 64\pi^4\(2m_0 - s_0t_0\)\frac{N^2}{\lambda'^2}, \hs{0.75} t = 4\pi\sqrt{2 s_0 t_0^3}\frac{1}{\sqrt{\lambda'}}, \hs{0.75} s = 16\sqrt{2}\pi^3\sqrt{\frac{s_0}{t_0}}\frac{N^2}{{\lambda'}^{3/2}}.
\enq
Applying this map to the neutral UBS one gets the well-known results \cite{Aharony:2004ig}:
\beq
\varepsilon_{\text{UBS}}(r_0) = \frac{32\pi^7}{3}\(\frac{r_0}{\bar{L}}\)^6\frac{N^2}{{\lambda'}^2}, \hs{0.75} t_{\text{UBS}}(r_0) = 3\pi^{3/2}\(\frac{r_0}{\bar{L}}\)^2 \frac{1}{\sqrt{\lambda'}}.
\enq
At $r_0 = r_0^{\text{GL}}$, these expressions correspond to the values $\varepsilon_{\text{GL}}$ and $t_{\text{GL}}$ previously discussed.

%~~~~~~~~~~~~~~~~~~~~~~~~~~~~~~~~~~~~~~~~~~~~~~~
\subsection{Thermodynamics}
\label{thermoD0}
%~~~~~~~~~~~~~~~~~~~~~~~~~~~~~~~~~~~~~~~~~~~~~~
In this subsection we construct the phase diagrams in the microcanonical and the canonical ensemble describing the thermodynamics of SYM gauge theory on $\mathbb{T}^2$ using the supergravity approximation. Both diagrams are shown below in Fig.\ \ref{microcanod0} and reproduce and complete those in \cite{Dias:2017uyv}. In addition, we determine for first time the merger point between charged NUBS and LOC.

In Fig.\ \ref{microcanod0} we plot dimensionless entropy or free energy difference between a given phase and the uniform one: $(s_i(\varepsilon)-s_{\text{UBS}}(\varepsilon))\times({\lambda'}^{3/2}/N^2)$, $(f_i(t)-f_{\text{UBS}}(t))\times({\lambda'}^2/N^2)$ with $i = \text{NUBS, LOC}$. Then the uniform phase is represented by a simple horizontal (black) line at the origin of the vertical axis. UBS are unstable for $\varepsilon < \varepsilon_{\text{GL}}$ in the microcanonical ensemble and for $t < t_{\text{GL}}$ in the canonical ensemble. The non-uniform phase, which exists beyond this point, is never dominant. On the other hand, the localized phase intercepts the uniform one at \beq
\varepsilon_{\text{PT}} = 97.29477\frac{N^2}{{\lambda'}^2}, \hs{0.25} \text{ or } \hs{0.25} t_{\text{PT}} = 2.45442\frac{1}{\sqrt{\lambda'}}.
\enq
The ratios are: $\varepsilon_{\text{PT}}/\varepsilon_{\text{GL}} = 1.24181$ and $t_{\text{PT}}/t_{\text{GL}} = 1.09257$, and they are consistent with the predictions from the studies of SYM on the lattice. For energies or temperatures greater than this value, the uniform phase is dominant, and the LOC dominate the corresponding ensemble otherwise. The phase transition is first order, and it is thought to correspond to a confinement/deconfinement phase transition in the gauge side.

Our results allows us to determine, for first time, the merger between localized black holes and non-uniform black strings with D0-brane charge. From Table \ref{tablecrit} we can read off $t_0^{\text{Merger}}$, $m_0^{\text{Merger}}$ and $s_0^{\text{Merger}}$, and using (\ref{mapT2}) one finds that the merger occurs at 
\beq\begin{split}
\varepsilon_{\text{Meger}}^{\text{NUBS}} &= 143.42647\frac{N^2}{{\lambda'}^2}, \hs{0.25} \text{ or } \hs{0.25} t_{\text{Merger}}^{\text{NUBS}} = 2.85934\frac{1}{\sqrt{\lambda'}}, \\
\varepsilon_{\text{Meger}}^{\text{LOC}} &= 143.41301\frac{N^2}{{\lambda'}^2}, \hs{0.25} \text{ or } \hs{0.25} t_{\text{Merger}}^{\text{LOC}} = 2.85949\frac{1}{\sqrt{\lambda'}}.
\end{split}\enq

Note that since the non-uniform phase never dominates any of the ensembles, it may be difficult to test these numbers using lattice simulations.

\begin{center}
\begin{minipage}{\textwidth}
\begin{minipage}[h!]{0.5\textwidth}
% GNUPLOT: LaTeX picture with Postscript
\begingroup
  \makeatletter
  \providecommand\color[2][]{%
    \GenericError{(gnuplot) \space\space\space\@spaces}{%
      Package color not loaded in conjunction with
      terminal option `colourtext'%
    }{See the gnuplot documentation for explanation.%
    }{Either use 'blacktext' in gnuplot or load the package
      color.sty in LaTeX.}%
    \renewcommand\color[2][]{}%
  }%
  \providecommand\includegraphics[2][]{%
    \GenericError{(gnuplot) \space\space\space\@spaces}{%
      Package graphicx or graphics not loaded%
    }{See the gnuplot documentation for explanation.%
    }{The gnuplot epslatex terminal needs graphicx.sty or graphics.sty.}%
    \renewcommand\includegraphics[2][]{}%
  }%
  \providecommand\rotatebox[2]{#2}%
  \@ifundefined{ifGPcolor}{%
    \newif\ifGPcolor
    \GPcolortrue
  }{}%
  \@ifundefined{ifGPblacktext}{%
    \newif\ifGPblacktext
    \GPblacktexttrue
  }{}%
  % define a \g@addto@macro without @ in the name:
  \let\gplgaddtomacro\g@addto@macro
  % define empty templates for all commands taking text:
  \gdef\gplbacktext{}%
  \gdef\gplfronttext{}%
  \makeatother
  \ifGPblacktext
    % no textcolor at all
    \def\colorrgb#1{}%
    \def\colorgray#1{}%
  \else
    % gray or color?
    \ifGPcolor
      \def\colorrgb#1{\color[rgb]{#1}}%
      \def\colorgray#1{\color[gray]{#1}}%
      \expandafter\def\csname LTw\endcsname{\color{white}}%
      \expandafter\def\csname LTb\endcsname{\color{black}}%
      \expandafter\def\csname LTa\endcsname{\color{black}}%
      \expandafter\def\csname LT0\endcsname{\color[rgb]{1,0,0}}%
      \expandafter\def\csname LT1\endcsname{\color[rgb]{0,1,0}}%
      \expandafter\def\csname LT2\endcsname{\color[rgb]{0,0,1}}%
      \expandafter\def\csname LT3\endcsname{\color[rgb]{1,0,1}}%
      \expandafter\def\csname LT4\endcsname{\color[rgb]{0,1,1}}%
      \expandafter\def\csname LT5\endcsname{\color[rgb]{1,1,0}}%
      \expandafter\def\csname LT6\endcsname{\color[rgb]{0,0,0}}%
      \expandafter\def\csname LT7\endcsname{\color[rgb]{1,0.3,0}}%
      \expandafter\def\csname LT8\endcsname{\color[rgb]{0.5,0.5,0.5}}%
    \else
      % gray
      \def\colorrgb#1{\color{black}}%
      \def\colorgray#1{\color[gray]{#1}}%
      \expandafter\def\csname LTw\endcsname{\color{white}}%
      \expandafter\def\csname LTb\endcsname{\color{black}}%
      \expandafter\def\csname LTa\endcsname{\color{black}}%
      \expandafter\def\csname LT0\endcsname{\color{black}}%
      \expandafter\def\csname LT1\endcsname{\color{black}}%
      \expandafter\def\csname LT2\endcsname{\color{black}}%
      \expandafter\def\csname LT3\endcsname{\color{black}}%
      \expandafter\def\csname LT4\endcsname{\color{black}}%
      \expandafter\def\csname LT5\endcsname{\color{black}}%
      \expandafter\def\csname LT6\endcsname{\color{black}}%
      \expandafter\def\csname LT7\endcsname{\color{black}}%
      \expandafter\def\csname LT8\endcsname{\color{black}}%
    \fi
  \fi
    \setlength{\unitlength}{0.0500bp}%
    \ifx\gptboxheight\undefined%
      \newlength{\gptboxheight}%
      \newlength{\gptboxwidth}%
      \newsavebox{\gptboxtext}%
    \fi%
    \setlength{\fboxrule}{0.5pt}%
    \setlength{\fboxsep}{1pt}%
\begin{picture}(5102.00,3400.00)%
    \gplgaddtomacro\gplbacktext{%
      \csname LTb\endcsname%%
      \put(946,704){\makebox(0,0)[r]{\strut{}$-0.8$}}%
      \csname LTb\endcsname%%
      \put(946,979){\makebox(0,0)[r]{\strut{}$-0.6$}}%
      \csname LTb\endcsname%%
      \put(946,1254){\makebox(0,0)[r]{\strut{}$-0.4$}}%
      \csname LTb\endcsname%%
      \put(946,1529){\makebox(0,0)[r]{\strut{}$-0.2$}}%
      \csname LTb\endcsname%%
      \put(946,1804){\makebox(0,0)[r]{\strut{}$0$}}%
      \csname LTb\endcsname%%
      \put(946,2079){\makebox(0,0)[r]{\strut{}$0.2$}}%
      \csname LTb\endcsname%%
      \put(946,2354){\makebox(0,0)[r]{\strut{}$0.4$}}%
      \csname LTb\endcsname%%
      \put(946,2629){\makebox(0,0)[r]{\strut{}$0.6$}}%
      \csname LTb\endcsname%%
      \put(946,2904){\makebox(0,0)[r]{\strut{}$0.8$}}%
      \csname LTb\endcsname%%
      \put(946,3179){\makebox(0,0)[r]{\strut{}$1$}}%
      \csname LTb\endcsname%%
      \put(1195,484){\makebox(0,0){\strut{}$0$}}%
      \csname LTb\endcsname%%
      \put(1663,484){\makebox(0,0){\strut{}$20$}}%
      \csname LTb\endcsname%%
      \put(2131,484){\makebox(0,0){\strut{}$40$}}%
      \csname LTb\endcsname%%
      \put(2599,484){\makebox(0,0){\strut{}$60$}}%
      \csname LTb\endcsname%%
      \put(3067,484){\makebox(0,0){\strut{}$80$}}%
      \csname LTb\endcsname%%
      \put(3535,484){\makebox(0,0){\strut{}$100$}}%
      \csname LTb\endcsname%%
      \put(4003,484){\makebox(0,0){\strut{}$120$}}%
      \csname LTb\endcsname%%
      \put(4471,484){\makebox(0,0){\strut{}$140$}}%
      \put(3059,1835){\makebox(0,0)[l]{\strut{}}}%
    }%
    \gplgaddtomacro\gplfronttext{%
      \csname LTb\endcsname%%
      \put(198,1941){\rotatebox{-270}{\makebox(0,0){\strut{}$\Delta s \times {\lambda'}^{3/2}/N^2$}}}%
      \put(2891,154){\makebox(0,0){\strut{}$\varepsilon \times (\lambda'/N)^2$}}%
      \csname LTb\endcsname%%
      \put(4114,2951){\makebox(0,0)[r]{\strut{}NUBS}}%
      \csname LTb\endcsname%%
      \put(4114,2621){\makebox(0,0)[r]{\strut{}LOC}}%
      \csname LTb\endcsname%%
      \put(4114,2291){\makebox(0,0)[r]{\strut{}UBS}}%
    }%
    \gplbacktext
    \put(0,0){\includegraphics{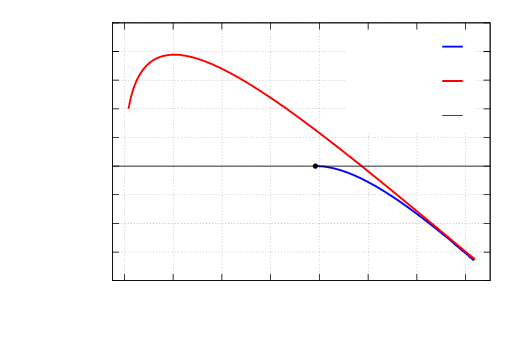}}%
    \gplfronttext
  \end{picture}%
\endgroup

\end{minipage}
\hfill
\begin{minipage}[h!]{0.5\textwidth}
% GNUPLOT: LaTeX picture with Postscript
\begingroup
  \makeatletter
  \providecommand\color[2][]{%
    \GenericError{(gnuplot) \space\space\space\@spaces}{%
      Package color not loaded in conjunction with
      terminal option `colourtext'%
    }{See the gnuplot documentation for explanation.%
    }{Either use 'blacktext' in gnuplot or load the package
      color.sty in LaTeX.}%
    \renewcommand\color[2][]{}%
  }%
  \providecommand\includegraphics[2][]{%
    \GenericError{(gnuplot) \space\space\space\@spaces}{%
      Package graphicx or graphics not loaded%
    }{See the gnuplot documentation for explanation.%
    }{The gnuplot epslatex terminal needs graphicx.sty or graphics.sty.}%
    \renewcommand\includegraphics[2][]{}%
  }%
  \providecommand\rotatebox[2]{#2}%
  \@ifundefined{ifGPcolor}{%
    \newif\ifGPcolor
    \GPcolortrue
  }{}%
  \@ifundefined{ifGPblacktext}{%
    \newif\ifGPblacktext
    \GPblacktexttrue
  }{}%
  % define a \g@addto@macro without @ in the name:
  \let\gplgaddtomacro\g@addto@macro
  % define empty templates for all commands taking text:
  \gdef\gplbacktext{}%
  \gdef\gplfronttext{}%
  \makeatother
  \ifGPblacktext
    % no textcolor at all
    \def\colorrgb#1{}%
    \def\colorgray#1{}%
  \else
    % gray or color?
    \ifGPcolor
      \def\colorrgb#1{\color[rgb]{#1}}%
      \def\colorgray#1{\color[gray]{#1}}%
      \expandafter\def\csname LTw\endcsname{\color{white}}%
      \expandafter\def\csname LTb\endcsname{\color{black}}%
      \expandafter\def\csname LTa\endcsname{\color{black}}%
      \expandafter\def\csname LT0\endcsname{\color[rgb]{1,0,0}}%
      \expandafter\def\csname LT1\endcsname{\color[rgb]{0,1,0}}%
      \expandafter\def\csname LT2\endcsname{\color[rgb]{0,0,1}}%
      \expandafter\def\csname LT3\endcsname{\color[rgb]{1,0,1}}%
      \expandafter\def\csname LT4\endcsname{\color[rgb]{0,1,1}}%
      \expandafter\def\csname LT5\endcsname{\color[rgb]{1,1,0}}%
      \expandafter\def\csname LT6\endcsname{\color[rgb]{0,0,0}}%
      \expandafter\def\csname LT7\endcsname{\color[rgb]{1,0.3,0}}%
      \expandafter\def\csname LT8\endcsname{\color[rgb]{0.5,0.5,0.5}}%
    \else
      % gray
      \def\colorrgb#1{\color{black}}%
      \def\colorgray#1{\color[gray]{#1}}%
      \expandafter\def\csname LTw\endcsname{\color{white}}%
      \expandafter\def\csname LTb\endcsname{\color{black}}%
      \expandafter\def\csname LTa\endcsname{\color{black}}%
      \expandafter\def\csname LT0\endcsname{\color{black}}%
      \expandafter\def\csname LT1\endcsname{\color{black}}%
      \expandafter\def\csname LT2\endcsname{\color{black}}%
      \expandafter\def\csname LT3\endcsname{\color{black}}%
      \expandafter\def\csname LT4\endcsname{\color{black}}%
      \expandafter\def\csname LT5\endcsname{\color{black}}%
      \expandafter\def\csname LT6\endcsname{\color{black}}%
      \expandafter\def\csname LT7\endcsname{\color{black}}%
      \expandafter\def\csname LT8\endcsname{\color{black}}%
    \fi
  \fi
    \setlength{\unitlength}{0.0500bp}%
    \ifx\gptboxheight\undefined%
      \newlength{\gptboxheight}%
      \newlength{\gptboxwidth}%
      \newsavebox{\gptboxtext}%
    \fi%
    \setlength{\fboxrule}{0.5pt}%
    \setlength{\fboxsep}{1pt}%
\begin{picture}(5102.00,3400.00)%
    \gplgaddtomacro\gplbacktext{%
      \csname LTb\endcsname%%
      \put(946,704){\makebox(0,0)[r]{\strut{}$-1.5$}}%
      \csname LTb\endcsname%%
      \put(946,1021){\makebox(0,0)[r]{\strut{}$-1$}}%
      \csname LTb\endcsname%%
      \put(946,1339){\makebox(0,0)[r]{\strut{}$-0.5$}}%
      \csname LTb\endcsname%%
      \put(946,1656){\makebox(0,0)[r]{\strut{}$0$}}%
      \csname LTb\endcsname%%
      \put(946,1973){\makebox(0,0)[r]{\strut{}$0.5$}}%
      \csname LTb\endcsname%%
      \put(946,2291){\makebox(0,0)[r]{\strut{}$1$}}%
      \csname LTb\endcsname%%
      \put(946,2608){\makebox(0,0)[r]{\strut{}$1.5$}}%
      \csname LTb\endcsname%%
      \put(946,2925){\makebox(0,0)[r]{\strut{}$2$}}%
      \csname LTb\endcsname%%
      \put(1078,484){\makebox(0,0){\strut{}$0.5$}}%
      \csname LTb\endcsname%%
      \put(1803,484){\makebox(0,0){\strut{}$1$}}%
      \csname LTb\endcsname%%
      \put(2529,484){\makebox(0,0){\strut{}$1.5$}}%
      \csname LTb\endcsname%%
      \put(3254,484){\makebox(0,0){\strut{}$2$}}%
      \csname LTb\endcsname%%
      \put(3980,484){\makebox(0,0){\strut{}$2.5$}}%
      \csname LTb\endcsname%%
      \put(4705,484){\makebox(0,0){\strut{}$3$}}%
      \put(3643,1687){\makebox(0,0)[l]{\strut{}}}%
    }%
    \gplgaddtomacro\gplfronttext{%
      \csname LTb\endcsname%%
      \put(198,1941){\rotatebox{-270}{\makebox(0,0){\strut{}$\Delta f \times (\lambda'/N)^2$}}}%
      \put(2891,154){\makebox(0,0){\strut{}$t \times \sqrt{\lambda'}$}}%
      \csname LTb\endcsname%%
      \put(2002,2951){\makebox(0,0)[r]{\strut{}NUBS}}%
      \csname LTb\endcsname%%
      \put(2002,2621){\makebox(0,0)[r]{\strut{}LOC}}%
      \csname LTb\endcsname%%
      \put(2002,2291){\makebox(0,0)[r]{\strut{}UBS}}%
    }%
    \gplbacktext
    \put(0,0){\includegraphics{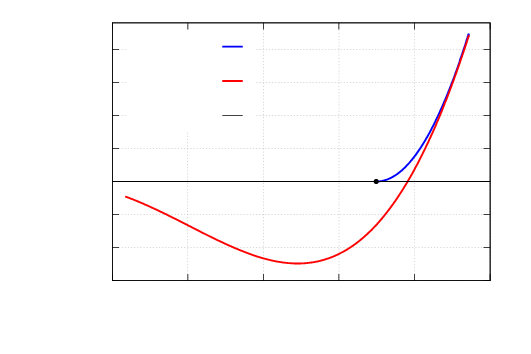}}%
    \gplfronttext
  \end{picture}%
\endgroup

\end{minipage}
\end{minipage}
\captionsetup{width=0.9\textwidth}
\captionof{figure}{\textsl{Phase diagram in the microcanonical ensemble (left) and canonical ensemble (right), of UBS, NUBS and LOC with D0-charge in the decoupling limit. The GL threshold point $\varepsilon_{\text{GL}}$ or $t_{\text{GL}}$ is indicated with a black solid disc.}}
\label{microcanod0}
\end{center}

%~~~~~~~~~~~~~~~~~~~~~~~~~~~~~~~~~~~~~~~~~~~~~~~
\section{Discussion and outlook}
\label{disc}
%~~~~~~~~~~~~~~~~~~~~~~~~~~~~~~~~~~~~~~~~~~~~~~
In this paper we have constructed NUBS and LOC in $D=10$ and followed these two branches very close to the merger point. $D=10$ is special from the point of view of this system since this is the critical dimension for the Ricci-flat double-cone metric that was conjectured to control the merger \cite{Kol:2002xz}. By fitting the physical quantities of both NUBS and LOC close to the merger point, we have shown that in $D=10$, their approach to their critical values is governed by a power law plus a logarithmic correction, in accordance to the double-cone model. This result should be contrasted to the results in $D=5,6$ obtained in \cite{Kalisch:2017bin}, which exhibit a spiraling behavior of the physical quantities as they approach their critical values. Moreover, we have found evidence that in $D=10$, the merger happens at cusp in the phase diagram, and physical quantities belonging to the NUBS and the LOC emerge from the critical point in the `same direction'. This feature should be related to the fact that $D=10$ is the critical dimension for the double-cone metric and it would be very interesting to understand it in detail. To further confirm the double-cone model of the merger, one should construct NUBS and LOC very close to the critical point in $D>10$ and verify that the physical quantities approach their critical values according to the predictions of the double-cone model. Work in this direction is underway. 

In this paper we have not discussed the dynamical stability of NUBS and LOC. Ref.\ \cite{Emparan:2015gva} considered the evolution of the GL instability of black strings in the large-$D$ expansion and showed that they settle on a stable NUBS. More recently, \cite{Emparan:2018bmi} included corrections beyond the leading order term in the large-$D$ expansion, and found that the endpoint depends on the thickness of the initial black string. For the cases where the black string is expected to pinch off, \cite{Emparan:2015gva} could not follow the evolution all the way to the end. It would be very interesting to study the evolution of the instability of uniform black strings for large yet finite $D$. The techniques used in \cite{Figueras:2017zwa} seem appropriate and we are currently investigating this problem.  

With the methods of \cite{Kalisch:2017bin} and those used in this paper, one can study the details of the mergers of other black hole systems of interest. In particular, for lumpy and localized black holes in AdS$_5\times S^5$. Moreover, recently \cite{Catterall:2017lub} obtained accurate results of the thermal phase diagram of $1+1$ SYM on a twisted torus using lattice simulations. It would be very interesting to compare their results using the supergravity approximation, but to do so one needs to consider NUBS that are electrically charged with respect to a 2-form.

%%%%%%%%%%%%%%%%%%%%%%%%%%%%%%%%%%%%%%%%%%%%%%%
\acknowledgments\addcontentsline{toc}{section}{Acknowledgements}
%%%%%%%%%%%%%%%%%%%%%%%%%%%%%%%%%%%%%%%%%%%%%%%
We would like to thank Toby Wiseman for enlightening discussions. PF would like to thank the Institute for Theoretical Physics (University of Amsterdam) for hospitality during the final stages of this work.  BC and PF are supported by the European Research Council grant ERC-2014-StG 639022-NewNGR \textit{``New frontiers in numerical general relativity''}. PF is also supported  by a Royal Society University Research Fellowship (Grant No. UF140319).

\appendix

%~~~~~~~~~~~~~~~~~~~~~~~~~~~~~~~~~~~~~~~~~~~~~~~
\section{Generic integration domain for localized solutions}
\label{intdom}
%~~~~~~~~~~~~~~~~~~~~~~~~~~~~~~~~~~~~~~~~~~~~~~
In this appendix we describe the integration domain we have used to construct the localized black holes. Due to their nature, the numerical construction involves to work in two separate coordinates systems, one adapted to the asymptotic behavior and another one adapted to the near horizon behavior.

\begin{figure}[h!]
\centering
\input{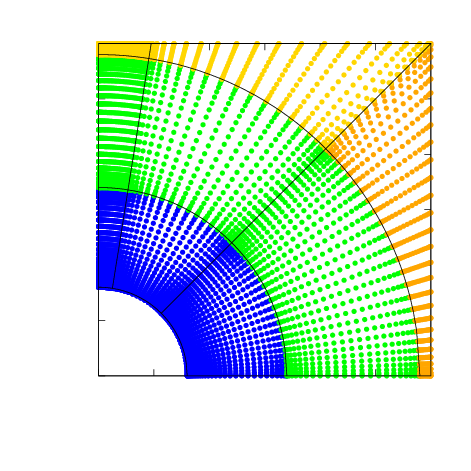}
\input{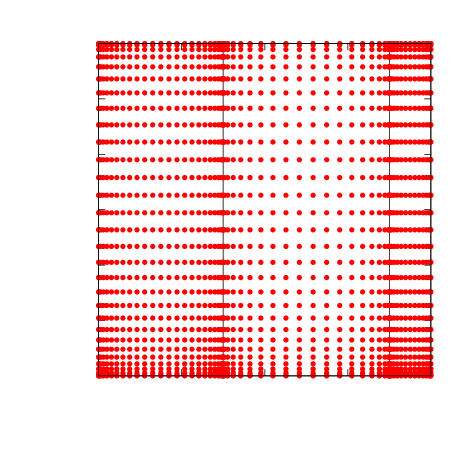}
\captionsetup{width=.9\linewidth}
\captionof{figure}{\textsl{Physical grid with parameters: $L = 6$, $r_0 = 0.8$, $r_* = 1.7$ and $r_1 = 2.9$, $a_* = 0.9\times (\pi/2)$, $\xi_* = -0.25$, $\xi_{**} = 0.75$, $\chi_1 = 2.5$ and $\chi_2 = 3$. Near patch (left) and far patch (right) in terms of the compactified $\xi$-coordinate.}}
\label{lattice}
\end{figure}

The construction considered in \cite{Kalisch:2017bin} takes the near chart ($r,a$) with five boundaries and divides it into three different subdomains. This encompasses the horizon, the axis, the boundaries of the internal space and a shared boundary with the far patch. The blue and green regions in Fig.\ \ref{lattice}, say region 1 in the near patch, are covered by polar coordinates $r\in[r_0,r_1]$, $a\in[0,\pi/2]$ whose relation with the far patch is simply given by $x(r,a) = r\cos a$, $y(r,a) = r\sin a$. The orange and yellow regions, say regions 2 and 3, are covered by polar-like coordinates with a modified radial coordinate which is parametrized in terms of $v\in[r_1,L/2]$. In the orange region $a\in[0,\pi/4]$, whereas in the yellow one $a\in[\pi/4,\pi/2]$. The precise relation with the far coordinates is: $x(v,a) = r_k(v,a)\cos a$, $y(v,a) = r_k(v,a)\sin a$, with \beq
r_k(v,a) = r_1\frac{L/2-v}{L/2-r_1} + \frac{L}{2}\,\frac{v-r_1}{L/2-r_1}\frac{1}{\delta_{k2}\,\sin a + \delta_{k3}\,\cos a}\,,\hs{0.75} k=2,3\,.
\enq
Here $r_0$ and $r_1$ ($ < L/2$) are parameters that we are free to specify, and $L$ is the asymptotic length of the Kaluza-Klein circle. Notice that this construction assumes that the angular coordinate in the near patch 1 is further divided into two subregions: one patch where $0\leq a \leq \pi/4$, to match the density of grid points with that in the near patch 2, and another one where $\pi/4\leq a \leq \pi/2$ to match the density of grid points with region 3.

The far chart $(x,y)$ covers the ranges $L/2 \leq x < \infty$ and $0\leq y\leq L/2$. To deal with the infinity, the coordinate $x$ is compactified introducing a new coordinate $-1\leq \xi \leq 1$. Ref.\ \cite{Kalisch:2017bin} considers $x(\xi) = L/(1-\xi)$, such that $\xi = -1$ corresponds to the shared boundary $x = L/2$ with the near patch and $\xi = 1$ corresponds to asymptotic infinity. The problem with this is that to find the charges $C_\tau$ and $C_y$, i.e.\ the mass and the tension, one needs to consider the asymptotic expansion of the metric components up to $(D-4)$th order, which implies to take $D-4$ derivatives. Of course, this is problematic for $D>5,6$. We overcome this issue by considering the compactification 
\beq
x(\xi) = \frac{L}{2}\(\frac{2}{1-\xi}\)^\Delta,
\enq
where $\Delta = (D-4)^{-1}$ has been defined in \S\ref{nubs}. This way we still have $x(\xi = -1) = L/2$ and $x(\xi = 1) = \infty$, but it is sufficient to consider the metric expansion at infinity up to 1st order. In particular, the charges are given by 1st derivatives of the metric, \beq\label{ctcyloc}
C_\tau = 2\(\frac{L}{2}\)^{D-5}\int_0^{L/2}\dd y\parcial[Q_1]{\xi}\bigg|_{\xi = 1}, \hs{0.75} C_y = -2\(\frac{L}{2}\)^{D-5}\int_0^{L/2}\dd y\parcial[Q_4]{\xi}\bigg|_{\xi = 1}.
\enq

Additionally, each patch has been further divided into other small subregions in order to be able to increase the grid resolution just where it is necessary. This is of particular interest since as we increase $D$ gravity turns out to be more localized and the spacetime region close to the horizon needs special care. Moreover, close to the merger point with NUBS, some functions develop steep gradients. In practice, the radial coordinate in the near patch 1 is divided into two subdomains, and the compactified coordinate $\xi$ in the far patch is divided into three subdomains. In the near patches containing the axis, the angular coordinate is also divided into two subregions. In total, this introduces four new parameters in the integration domain: $r_*$, $\xi_*$, $\xi_{**}$ and $a_*$, corresponding to the values where the different patches meet. At each shared boundary, either near-near, near-far or far-far patch, one must impose continuity of the functions and their first normal derivatives.

To impose these matching conditions one may consider the same grid point densities from both sides of a given shared boundary. Alternatively, one can still require continuity of the function and its normal derivative by performing the matching on an interpolation function. Unlike \cite{Kalisch:2017bin}, we have opted to work with the same grid point densities. They are naturally always the same except at the shared boundary between near and far patches. We fix this by considering the coordinate $y$ given in terms of a coordinate $\sigma$ lying in the unit interval $\sigma\in[0,1]$: \beq
y(\sigma) = \frac{L}{2}\tan\(\frac{\pi}{4}\sigma\).
\enq
Using Chebyshev grid points for $\sigma$, then the grid points along the $y$-direction are properly distributed.

Finally, we consider the same mesh-refinement as in \cite{Kalisch:2017bin}, properly modified to take into account our redefinition of the angle $a$, near the axis $\tilde{a} = \text{mesh}(a;\pi/2,a_*,\chi_1)$. We also use this type of mesh-refinement near the horizon $\tilde{r} = \text{mesh}(r;r_0,r_*,\chi_2)$, with the function mesh(\dots) given by \beq\label{meshh}
\text{mesh}(X; A,B,C) = A + \frac{B-A}{\sinh C}\sinh\(C\frac{X-A}{B-A}\).
\enq

To check whether our code with the described modifications gives rise to reasonable solutions and, in particular, accurate values for the mass, we compare the obtained results, (i) for small localized black holes, with the mass of a Schwarzschild black hole in $D = 10$, or (ii) with the perturbative results. For small localized black holes one expects that the spacetime metric can be systematically expanded in a perturbation series with a small parameter $\rho_0/L$, being $\rho_0$ the location of the horizon. The best available perturbative approximation for the thermodynamic quantities, with $D$ arbitrary, are given in \cite{Harmark:2003yz}. (We did not include these curves in our plots in \S \ref{thermo0} or \S\ref{thermoD0} since they were not much clarifying.) For small enough black holes, i.e.\ with eccentricity $\epsilon < 10^{-3}$, our numerical values differ by less than a $0.05\%$ when compared to those obtained by (i) or (ii). From the geometrical point of view, another check is to compare $L_{\text{polar}}$ defined in (\ref{Lss}), with one half of the perimeter of a Schwarzschild black hole of the same temperature in $D = 10$. In this case the deviations are always less than $0.01\%$.

%~~~~~~~~~~~~~~~~~~~~~~~~~~~~~~~~~~~~~~~~~~~~~~~
\section{Convergence tests}
\label{convtest}
\begin{figure}[h!]
\centering
\input{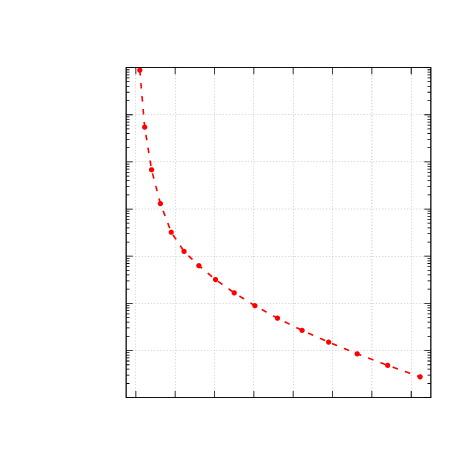}
\input{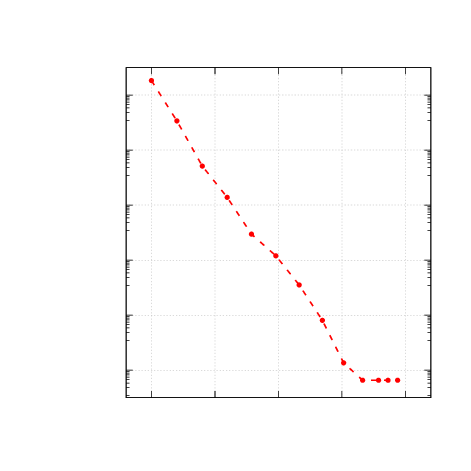}
\captionsetup{width=.9\linewidth}
\captionof{figure}{\textsl{Logarithmic plot of $\xi^2_{\text{max}} = \text{max}[\xi^2]$ as a function of the grid size. In both cases, the error decays exponentially, as expected.}}
\label{fig:convergence}
\end{figure}

%~~~~~~~~~~~~~~~~~~~~~~~~~~~~~~~~~~~~~~~~~~~~~~~
In this appendix we check that our numerical solutions converge to the continuum limit according to our discretization scheme. In the case of using pseudo-spectral methods, the error should be exponentially suppressed with increasing the grid size. To monitor it we use the squared norm of the DeTurck vector $\xi^2$. We expect it to become zero in the continuum limit. Indeed, Fig.\ \ref{fig:convergence} shows that our numerical implementation exhibits the expected behavior. 

To produce this figure we picked up a reference solution of each branch, we interpolated it at different resolutions and then we filtered through the Newton-Raphson loop. For each output we computed the quantity of interest. For NUBS we just considered one single patch with resolution $N = N_xN_y$, being $N_x$ and $N_y$ the number of grid points in each direction. In the case of LOC, we considered the usual 12 patches and varied the mean resolution $\bar{N}$.

%~~~~~~~~~~~~~~~~~~~~~~~~~~~~~~~~~~~~~~~~~~~~~~~
\section{D0-charge via Uplifting + Boosting + KK reduction}
\label{d0charge}
%~~~~~~~~~~~~~~~~~~~~~~~~~~~~~~~~~~~~~~~~~~~~~~
In this appendix we derive the mapping between the thermodynamics of neutral Kaluza-Klein black solutions and the thermodynamics of near-extremal D0-black branes on a circle of type IIA supergravity \cite{Dias:2017uyv,1126-6708-2002-05-032,Harmark:2004ws,Aharony:2004ig}. This involves a M-theory lift-boost-reduce procedure.

Consider any static, axially symmetric metric solving $R_{ab} = 0$ in $D$ spacetime dimensions and approaching the direct product manifold $\R^{1,D-2}\times \bar{S}^1$ asymptotically. The bar notation in $\bar{S}^1$ (of length $\bar{L}$) is to distinguish, in $D = 10$, the T-dual circle of type IIA supergravity from the original circle of length $L$ of the type IIB theory. This solution can be written using isotropic coordinates: \beq\label{initmetric}
\dd s^2 = -f^2\dd t^2 + g^2\(\dd \rho^2 + \rho^2\dd\Omega_{D-3}^2\) + h^2\dd y^2,
\enq
with the generic functions $f, g$ and $h$ approaching 1 at $\rho\riga\infty$, and $y \sim y+\bar{L}$ is the coordinate of $\bar{S}^1$. If (\ref{initmetric}) is a black hole with a Killing horizon located at $\rho = \rho_0$, then $f(\rho_0,y) = 0$. We can construct the dimensionless quantity $p_0\equiv\rho_0/\bar{L}$ to label a given family of such metrics.

Now uplift the solution adding a compact coordinate $z$, $\dd \hat{s}_{D+1}^2 = \dd s^2+\dd z^2$. Boosting along $z$ with rapidity parameter $\alpha$ yields a solution to vacuum general relativity in $D+1$ dimensions. Upon dimensional reduction with respect to $z$ we rewrite the metric as \beq
\dd \hat{s}_{D+1}^2 = e^{-2\eta\phi}\dd s^2_{D} + e^{2\zeta\phi}\(\dd z - A_t\dd t\)^2,
\enq
with $\eta^2 = (2(D-1)(D-2))^{-1}$ and $\zeta = (D-2)\eta$. This choice of constants ensures an Einstein framed dimensionally reduced action and canonical normalization for the dilaton kinetic term. For $D = 10$, this procedure allows us to construct solutions of type IIA supergravity with only graviton, dilaton and 1-form field excitations turned on.

The new metric, the non-trivial dilaton field and the 1-form gauge field are identified to be:
\beq\begin{split}\label{chargedmetric}
\dd s^2_{D} &= H^{\frac{2}{D-2}}\(-\frac{f^2}{H^2}\dd t^2 + g^2\(\dd\rho^2 + \rho^2\dd\Omega_{D-3}^2\) + h^2\dd y^2\), \\
e^{\phi} &= H^{1/\zeta}, \\
A_{(1)} &= \(H^{-2}-1\)\coth\alpha\hs{0.05} \dd t\, ,
\end{split}\enq
where \beq
H = \sqrt{1+ (1- f^2)\sinh^2\alpha}\, .
\enq
This works because momentum around the circle in the $(D+1)$th dimension is reinterpreted as D0-brane charge from the lower-dimensional viewpoint \cite{Harmark:2004ws}.

We can start with (\ref{initmetric}) being a neutral non-uniform black string or localized black hole, and obtain the charged solution using (\ref{chargedmetric}) (which depends on the parameter $\alpha$ in addition to $\rho_0$). Since we are interested in their thermodynamics rather than the solutions themselves, we will proceed by expressing the quantities of interest of the new charged solutions in terms of the uncharged ones. For instance, it easy to see that the temperature and the entropy of the charged solution are simply shifted by a factor of $\cosh\alpha$ with respect to the uncharged ones. To be precise,
\beq\label{tscharg}
T = \frac{1}{\bar{L}\cosh\alpha}t(p_0), \hs{0.75} S = \frac{1}{4G_D}\bar{L}^{D-2}\Omega_{D-3}\cosh\alpha\, s(p_0),
\enq
where $t(p_0)$, $s(p_0)$ encode the parametric dependence of dimensionless temperature and entropy of neutral solutions. The mass and the charge can be obtained from the asymptotic expansion of the metric and the gauge field. Because (\ref{initmetric}) is asymptotically KK, for large $\rho$ we may expand \beq
f(\rho,y) \simeq 1 - c_t(p_0)\frac{\bar{L}^{D-4}}{\rho^{D-4}}, \hs{0.75}  h(\rho,y) \simeq 1+c_y(p_0)\frac{\bar{L}^{D-4}}{\rho^{D-4}}.
\enq
Taking the square and considering the factors of $H$, one obtains the effective charges for the charged solution (\ref{chargedmetric}), $C_t(p_0)$ and $C_y(p_0)$, entering in the expression (\ref{masstension}). The D0-charge may be obtained from the flux or from the asymptotic behavior of the gauge field, and the chemical potential is given by $\mu = -A_t\big|_H$. The result is:
\beq\begin{split}\label{mqcharg}
M &= \frac{\bar{L}^{D-3}\Omega_{D-3}}{8\pi G_D}\((D-3)c_t(p_0) - c_y(p_0) + (D-4)c_t(p_0)\sinh^2\alpha\), \\
Q &= \frac{\bar{L}^{D-3}\Omega_{D-3}}{8\pi G_D}(D-4)c_t(p_0)\sinh\alpha\cosh\alpha, \\
\mu &= \tanh\alpha.
\end{split}\enq

Now set $D = 10$. The derived quantities so far correspond to the thermodynamic quantities of the charged solution (\ref{chargedmetric}) of type IIA supergravity with a metric, a dilaton and a 1-form. At this point note that all physical quantities appearing in (\ref{tscharg}) and (\ref{mqcharg}), reduce to those of the uncharged solution in the limit $\alpha \riga 0$. The opposite limit $\alpha\riga\infty$ corresponds to take the near-extremal limit. To obtain the desired mapping we have to take the decoupling limit of Ref.\ \cite{PhysRevD.58.046004} of near-extremal configurations, which sends the string length $\ell_s$ to zero while keeping $g^2_{\text{\tiny YM}}$ fixed. According to holography, these are dual to the decoupled field theory at finite temperature. To do so, one needs the relation between the 10-dimensional Newton's constant and the string length and coupling constant, $16\pi G_{10} = (2\pi)^7\bar{g}_s^2\ell_s^8$, the T-dual relations \cite{polchinski_1998}: $\bar{L} = (2\pi\ell_s)^2/L$, $\bar{g}_s = (2\pi\ell_s/L)g_s$, and the relation between the string coupling constant and SYM coupling. In the case of IIB string theory with D1-branes, this is $g_{\text{\tiny YM}}^2 = (2\pi)^{p-2}g_s\ell_s^{p-3}$ with $p = 1$ \cite{PhysRevD.58.046004}; to translate back the quantities above in terms of SYM variables recall that $\lambda' = \lambda L^2 = Ng_{\text{\tiny YM}}^2L^2$.

The dimensionless energy above extremality, $\varepsilon = L\mc{E} = L(M - Q)$, in these limits corresponds to the energy density of the SYM theory. Since this becomes independent of $\ell_s$, the decoupling limit is trivial. The decoupling limit of the temperature and the entropy needs to be taken with more care. At the end, one finds that the dimensionless energy, temperature and entropy associated to a stack of $N$-coincident near-extremal D0-branes in the decoupling limit are: \beq\begin{split}\label{symdim}
\varepsilon &= \frac{16}{3}\pi^7\(4c_t(p_0) - c_y(p_0)\)\frac{N^2}{{\lambda'}^2}, \hs{0.75} t = 2\pi^{5/2}\sqrt{2c_t(p_0)}t(p_0)\frac{1}{\sqrt{\lambda'}}, \\
s &= \frac{16}{3}\pi^{11/2}\frac{s(p_0)}{\sqrt{2c_t(p_0)}}\frac{N^2}{{\lambda'}^{3/2}}.
\end{split}\enq

These expressions explicitly depend on functions that can be obtained from the KK vacuum solution (\ref{initmetric}). Clearly, the numerical solutions found in \S\ref{numconst} are not written in isotropic coordinates which difficult the computation of such functions. It is then convenient to write (\ref{symdim}) in terms of gravitational variables which are intrinsic of the solution instead of the coordinates. To this end, set $\alpha = 0$ in (\ref{tscharg}) and (\ref{mqcharg}) and solve for the functions $t(p_0), s(p_0), c_t(p_0)$ and $c_y(p_0)$. The Smarr's relation closes the system of equations. The solution is: \beq\begin{split}
t(p_0) &= t_0, \hs{2.79} s(p_0) = \frac{4s_0}{\Omega_{D-3}}, \\
c_t(p_0) &= \frac{8\pi}{\Omega_{D-3}}\frac{t_0 s_0}{D-4}, \hs{0.75} c_y(p_0) = \frac{8\pi}{\Omega_{D-3}}\(\frac{D-3}{D-4}t_0 s_0 - m_0\), \\
\end{split}\enq
where $t_0 = \bar{L}T$, $m_0 = M/\bar{L}^{D-3}$ and $s_0 = S/\bar{L}^{D-2}$ are the dimensionless temperature, mass and entropy of the neutral gravity solutions. Setting $D = 10$ and inserting these expressions back into (\ref{symdim}), gives the final mapping: \beq\label{SYMmap}
\varepsilon = 64\pi^4\(2m_0 - s_0t_0\)\frac{N^2}{{\lambda'}^2}, \hs{0.75} t = 4\pi\sqrt{2 s_0 t_0^3}\frac{1}{\sqrt{\lambda'}}, \hs{0.75} s = 16\sqrt{2}\pi^3\sqrt{\frac{s_0}{t_0}}\frac{N^2}{{\lambda'}^{3/2}}.
\enq

%-------------------------------------------------------
% Bibliography
\addcontentsline{toc}{section}{References}
%-------------------------------------------------------

\bibliography{refs}
\bibliographystyle{jhep}

\end{document}